%% file: manuscript.tex
\documentclass[ 
reprint,
 amsmath,amssymb,
 aps,
 pra,
floatfix,superscriptaddress
]{revtex4-2}

\usepackage{orcidlink}
\usepackage{subfiles}
\usepackage{tikz}
\usetikzlibrary{decorations.pathmorphing,patterns}

\usepackage{pgfplots}
\pgfplotsset{compat=1.17} 
\usepgfplotslibrary{dateplot}

\usepackage{caption}
\usepackage{subcaption}

\usepackage{graphicx}
\usepackage{dcolumn}
\usepackage{amsmath,amsfonts}
\usepackage{bm}

\newcommand{\crosslinkerV}[5]{
\begin{scope}
    \clip (#1+0.5,#2-0.5) rectangle (#3-0.5,#4+0.5);
    \draw[ultra thick] (#1,#2-0.5) circle (0.5);
    \draw[ultra thick] (#3,#4+0.5) circle (0.5);
\end{scope}    
\draw[thick,decoration={aspect=0.3, segment length=#5,amplitude=6pt,coil, pre length = 1 pt, post length = 1 pt},
    decorate] (#1,#2) -- (#3,#4);
}
\newcommand{\crosslinkerH}[5]{
\begin{scope}
    \clip (#1-0.5,#2-0.5) rectangle (#3+0.5,#4+0.5);
    \draw[ultra thick] (#1-0.5,#2) circle (0.5);
    \draw[ultra thick] (#3+0.5,#4) circle (0.5);
\end{scope}
\draw[thick, decoration={aspect=0.3, segment length=#5,amplitude=6pt,coil, pre length = 1 pt, post length = 1 pt},
    decorate] (#1,#2) -- (#3,#4);
}

\newcommand{\SUPhys}{Department of Physics, Stellenbosch University, Stellenbosch 7602, South Africa}
\newcommand{\NITheCS}{National Institute for Theoretical and Computational Sciences, Stellenbosch 7602, South Africa}
\newcommand{\UKZN}{School of Chemistry \& Physics, University of KwaZulu-Natal, Pietermaritzburg, Scottsville 3209, South Africa}
\newcommand{\Messina}{Dipartimento di Scienze Biomediche, Odontoiatriche e delle Immagini Morfologiche e Funzionali,
Università degli Studi di Messina, I-98125 Messina, Italy}

\begin{document}

\title{Dynamical Networking of Polymer Networks with Dedicated Cross-linker Particles}

\author{Nadine du Toit \orcidlink{0000-0003-0262-7010}}
\email{Corresponding author: 24461989@sun.ac.za}
\affiliation{\SUPhys}
 
\author{Kristian K. M\"uller-Nedebock \orcidlink{0000-0002-1772-1504}}
 \email{kkmn@sun.ac.za}
\affiliation{\SUPhys}
\affiliation{\NITheCS}
\author{Giuseppe Pellicane \orcidlink{0000-0002-3805-830X}}
\email{gpellicane@unime.it}
\affiliation{\NITheCS}
\affiliation{\UKZN}
\affiliation{\Messina}

\date{\today}

\begin{abstract}
This paper extends a field-theoretical dynamical networking formalism for mesoscopic polymer dynamics to explicitly include dedicated cross-linker particles. Cross-linkers are represented within a Martin–Siggia–Rose generating functional and reversibly coupled to polymers through Gaussian networking fields, enabling an approximation scheme that reduces their degrees of freedom while remaining compatible with polymer dynamics. The framework is applied to a two-species polymer system in which intra- and inter-species cross-linking are assigned different statistical advantages. Effective networking potentials are derived and used to calculate correlation functions and dynamic structure factors. To validate these results, molecular dynamics simulations of semi-flexible polymers with reversible intra- and inter-species cross-linking are performed. Simulations show that cross-linking decreases polymer persistence lengths and local alignment, and the resulting trajectories yield dynamic structure factors consistent with theoretical predictions. In both approaches, cross-linking broadens the diffusive peaks and enhances the high-frequency tails of the structure factors. Together, theory and simulation provide complementary insights into the dynamics of cross-linked polymers, establishing a tractable framework that captures essential features observed in experiments and offering a basis for exploring more complex synthetic and biological networks.
\end{abstract}

\maketitle

\section{Introduction}
\label{sec:intro}

Dynamic polymer networks underpin the behaviour of both synthetic gels and biological structures such as the cytoskeleton, where cross-linking strongly influences viscoelastic and dynamical properties. Understanding these effects at the mesoscopic scale is challenging: analytical models are often intractable, while simulations are computationally intensive. The present work addresses the dynamics associated with reversible cross-linking and the resulting network formation, rather than hydrodynamic or entanglement effects (see, for example \cite{mateyisiInfluenceWeakReversible2018a}).

This paper builds on a theoretical formalism presented in Ref.~\cite{dutoitDynamicalNetworkingUsing2025}, in which Gaussian fields are used to model and introduce reversible cross-linking in a polymer mixture. The approach requires introducing a networking functional, which intermittently constrains particle positions to be equal to one another, into a Martin-Siggia-Rose generating functional describing the collective dynamics of solutions of flexible polymers. This results in dynamic structure factors for a polymer mixture wherein the networking formalism acts as effective interaction potentials between the polymers. In Ref.~\cite{dutoitDynamicalNetworkingUsing2025}, the networking formalism has been shown to result in tractable analytical expressions with adjusted diffusive behaviour in both an example system of Brownian particles and a solution of flexible polymers. The development of the formalism was motivated by the need for a deeper understanding of the effects of cross-linking in both synthetic and biological polymer networks. In both these types of systems, cross-linking can occur not only via direct bonding of the monomers of polymers to one another, \textit{i.e.} chemical cross-linking, but also via physical cross-linking where another molecule or protein  bonds to one monomer on each side to create a bridge between the polymers \cite{gennesScalingConceptsPolymer1991}.  The present work therefore extends the previous theoretical modelling approach by introducing cross-linker particles with their own associated dynamics and applying the dynamical network formalism to create reversible bonds between the ends of the cross-linker particles and the polymers.

Apart from theoretical modelling, significant progress has been made in characterising the dynamics of polymer networks through advanced experimental and computational techniques. Super-resolution imaging and neutron scattering have provided insight into the spatio-temporal organisation of cytoskeletal components \cite{finkenstaedt-quinnSuperresolutionImagingMonitoring2016,chaubetDynamicActinCrosslinking2020,aomuraQuasielasticNeutronScattering2023}, while simulations continue to reveal the effects of polymer flexibility, topology, and confinement \cite{garduno-juarezMolecularDynamicSimulations2024}. Thus, the present work not only expands upon the theoretical modelling approach, but also presents small-scale molecular dynamics simulations of a comparable system for validation and comparison. The simulations presented here utilise open source tools such as LAMMPS \cite{thompsonLAMMPSFlexibleSimulation2022}, REACTER \cite{gissingerREACTERHeuristicMethod2020a} and Dynasor \cite{franssonDynasorToolExtracting2021} in order to obtain dynamic structure factors of a polymer network with reversible cross-linking between polymers and cross-linker particles. Qualitative trends observed in both the analytical model and simulations are also discussed in the context of recent experimental work presented in Ref.~\cite{aomuraQuasielasticNeutronScattering2023}. The present work provides a tractable analytical framework for cross-linked polymer networks with dedicated cross-linker particles and is validated against molecular dynamics simulations and recent experimental work.

The paper begins by introducing the dynamics of generic cross-linker particles and applying the networking formalism to create reversible bonds between cross-linkers and polymers in Sec.~\ref{sec:CL-Bnetworking}. This allows the introduction of relevant degrees of freedom and approximation schemes in preparation for the main analytical model presented in Sec.~\ref{sec:polymerCL}. This describes a two species polymer model with dynamical networking applied to implement both intra- and inter-species cross-linking between polymers and cross-linker particles. Molecular dynamics simulations of a comparable system are presented in Sec.~\ref{sec:MD}, with details of the modelling approach discussed in Sec.~\ref{sec:LAMMPSmodel} and results presented in Sec.~\ref{sec:MDresults}.

\section{Cross-linkers: dynamics and networking}
\begin{figure}[ht]
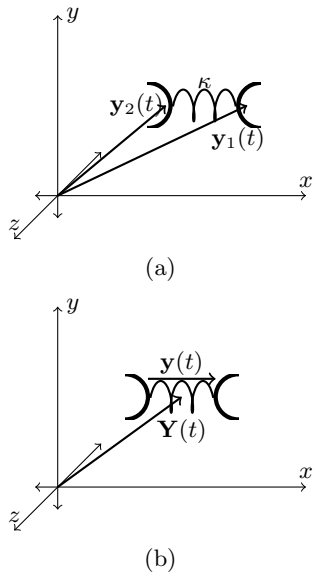

    \centering
    \begin{subfigure}[b]{0.45\textwidth}
        \centering
        \subfile{./diagrams/crosslinkeranatomy}
        \caption{}
        \label{fig:crosslinkeranatomy}
    \end{subfigure}
    \hfill
    \begin{subfigure}[b]{0.45\textwidth}
        \centering
        \subfile{./diagrams/crosslinkerCOM}
        \caption{}
        \label{fig:crosslinkerCOM}
    \end{subfigure}

    \caption[Schematic diagrams of cross-linkers]{Schematic diagrams of cross-linkers with spring constant $\kappa$:  
(a) represented by endpoint coordinates $\mathbf{y}_1(t)$ and $\mathbf{y}_2(t)$;  
(b) represented by centre of mass $\mathbf{Y}(t)$ and extension $\mathbf{y}(t)$.}
    \label{fig:crosslinkerCombined}
\end{figure}

In order to dynamically cross-link two freely diffusing particles to one another, we will introduce a third \textit{mediator} particle. This mediator particle will have two endpoints that are able to attach to the beads, with its endpoints being joined by a spring as depicted in Fig. \ref{fig:crosslinkeranatomy}.

The eventual goal is to utilise this cross-linker to join polymer chains to one another, by allowing the endpoints to attach to polymer chains as the cross-linker diffuses. The model of the dynamics of the cross-linker will be outlined first, deferring the mechanism of attachment of the cross-linker end-points to Sec.~\ref{sec:CL-Bnetworking}. Thus, we may write down the following coupled Langevin equations for the endpoints of the cross-linker:
\begin{eqnarray}
    -\gamma_y \dot{\mathbf{y}}_1 (t)  - \kappa \left(\mathbf{y}_1(t) - \mathbf{y}_2(t)\right) + \mathbf{f}_1(t) =0\, ,\\
     -\gamma_y \dot{\mathbf{y}}_2 (t)  + \kappa \left(\mathbf{y}_1(t) - \mathbf{y}_2(t)\right) + \mathbf{f}_2(t)=0\, .
\end{eqnarray}
where $\gamma_y$ is the drag coefficient for each of the endpoints and $\mathbf{f}_1(t)$ and $\mathbf{f}_2(t)$ are the stochastic forces acting on each of the endpoints. Both stochastic forces are Gaussian correlated with strength $\lambda_y$. 

The cross-linker may perhaps be better expressed using centre of mass coordinates $\mathbf{y}(t) = \mathbf{y}_1(t)- \mathbf{y}_2(t)$ and $\mathbf{Y}(t) = \frac{1}{2}(\mathbf{y}_1(t)+ \mathbf{y}_2(t))$ as depicted in Fig. \ref{fig:crosslinkerCOM}. With this, the coupled Langevin equations become:
\begin{eqnarray}
    -\gamma_y \dot{\mathbf{ y}} (t)  - 2\kappa \mathbf{y}(t)+ \mathbf{f}_1(t)-\mathbf{f}_2(t)=0\, ,\label{eq:crosslinkerLangeviny}\\
     -\gamma_y \dot{\mathbf{Y}}(t)  + \frac{1}{2}\left(\mathbf{f}_1(t)+ \mathbf{f}_2(t)\right)=0\, .
     \label{eq:crosslinkerLangevinY}
\end{eqnarray}
Eqs. \eqref{eq:crosslinkerLangevinY} and \eqref{eq:crosslinkerLangeviny} can be rewritten in the Martin-Siggia-Rose \cite{Martin:1973zz} formalism as a generating functional as follows:
\begin{multline}
    \mathbb{Z} = \mathcal{N}\int [\mathrm{d} \mathbf{y}][\mathrm{d} \mathbf{Y}][\mathrm{d} \hat{\mathbf{y}}][\mathrm{d} \hat{\mathbf{Y}}][\mathrm{d} \mathbf{f}_1][\mathrm{d} \mathbf{f}_2]\mathrm{e}^{- \frac{1}{2 \lambda_y} \int_t \left( |\mathbf{f}_1(t)|^2 + |\mathbf{f}_2(t)|^2\right) }    \\
    \times \mathrm{e}^{-\mathrm{i} \int_t \hat{\mathbf{y}} \cdot \left( -\gamma_y \dot{\mathbf{y}} (t)  - 2\kappa \mathbf{y}(t)+ \mathbf{f}_1(t)-\mathbf{f}_2(t)\right) - \mathrm{i} \int_t \hat{\mathbf{Y}} \cdot \left( -\gamma_y \dot{\mathbf{Y}}(t)  + \frac{1}{2}\left(\mathbf{f}_1(t)+ \mathbf{f}_2(t)\right)\right)}\, .
    \label{eq:CL_MSR1}
\end{multline}
Implementing the Gaussian functional integrals over both stochastic forces yields
\begin{multline}
    \mathbb{Z} = \mathcal{N}\int [\mathrm{d} \mathbf{y}][\mathrm{d} \mathbf{Y}][\mathrm{d} \hat{\mathbf{y}}][\mathrm{d} \hat{\mathbf{Y}}]\mathrm{e}^{-\mathrm{i} \int_t \hat{\mathbf{y}} \cdot \left( -\gamma_y \dot{\mathbf{y}} (t)  - 2\kappa \mathbf{y}(t) - \lambda_y \hat{\mathbf{y}}(t)\right)} \\ 
    \times \mathrm{e}^{- \mathrm{i} \int_t \hat{\mathbf{Y}} \cdot \left( -\gamma_y \dot{\mathbf{Y}}(t)  + \frac{\lambda_y}{4}\hat{\mathbf{Y}}(t)\right) } \, .
    \label{eq:CL_MSR2}
\end{multline}

\subsection{Applying the networking formalism to cross-linkers}
\label{sec:CL-Bnetworking}
\begin{figure}
     \centering
    \begin{tikzpicture}
        \draw  node[fill= blue,circle,scale=2] at (6,0) {};
        \draw  node[fill= blue,circle,scale=2] at (6,2) {};
        \draw  node[fill= blue,circle,scale=2] at (6,4) {};
        \draw  node[fill= blue,circle,scale=2] at (6,6) {};
        
        \draw  node[fill= blue,circle,scale=2] at (4,0) {};
        \draw  node[fill= blue,circle,scale=2] at (4,2) {};
        \draw  node[fill= blue,circle,scale=2] at (4,4) {};
        \draw  node[fill= blue,circle,scale=2] at (4,6) {};
        
        \draw  node[fill= blue,circle,scale=2] at (2,0) {};
        \draw  node[fill= blue,circle,scale=2] at (2,2) {};
        \draw  node[fill= blue,circle,scale=2] at (2,4) {};
        \draw  node[fill= blue,circle,scale=2] at (2,6) {};
        
        \draw  node[fill= blue,circle,scale=2] at (0,0) {};
        \draw  node[fill= blue,circle,scale=2] at (0,2) {};
        \draw  node[fill= blue,circle,scale=2] at (0,4) {};
        \draw  node[fill= blue,circle,scale=2] at (0,6) {};
        
        \crosslinkerH{4.5}{0}{5.5}{2}{20pt};
        \crosslinkerH{2.5}{2}{3.5}{2}{8pt};
        \crosslinkerH{4.5}{6}{5.5}{6}{8pt};
        
        \crosslinkerV{0}{0.5}{0}{1.5}{8pt};
        \crosslinkerV{0}{4.5}{0}{5.5}{8pt};
        \crosslinkerV{4}{4.5}{2}{5.5}{20pt};

    \end{tikzpicture}
    
     \caption[Schematic diagram of networking with cross-linkers]{A diagram of $M=16$ blue beads  with positions $\mathbf{r}_1,\mathbf{r}_2,\mathbf{r}_3, ... ,\mathbf{r}_M$, being cross-linked to one another  by $N=6$ cross-linkers. The diagram depicts an instantaneous snapshot of this dynamic process, at one of the timesteps $t=t_j \,, \,j\, \epsilon \,\mathbb{Z}$ at which networking is required by eq.~\eqref{eq:Qcl_b}.}
     \label{fig:cl-beads}
 \end{figure}
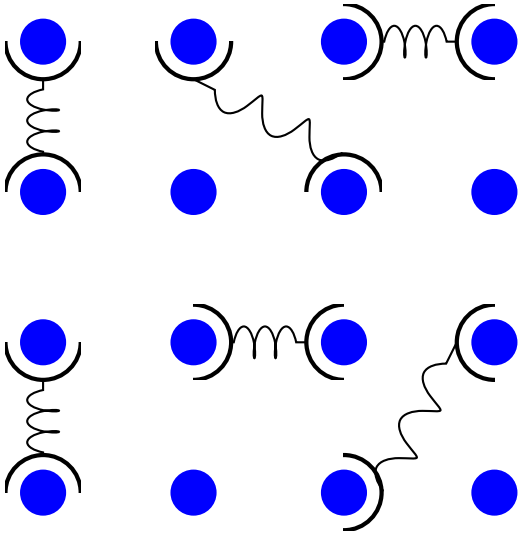

Following the approach developed in Ref.~\cite{dutoitDynamicalNetworkingUsing2025}, one can construct a networking functional which, at incremental time-steps, networks the end points of $N$ cross-linkers to some other objects. Let there be $M$ objects  with positions $\mathbf{r}_1,\mathbf{r}_2,\mathbf{r}_3, ... ,\mathbf{r}_M$ to which the cross-linker endpoints may attach. For the moment, these objects can be thought of as stationary beads, as depicted in Fig.~\ref{fig:cl-beads}.  The networking functional at discrete time steps $t_j$ may be written as a product over $j$ as follows:

\begin{multline}
Q = \mathcal{N} \prod_{j} \int[\mathrm{d} \Phi][\mathrm{d} \Phi^*]\, \left( \prod_{n=1}^N \Phi(\mathbf{y}_{1,n},t_j)\Phi(\mathbf{y}_{2,n},t_j) \right)  \\
\times\left( \prod_{m=1}^{M} (1+\Phi^*(\mathbf{r}_{m},t_j)) \right) \, \mathrm{e}^{- \frac{1}{\ell}\int_r \Phi(\mathbf{r},t_j) \Phi^*(\mathbf{r}, t_j)}
\label{eq:Qcl_b}
\end{multline}
where $\mathbf{y}_{1,n} (t) = \mathbf{Y}_{n} (t)+ \frac{1}{2} \mathbf{y}_{n} (t)  $ and $\mathbf{y}_{2,n} (t) = \mathbf{Y}_{n} (t)- \frac{1}{2} \mathbf{y}_{n} (t)  $ are the positions of the cross-linker endpoints. This networking functional requires that both ends of each of the $N$ cross-linkers are attached to one of the $M$ objects at all discrete time steps. This means that there must be at least as many cross-linker endpoints as there are objects for them to attach to, \textit{i.e.} $M\geq 2 N$ in order for this expression to hold. Assuming that this is the case, all $M$ cross-linkers will be networked to the beads, as in Fig. \ref{fig:cl-beads},instantaneously at all times $t=t_j \,\forall \,j\, \epsilon \,\mathbb{Z}$,  whilst diffusing unconstrained at all other times. In this way, the cross-linkers are intermittently constrained to attach to two objects, allowing them to rearrange and bind to different objects throughout their diffusion, while remaining in the vicinity of the objects and unable to diffuse away entirely. 

Collective variables 
\begin{equation}
    C(\mathbf{r},t)= \sum_{m=1}^M \delta (\mathbf{r}-\mathbf{r}_m(t))
\end{equation}
and
\begin{equation}
    \rho(\mathbf{Y}, \mathbf{y},t) = \sum_{n=1}^N \delta ( \mathbf{Y} - \mathbf{Y}_n (t))\delta (\mathbf{y}- \mathbf{y}_n(t))
    \label{eq:CLrhodef}
\end{equation}
for the $M$ objects and $N$ cross-linkers, respectively, may be introduced into the networking functional.
Since $\rho(\mathbf{Y}, \mathbf{y},t) $ is a probability distribution rather than a physical density, we may define a physical density for the cross-linkers. This density represents the number of cross-linker endpoints at a position $z$ and time $t$
\begin{equation}
P(\mathbf{z},t) = \int\mathrm{d}\mathbf{y} \left( \rho (\mathbf{z}- \frac{ \mathbf{y}}{2},\mathbf{y},t) +\rho (\mathbf{z}+ \frac{ \mathbf{y}}{2},\mathbf{y},t)\right)\, . 
\label{eq:Pdef}
\end{equation}
Now noting that
\begin{equation}
P(\mathbf{z},t) = \sum_{i=1}^N \left(\delta \left(  \mathbf{z}- \tfrac{\mathbf{y}_i(t)}{2}- \mathbf{Y}_i(t)\right)+ \delta \left(\mathbf{z}+ \tfrac{\mathbf{y}_i(t)}{2}- \mathbf{Y}_i(t)\right)\right)\, , 
\label{eq:PitoSums}
\end{equation}
we can use this new density to exponentiate the product over $n$. This results in the following expression for the networking functional:

\begin{multline}
Q = \mathcal{N}  \int[\mathrm{d} \Phi][\mathrm{d} \Phi^*]\, \mathrm{e}^{\frac{1}{\tau} \int_{\mathbf{r},t} P(\mathbf{r},t)  \mathbf{ln} \left( \Phi(r,t) \right) +\frac{1}{\tau} \int_{\mathbf{r},t} C(\mathbf{r},t) \mathbf{ln}(1+\Phi^*(\mathbf{r},t)) }\\
\times \mathrm{e}^{- \alpha \int_{\mathbf{r},t} \Phi(\mathbf{r},t) \Phi^*(\mathbf{r}, t)}\, .
\label{eq:CLnetworkingitoP}
\end{multline}
In this case, the form of the networking functional is the same as for diffusing beads and stationary attachment points presented in Ref.~\cite{dutoitDynamicalNetworkingUsing2025}, as one would expect. 

\begin{equation}
    P(\mathbf{z},t) = \sum_{m=1}^M \left(2 \,\delta \left(  \mathbf{z}- \mathbf{Y}_m(t)\right)+ 2 \mathbf{y}^2_m(t)\,\delta ''\left(2\mathbf{z}- 2\mathbf{Y}_m(t)\right)\right)\, .
\label{eq:PapproxitoSums}
\end{equation}
Let $\rho_0(z,t)$ denote the density of cross-linker centres of mass, \textit{i.e.} $\rho(z,0,t)$ then this can be re-written as
    \begin{equation}
    P(\mathbf{z},t) = 2 \,\rho_0(z,t)+\sum_{m=1}^M \left( \mathbf{y}^2_m(t)\,\delta ''\left(\mathbf{z}- \mathbf{Y}_m(t)\right)\right)\, .
\label{eq:Papproxitorho0}
\end{equation}
If the cross-linker extension is assumed to be short, \textit{i.e.} $\mathbf{y}_m(t)\approx 0$, then $ P(\mathbf{z},t) \approx 2 \,\rho_0(z,t)$, which can be used as a lowest order approximation. Alternatively, $\mathbf{y}^2_m(t)$ can be replaced with its average  $\langle \mathbf{y}^2_m(t) \rangle$ thereby pre-averaging over the cross-linker extension. To obtain this average, recall the following result of the Random Phase Approximation (RPA)  (eq.~\eqref{eq:CLRPA-A}) 
    \begin{equation}
    \mathcal{A}_{\mathbf{K}, \mathbf{k}}(t,t') = N \mathrm{e}^{-\frac{\lambda_y}{4 \gamma_y^2 }|\mathbf{K}|^2(t-t') -\frac{\lambda_y}{\kappa \gamma_y }|\mathbf{k}|^2\left(1 + \coth{\left( \tfrac{\kappa}{\gamma_y}(t-t') \right)}\right)}\, \,.
    \label{eq:CLRPA-A2}
\end{equation}
Setting $\mathbf{K}=0$ and $t'=t_0$, this can be viewed as a probability distribution for the Fourier transform of the spring extension length as a function of time
\begin{equation}
    \mathbb{P}_\mathbf{k}(t)  \propto \mathcal{A}_{\mathbf{0}, \mathbf{k}}(t,t_0) = N \mathrm{e}^{-\frac{\lambda_y}{\kappa \gamma_y }|\mathbf{k}|^2\left(1 + \coth{\left( \tfrac{\kappa}{\gamma_y}(t-t_0) \right)}\right)}
    \label{eq:extensionProb}
\end{equation}
such that the average of $\mathbf{y}^2_m(t)$ is given by the second moment of $\mathbb{P}_\mathbf{k}(t)$, \textit{i.e.}
\begin{equation}
    \langle\mathbf{y}^2_m(t) \rangle = - \left. \frac{\mathrm{d}^2 \mathbb{P}_\mathbf{k}(t) }{\mathrm{d}\mathbf{k}^2}\right|_ {\mathbf{k}=\mathbf{0}}. 
\end{equation}
The only time dependence in eq.~\eqref{eq:extensionProb},  lies within the argument of the hyperbolic cotangent. Consequently the probability distribution decays to a steady state value when $t-t_0 \gg \tfrac{\kappa}{\gamma_y}$. Keeping in mind the eventual goal of utilising this networking formalism to create cross-links between polymer chains, it is reasonable to assume that the dynamics of the chains, which are far larger molecules than the cross-linker particles, will occur on timescales longer than $\tfrac{\kappa}{\gamma_y}$. Thus, the steady state average $\langle \mathbf{y}^2_m(t) \rangle_{t \rightarrow \infty}$ will be used to approximate the spring extension. With this the density of cross-linker end-points can be written as
\begin{align}
    P(\mathbf{z},t) &= 2 \,\rho_0(z,t)+ \sum_{m=1}^M \left(\langle \mathbf{y}^2_m(t) \rangle_{t \rightarrow \infty} \,\,\, \,\delta ''\left(\mathbf{z}- \mathbf{Y}_m(t)\right)\right) \\
    &=  2 \,\rho_0(z,t)+ \frac{\lambda}{\kappa \gamma_y} \rho''_0(z,t) \, 
\label{eq:Papproxitorho0-1}
\end{align}
and substituted into eq.~\eqref{eq:CLnetworkingitoP}, to obtain a pre-averaging approximation for the cross-linker extension.

\section{Cross-linking two species of polymers}
\label{sec:polymerCL}

\begin{figure*}[ht]
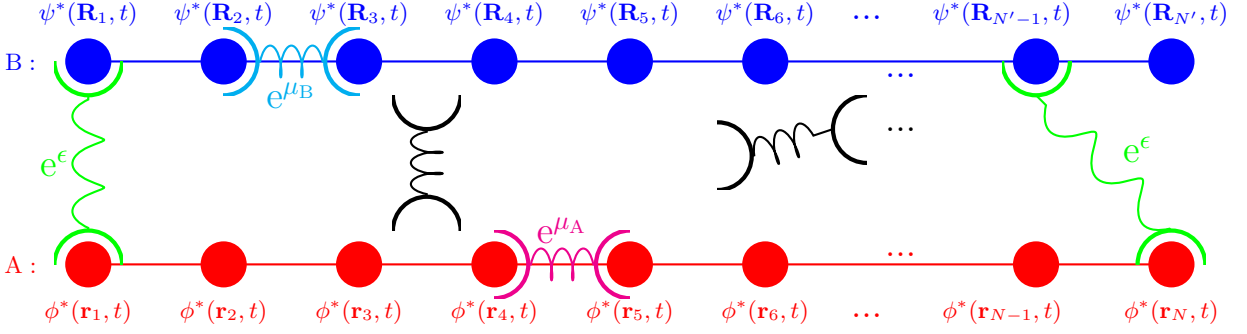

\subfile{./diagrams/crosslinker_tracksDiagram}
\caption[Schematic diagram of intra- and inter-species cross-linking]{Schematic diagram of cross-linkers diffusing between two sets of attachment points, those in red with positions $\mathrm{A}=\{\mathbf{r}_1,\mathbf{r}_2, \,...\,\mathbf{r}_N\}$ and  in blue with positions $\mathrm{B}=\{\mathbf{R}_1,\mathbf{R}_2, \,...\,\mathbf{R}_{N'}\}$. Freely Diffusing cross-linkers are depicted in black. Cross-links between an $\mathbf{r}_n$ and an $\mathbf{R}_{n'}$ are depicted in green and are assigned an advantage of $\mathrm{e}^\epsilon$.  Cross-links between two $\mathbf{r}_n$'s (in magenta) are assigned $\mathrm{e}^{\mu_\mathrm{A}}$ whilst $\mathrm{e}^{\mu_\mathrm{B}}$ is assigned to cross-links between two $\mathbf{R}_{n'}$'s (in cyan), respectively.}
\label{fig:crosslinker&tracks}
\end{figure*}

Drawing the attention of the reader towards Fig. \ref{fig:crosslinker&tracks}, consider two sets of attachment points with positions $\mathrm{A}=\{\mathbf{r}_1,\mathbf{r}_2, \,...\,,\mathbf{r}_N\}$ and  $\mathrm{B}=\{\mathbf{R}_1,\mathbf{R}_2, \,...\,,\mathbf{R}_{N'}\}$. For the moment, it is simplest to think of these points as stationary positions along two straight lines, as depicted. These coordinates are placeholders for two polymer chains, of which the dynamics will be dealt with separately. In addition, the system contains $M$ cross-linkers, with the position of each given in terms of its centre of mass position vector $\mathbf{Y}_m(t) $ and  end-to-end vector $\mathbf{y}_m (t)$, as defined previously.

As can also be seen in Fig. \ref{fig:crosslinker&tracks}, an advantage of $\mathrm{e}^\epsilon$, $\mathrm{e}^{\mu_\mathrm{A}}$ or $\mathrm{e}^{\mu_\mathrm{B}}$ is assigned to each networked cross-linker. A networked cross-linker being, in this system, a cross-linker which is doubly networked, \textit{i.e.} networked at \textit{both} of its endpoints. The different advantages are assigned, depending onto which pairs of attachment points the cross-linker networks. In order to introduce these advantages into a suitable networking expression, two sets of networking fields $\phi$ and $\phi^*$ and also $\psi$ and $\psi^*$  may be utilised. The field $\phi^*$ is associated with points in $\mathrm{A}$ whilst the field $\psi^*$ is associated with points in $\mathrm{B}$, allowing one to use the fields $\phi$ and $\psi$ to model the attachment of the cross-linker ends to points in $\mathrm{A}$ and $\mathrm{B}$ as follows:
\begin{widetext}
\begin{multline}
    \mathbb{Q}= \int [ \mathrm{d}\phi] [\mathrm{d} \phi^*][ \mathrm{d}\psi] [\mathrm{d} \psi^*]\,\, \prod_j 
\,  \left( \prod_{n=1}^{N} (1 +\phi^*(\mathbf{r}_n,t_j')) \,\prod_{n'=1}^{N'} (1+\psi^*(\mathbf{R}_{n'},t_j))\right.\\
\left.\times \,\prod_{m=1}^M \left(1+\phi\left(\mathbf{y}_{m,1},t_j\right)\psi\left(\mathbf{y}_{m,2} ,t_j\right)\mathrm{e}^\epsilon+\phi\left(\mathbf{y}_{m,1},t_j\right)\psi\left(\mathbf{y}_{m,2},t_j\right)\mathrm{e}^\epsilon +\phi\left(\mathbf{y}_{m,1},t_j\right)\phi\left(\mathbf{y}_{m,2},t_j\right)\mathrm{e}^{\mu_\mathrm{A}}+\psi\left(\mathbf{y}_{m,1},t_j\right)\psi\left(\mathbf{y}_{m,2},t_j\right)\mathrm{e}^{\mu_\mathrm{B}}\right)\right.\\
\left.\times \mathrm{e}^{-  \frac{1}{ \ell}\int_\mathbf{r} \, \phi(\mathbf{r},t_j)\,\phi^*(\mathbf{r},t_j)-  \frac{1}{ \ell}\int_\mathbf{r} \, \psi(\mathbf{r},t_j)\,\psi^*(\mathbf{r},t_j)} \right)\, .\label{eq:Qfullintrainter}
\end{multline}
\end{widetext}
where $\mathbf{y}_{m,1} =\mathbf{Y}_m-\frac{\mathbf{y}_m}{2}$ and $\mathbf{y}_{m,2} =\mathbf{Y}_m+\frac{\mathbf{y}_m}{2}$ are the positions of the end-points of each cross-linker. The networking advantages can be used to determine the number of each type of cross-link that is formed by taking derivatives as follows:
\begin{subequations}
\label{eq:cl-tAves}
\begin{equation}
z_\mathrm{AB} = \frac{1}{\mathbb{Q}}
  \frac{\partial \ln{\mathbb{Q}}}{\partial \epsilon}
\end{equation}
\begin{equation}
z_\mathrm{AA} = \frac{1}{\mathbb{Q}}
  \frac{\partial \ln{\mathbb{Q}}}{\partial \mu_\mathrm{A}}
\end{equation}
\begin{equation}
z_\mathrm{BB} = \frac{1}{\mathbb{Q}}
  \frac{\partial \ln{\mathbb{Q}}}{\partial \mu_\mathrm{B}}
\end{equation}
\end{subequations}

Here, \(z_\mathrm{AB}\), \(z_\mathrm{AA}\), and \(z_\mathrm{BB}\) denote the number of
cross-links between an \(\mathbf{r}_n\) and \(\mathbf{R}_{n'}\),
between an \(\mathbf{r}_n\) and \(\mathbf{r}_{n'}\),
and between an \(\mathbf{R}_n\) and \(\mathbf{R}_{n'}\), respectively. Once again, collective coordinates can be introduced and the networking expression can be written in its continuous form. As before, the collective variable for the cross-linker positions is $\rho(\mathbf{Y}, \mathbf{y},t)$. Introducing 
\begin{subequations}
    \begin{equation}
        C_\mathrm{A}(\mathbf{r},t)=\sum_{n=1}^{N} \delta(\mathbf{r}-\mathbf{r}_n (t)) \mathrm{\,\,\,\,    and} 
    \end{equation}
    \begin{equation}
        C_\mathrm{B}(\mathbf{r},t)= \sum_{n'=1}^{N'} \delta(\mathbf{r}-\mathbf{R}_{n'} (t))
    \end{equation}
    \label{eq:CAandCBdef}
\end{subequations}
for the positions in $\mathrm{A}=\{\mathbf{r}_1,\mathbf{r}_2, \,...\,,\mathbf{r}_N\}$  and $\mathrm{B}=\{\mathbf{R}_1,\mathbf{R}_2, \,...\,,\mathbf{R}_{N'}\}$, respectively.

Assume, for the moment, that the length of each of the cross-linkers is small such that we may approximate $\mathbf{y}_{m,1}=\mathbf{y}_{m,2}=\mathbf{Y}_{m}$ for all $M$ cross-linkers. Then the networking functional may be written as:
\begin{equation}
    \mathbb{Q}= \int [ \mathrm{d}\phi] [\mathrm{d} \phi^*][ \mathrm{d}\psi] [\mathrm{d} \psi^*]\,\, 
\mathrm{e}^{\mathbb{F}[\phi, \phi^*, \psi, \psi^*]}
\label{eq:Qcl-tracks}
\end{equation}
with 
\begin{widetext}
\begin{multline}
    \mathbb{F}[\phi, \phi^*, \psi, \psi^*]={ \frac{1}{\tau}\int_{\mathbf{r},t} C_\mathrm{A}(\mathbf{r},t)\ln(1 +\phi^*(\mathbf{r},t)) \,+ \frac{1}{\tau}\int_{\mathbf{r},t}C_\mathrm{B}(\mathbf{r},t)\ln(1+\psi^*(\mathbf{r},t))}\\
+\frac{1}{\tau}\int_{\mathbf{Y},t}\rho(\mathbf{Y}, \mathbf{0},t) \ln \left(1+\phi(\mathbf{Y},t)\psi(\mathbf{Y} ,t)\mathrm{e}^\epsilon+\phi(\mathbf{Y},t)\psi(\mathbf{Y},t)\mathrm{e}^\epsilon +\phi(\mathbf{Y},t)\phi(\mathbf{Y},t)\mathrm{e}^{\mu_\mathrm{A}}+\psi(\mathbf{Y},t)\psi(\mathbf{Y},t)\mathrm{e}^{\mu_\mathrm{B}}\right)\\
-  \alpha\int_\mathbf{r} \, \phi(\mathbf{r},t_j)\,\phi^*(\mathbf{r},t_j)-  \alpha\int_\mathbf{r} \, \psi(\mathbf{r},t)\,\psi^*(\mathbf{r},t) \, .
\label{eq:Fcl-t}
\end{multline}
\end{widetext}
Following the sequence of calculations, as discussed in Ref.~\cite{dutoitDynamicalNetworkingUsing2025}, the next step is to implement a saddle point approximation. In order to find a suitable set of solutions $\{\bar{\phi}, \bar{\phi}^*,\bar{\psi}, \bar{\psi}^*\}$, one needs to solve the following set of simultaneous equations
\begin{subequations}
    \label{eq:SPidentities_cl-t}
    \begin{equation}
       0=\left.\frac{\delta\mathbb{F}}{\delta\phi(\mathbf{r},t)} \right|_{\bar{\phi},\bar{\phi}^*,\bar{\psi},\bar{\psi}^* }  = - \alpha \bar{\phi}(\mathbf{r},t) + \frac{1}{\tau}\frac{C_\mathrm{A}(\mathbf{r},t)}{1 + \bar{\phi}^*(\mathbf{z},t)} 
    \end{equation}
    \begin{equation}
        0=\left.\frac{\delta\mathbb{F}}{\delta\psi^*(\mathbf{r},t)} \right|_{\bar{\phi},\bar{\phi}^*,\bar{\psi},\bar{\psi}^* } = - \alpha \bar{\psi}(\mathbf{r},t) + \frac{1}{\tau}\frac{C_\mathrm{B}(\mathbf{r},t)}{ 1+\bar{\psi}^*(\mathbf{r},t)}\, ,
    \end{equation}
    \begin{equation}
         \!\!\!\!0=\frac{\delta\mathbb{F}}{\delta\psi(\mathbf{r},t)} \Big|_{\bar{\phi},\bar{\phi}^*,\bar{\psi},\bar{\psi}^* }  \!\!\!\!= - \alpha \bar{\phi}^*(\mathbf{r},t) + \tfrac{2\rho(\mathbf{r},\mathbf{0},t)\left(\bar{\psi}(\mathbf{r},t)\mathrm{e}^\epsilon+ \bar{\phi}(\mathbf{r},t)\mathrm{e}^{\mu_\mathrm{A}}\right)}{\tau \mathbb{A}[\bar{\phi}, \bar{\psi}]}\,, 
    \end{equation}
    \begin{equation}
         \!\!\!\!0=\frac{\delta\mathbb{F}}{\delta\psi^*(\mathbf{r},t)} \Big|_{\bar{\phi},\bar{\phi}^*,\bar{\psi},\bar{\psi}^* }  \!\!\!\!= - \alpha \bar{\psi}^*(\mathbf{r},t) + \tfrac{2\rho(\mathbf{r},\mathbf{0},t)\left(\bar{\phi}(\mathbf{r},t)\mathrm{e}^\epsilon+ \bar{\psi}(\mathbf{r},t)\mathrm{e}^{\mu_\mathrm{B}}\right)}{\tau \mathbb{A}[\bar{\phi}, \bar{\psi}]}\,
    \end{equation}
\end{subequations}
where
\begin{equation}
    \mathbb{A}[\bar{\phi}, \bar{\psi}]=\left(1+2\bar{\phi}(\mathbf{r},t)\bar{\psi(\mathbf{r},t)\mathrm{e}^\epsilon}+ \bar{\phi}^2(\mathbf{r},t)\mathrm{e}^{\mu_\mathrm{A}}+ \bar{\psi}^2(\mathbf{r},t)\mathrm{e}^{\mu_\mathrm{B}}\right)\,,
\end{equation}
gives the ratio of all free cross-linkers to those that are networked at both ends. This interpretation of $\mathbb{A}$ may be deduced since the manipulation of eqs. \eqref{eq:SPidentities_cl-t} allows one to rewrite this expression as
\begin{equation}
    \mathbb{A}[\bar{\phi}, \bar{\psi}]=\frac{\rho(\mathbf{r},\mathbf{0},t)}{\rho(\mathbf{r},\mathbf{0},t) - \frac{\alpha \tau}{2}(\bar{\phi}(\mathbf{r},t)\bar{\phi}^*(\mathbf{r},t)+\bar{\psi}(\mathbf{r},t)\bar{\psi}^*(\mathbf{r},t))}\, ,
\end{equation}
where
$\frac{\alpha \tau}{2}(\bar{\phi}(\mathbf{r},t)\bar{\phi}^*(\mathbf{r},t)+\bar{\psi}(\mathbf{r},t)\bar{\psi}^*(\mathbf{r},t))$ counts the total instances of networking that occurs at position $\mathbf{r}$ at time $t$.

Solving eqs. \eqref{eq:SPidentities_cl-t} for $\bar{\phi},\bar{\phi}^*,\bar{\psi}$ and $\bar{\psi}^*$, the saddle point approximation to eq. \eqref{eq:Qcl-tracks}, is given by
\begin{equation}
    \mathbb{Q}= \mathcal{N}\mathrm{e}^{\mathbb{F}[\bar{\phi}, \bar{\phi}^*,\bar{\psi}, \bar{\psi}^*]}.
    \label{eq:QSP_cl-t}
\end{equation}

The full solution of the coupled saddle point equations in eqs.~\eqref{eq:SPidentities_cl-t} is analytically intractable. To obtain tractable expressions, approximations for intra-species and inter-species cross-linking will be considered in Secs.~\ref{sec:intra} -- \ref{sec:inter}.

\subsection{Derivation of networking potentials}
\label{sec:derivationPotentials}
The following two subsections treat intra-species and inter-species 
cross-linking separately, specifically for scenarios where cross-linking is considered to be strong, \textit{i.e} frequent and likely. This yields simplified but physically meaningful expressions, that allow for the derivation of suitable networking potentials in each case. These effective potentials will later be combined to approximate the 
networking behaviour of the full system in Sec.~\ref{sec:CLpolyCorrs}.

\subsubsection{Intra-species cross-linking}
\label{sec:intra}
\begin{figure*}[!ht]
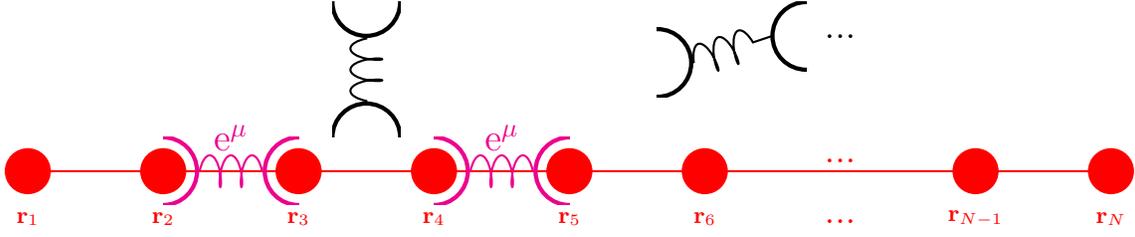

\subfile{./diagrams/crosslinker_tracksDiagramIntra}
\caption[Schematic diagram of intra-species cross-linking]{Schematic diagram of cross-linkers diffusing between a set of attachment points in red with positions $\{\mathbf{r}_1,\mathbf{r}_2, \,...\,\mathbf{r}_N\}$ . Freely Diffusing cross-linkers are depicted in black. Cross-linkers networked to an $\mathbf{r}_n$ and an $\mathbf{r}_{n}$ are depicted in magenta and are assigned an advantage of $\mathrm{e}^\mu$.}
\label{fig:crosslinker&tracks-intra}
\end{figure*}

Recalling, the networking functional given by eq.~\eqref{eq:Qfullintrainter}, consider the limiting case where same-species cross-linking to polymers of type $\mathrm{B}$ and intra-species cross-linking are both unfavourable or unlikely, \textit{i.e. }$\mu_B \rightarrow -\infty$ and $\epsilon \rightarrow -\infty$, such that the only types of cross-links which can form are those where both end of the cross-linker particle are networked to polymer chains of species $\mathrm{A}$. Since only one species of polymer chain is considered in this scenario and we can let $\mu_A = \mu$ for ease of notation, as depicted in Fig. \ref{fig:crosslinker&tracks-intra}.

The networking functional, now only accounting for one type of intra-species cross-linking, requires the use of a single pair of networking fields as follows
\begin{multline}
    \mathbb{Q}= \int [ \mathrm{d}\phi] [\mathrm{d} \phi^*]\,\, \prod_j 
\,  \left( \prod_{n=1}^{N} (1 +\phi^*(\mathbf{r}_n,t_j')) \,\right)\\
\times \,\left( \prod_{m=1}^M \left(1+\phi\left(\mathbf{y}_{m,1},t_j\right)\phi\left(\mathbf{y}_{m,2},t_j\right)\mathrm{e}^{\mu}\right)\right)\\
 \times\mathrm{e}^{-  \frac{1}{ \ell}\int_\mathbf{r} \, \phi(\mathbf{r},t_j)\,\phi^*(\mathbf{r},t_j) }\, .
\end{multline}
where $\mathbf{y}_{m,1} =\mathbf{Y}_m-\frac{\mathbf{y}_m}{2}$ and $\mathbf{y}_{m,2} =\mathbf{Y}_m+\frac{\mathbf{y}_m}{2}$ are the positions of the end-points of each cross-linker. Again utilising the collective variables
\begin{subequations}
    \begin{equation}
        \rho(\mathbf{Y}, \mathbf{y},t) = \sum_{m=1}^M \delta ( \mathbf{Y} - \mathbf{Y}_m (t))\delta (\mathbf{y}- \mathbf{y}_m(t))\,  \mathrm{\,\,\,\,    and} 
    \end{equation}
    \begin{equation}
        C_(\mathbf{r},t)=\sum_{n=1}^{N} \delta(\mathbf{r}-\mathbf{r}_n (t))
    \end{equation}
\end{subequations}
one can move the products into the exponents to become sums and write this in a continuous description, such that the networking functional may be written as:
\begin{equation}
    \mathbb{Q}= \int [ \mathrm{d}\phi] \,[ \mathrm{d}\phi^*]\, 
\mathrm{e}^{\mathbb{F}[\phi, \phi^*]}
\label{eq:Qcl-tracks_intra}
\end{equation}
with 

     \begin{multline}
    \mathbb{F}[\phi, \phi^*]= \frac{1}{\tau}\int_{\mathbf{r},t} C(\mathbf{r},t)\ln(1 +\phi^*(\mathbf{r},t)) \\
+\frac{1}{\tau}\int_{\mathbf{r},t} P(\mathbf{r},t) \ln \left(1+\phi(\mathbf{r},t)\mathrm{e}^\mu\right)\\
-  \alpha\int_{\mathbf{r},t} \, \phi(\mathbf{r},t)\,\phi^*(\mathbf{r},t) \, .
\label{eq:FIntraitoP}
\end{multline}
Here the approach of Sec.~\ref{sec:CL-Bnetworking} has been utilised, noting that the cross-linker density can be rewritten as a density of cross-linker ends $P(\mathbf{z},t) $ as defined in eq.~\eqref{eq:Pdef},
 such that the exponent of the networking functional can be rewritten to depend on $P(\mathbf{z},t) $ instead of $\rho(\mathbf{Y}, \mathbf{y},t)$ .

Taking the partial derivatives, as before, the following saddle point equations are obtained
\begin{subequations}
    \begin{equation}
        0=\left.\frac{\delta\mathbb{F}}{\delta\phi^*}\right|_{\bar{\phi},\bar{\phi}^* } = \frac{C(\mathbf{r}, t)}{\tau (1 + \bar{\phi}^*(\mathbf{r},t))} - \alpha \bar{\phi}( \mathbf{r},t)\, ,
    \end{equation}
        \begin{equation}
        0=\left.\frac{\delta\mathbb{F}}{\delta\phi}\right|_{\bar{\phi},\bar{\phi}^* } = \frac{P(\mathbf{r}, t) \bar{\phi}( \mathbf{r},t) \mathrm{e}^\mu}{\tau (1 + \bar{\phi}^2(\mathbf{r},t)\mathrm{e}^\mu)}- \alpha \bar{\phi}^*( \mathbf{r},t)\,.
    \end{equation}
    \label{eq:CL-intra_SPeqns}
\end{subequations}

In the scenario where intra-species cross-linking is very likely and occurs frequently, $\mathrm{e}^\mu$ will be large such that we can assume that $\mathrm{e}^\mu \gg 1$. Thus the approximation $1 + \bar{\phi}(\mathbf{r},t)\mathrm{e}^\mu \approx\bar{\phi}(\mathbf{r},t)\mathrm{e}^\mu$ can be utilised in \eqref{eq:CL-intra_SPeqns} such that 
\begin{subequations}
    \begin{equation}
         \bar{\phi}( \mathbf{r},t)= \frac{C(\mathbf{r}, t)}{\alpha \tau (1 + \bar{\phi}^*(\mathbf{r},t))}\, ,
    \end{equation}
        \begin{equation}
       \bar{\phi}^*( \mathbf{r},t)= \frac{P(\mathbf{r}, t)}{\alpha\tau  \bar{\phi}(\mathbf{r},t)} \,.
    \end{equation}
    \label{eq:CL-intra_SPeqn-strong}
\end{subequations}
One can solve \eqref{eq:CL-intra_SPeqn-strong} simultaneously to obtain the following saddle point solutions in the strong cross-linking limit
\begin{subequations}
    \begin{equation}
         \bar{\phi}( \mathbf{r},t)= \frac{C(\mathbf{r}, t)-P(\mathbf{r}, t)}{\alpha \tau }\, ,
    \end{equation}
        \begin{equation}
       \bar{\phi}^*( \mathbf{r},t)= \frac{P(\mathbf{r}, t)}{C(\mathbf{r}, t)-P(\mathbf{r}, t)} \,.
    \end{equation}
    \label{eq:CL-intra_SPsols-strong}
\end{subequations}
Substituting eqs.~\eqref{eq:CL-intra_SPsols-strong} back into eq.~\eqref{eq:FIntraitoP}, such that
\begin{equation}
    \mathbb{Q}= \mathrm{e}^{\mathbb{F}[\bar{\phi}, \bar{\phi^*}]}
\label{eq:Qcl-tracks_intra-SP_Fphiphistar}
\end{equation}
with 
\begin{multline}
    \mathbb{F}[\bar{\phi}, \bar{\phi^*}]= \frac{1}{\tau}\int_{\mathbf{r},t} C(\mathbf{r},t)\ln\left(\frac{C(\mathbf{r},t)}{C(\mathbf{r},t)-P(\mathbf{r}, t)}\right) \\
+\frac{1}{\tau}\int_{\mathbf{r},t}P(\mathbf{r}, t) \ln \left(1+\frac{(C(\mathbf{r},t)-P(\mathbf{r},t))\mathrm{e}^\mu}{\alpha^2 \tau^2}\right)\\
- \frac{2}{\tau}\int_{\mathbf{r},t} \, \rho(\mathbf{r}, \mathbf{0},t) \, .
\label{eq:Fcl-t_intra-SP-strong}
\end{multline}
amounts to the saddle point approximation of eq.~\eqref{eq:Qcl-tracks_intra}. Letting the densities once again be equal to a uniform background density plus a small fluctuation term and expanding up to second order in the background density, we find the networking functional in terms of the effective potentials. The magnitudes of the effective potentials due to strong intra-species networking are therefore given by

\begin{widetext}
\begin{subequations}
    \begin{equation}
        w_C(k) =-\frac{\bar{\rho}_0}{\tau } \left(\tfrac{4 \gamma_y  \kappa  \mathrm{e}^{2 \mu } \Gamma(k)}{\left(4 \alpha  \gamma_y  \kappa  \tau +\mathrm{e}^{\mu } \left(-8 \gamma_y  \kappa  \bar{\rho}_0+4 \gamma_y  \bar{C} \kappa +\lambda_y  k^2 \bar{\rho}_0\right)\right)^2}\right. \\
        \left.+\tfrac{\bar{\rho}_0 \Gamma^2(k)}{\bar{C} \left(-8 \gamma_y  \kappa  \bar{\rho}_0+4 \gamma_y  \bar{C} \kappa +\lambda_y  k^2 \bar{\rho}_0\right)^2}\right)
    \end{equation}
    \begin{multline}
        w_\rho(k)= \tfrac{\Gamma^2(k)\left(-64 \alpha ^2 \gamma_y ^3 \bar{C} \kappa ^3 \tau ^2+8 \alpha  \gamma_y  \kappa  \mathrm{e}^{\mu } \bar{\rho}_0 \tau  \Gamma(k) \left(-8 \gamma_y  \kappa  \bar{\rho}_0+4 \gamma_y  \bar{C} \kappa +\lambda_y  k^2 \bar{\rho}_0\right)+\mathrm{e}^{2 \mu } \left(-8 \gamma_y  \kappa  \bar{\rho}_0+4 \gamma_y  \bar{C} \kappa +\lambda_y  k^2 \bar{\rho}_0\right)^3\right)}{4 \gamma_y  \kappa  \tau  \left(-8 \gamma_y  \kappa  \bar{\rho}_0+4 \gamma_y  \bar{C} \kappa +\lambda_y  k^2 \bar{\rho}_0\right)^2 \left(4 \alpha  \gamma_y  \kappa  \tau +\mathrm{e}^{\mu } \left(-8 \gamma_y  \kappa  \bar{\rho}_0+4 \gamma_y  \bar{C} \kappa +\lambda_y  k^2 \bar{\rho}_0\right)\right)^2} \\
    \end{multline}
    \begin{multline}
        v_{C,\rho}(k)= -\frac{\Gamma(k)}{2 \tau }\Bigg(\tfrac{\mathrm{e}^{2 \mu } \bar{\rho}_0 \Gamma(k)}{\left(4 \alpha  \gamma_y  \kappa  \tau +\mathrm{e}^{\mu } \left(-8 \gamma_y  \kappa  \bar{\rho}_0+4 \gamma_y  \bar{C} \kappa +\lambda_y  k^2 \bar{\rho}_0\right)\right)^2}-\tfrac{2 \bar{\rho}_0 \Gamma^2(k)}{\left(-8 \gamma_y  \kappa  \bar{\rho}_0+4 \gamma_y  \bar{C} \kappa +\lambda_y  k^2 \bar{\rho}_0\right)^2}\\
        +\tfrac{\mathrm{e}^{\mu } (-\Gamma(k)) \left(8 \alpha  \gamma_y  \kappa  \tau +\mathrm{e}^{\mu } \left(-8 \gamma_y  \kappa  \bar{\rho}_0+8 \gamma_y  \bar{C} \kappa +\lambda_y  k^2 \bar{\rho}_0\right)\right)}{\left(4 \alpha  \gamma_y  \kappa  \tau +\mathrm{e}^{\mu } \left(-8 \gamma_y  \kappa  \bar{\rho}_0+4 \gamma_y  \bar{C} \kappa +\lambda_y  k^2 \bar{\rho}_0\right)\right)^2}\Bigg)\,.
    \end{multline}
    \label{eq:intraStrongPotentials}
\end{subequations}
\end{widetext}
In addition, we can take the partial derivative of eq.~\eqref{eq:Fcl-t_intra-SP-strong} w.r.t. $\mu$ as follows:
\begin{equation}
    \frac{\partial \mathrm{ln}{\mathbb{Q}}}{\partial \mu} =  \frac{\partial \mathbb{F}}{\partial \mu} = \frac{1}{\tau} \int_ {\mathbf{r},t}\frac{ P(\mathbf{r}, t)\bar{\phi}(\mathbf{r},t)\mathrm{e}^\mu}{\alpha\tau(1+  \bar{\phi}(\mathbf{r},t)\mathrm{e}^\mu)}\,.
    \label{eq:numberDensity-Intra}
\end{equation} 
This is an expression for the number of intra-species cross-links in terms of the saddle point solution $\bar{\phi}$. Applying the same assumptions and therefore also the approximation, \textit{i.e.} $1 + \bar{\phi}(\mathbf{r},t)\mathrm{e}^\mu \approx\bar{\phi}(\mathbf{r},t)\mathrm{e}^\mu$ to eq.~\eqref{eq:numberDensity-Intra}, leads to the following expression for the number of intra-species cross-links
\begin{equation}
    \frac{\partial \mathbb{F}_s}{\partial \mu} = \frac{1}{\alpha \tau^2} \int_ {\mathbf{r},t}P(\mathbf{r},t)\,.
\end{equation}
Thus, the number of intra-species cross-links is dependent on the total number of cross-linkers in the strong cross-linking limit.

\subsubsection{Inter-species cross-linking}
\label{sec:inter}
Again simplifying the networking functional given by eq.~\eqref{eq:Qfullintrainter}, consider the limit where both types of intra-spacies cross-linking are unfavourable or unlikely, \textit{i.e }$\mu_A \rightarrow -\infty$ and $\mu_B \rightarrow -\infty$, such that the only types of cross-links which can form are those where each end of the cross-linker particle is networked to a different species of polymer chain.  This would correspond to a physical system wherein each end of the cross-linker is only able to form bonds with one of the two polymer species. Thus, to avoid overcounting, one of the terms corresponding to an advantage of $\mathrm{e}^\epsilon$ in eq.~\eqref{eq:Qfullintrainter} is removed to obtain

\begin{widetext}
\begin{multline}
    \mathbb{Q}= \int [ \mathrm{d}\phi] [\mathrm{d} \phi^*][ \mathrm{d}\psi] [\mathrm{d} \psi^*]\,\, \prod_j 
\,  \left( \prod_{n=1}^{N} (1 +\phi^*(\mathbf{r}_n,t_j')) \,\prod_{n'=1}^{N'} (1+\psi^*(\mathbf{R}_{n'},t_j))\times \,\prod_{m=1}^M \left(1+\phi\left(\mathbf{y}_{m,1},t_j\right)\psi\left(\mathbf{y}_{m,2} ,t_j\right)\mathrm{e}^\epsilon \right)\right.\\
\left.\times \mathrm{e}^{-  \frac{1}{ \ell}\int_\mathbf{r} \, \phi(\mathbf{r},t_j)\,\phi^*(\mathbf{r},t_j)-  \frac{1}{ \ell}\int_\mathbf{r} \, \psi(\mathbf{r},t_j)\,\psi^*(\mathbf{r},t_j)} \right)\, .
\end{multline}
\end{widetext}

\begin{figure*}[!ht]
\subfile{./diagrams/crosslinker_tracksDiagramInterOnly}
\label{fig:crosslinker&tracks-interOnly}
\caption[Schematic diagram of inter-species cross-linking]{Schematic diagram of cross-linkers diffusing between two sets of attachment points, those in red with positions $\{\mathbf{r}_1,\mathbf{r}_2, \,...\,\mathbf{r}_N\}$ and  in blue with positions $\{\mathbf{R}_1,\mathbf{R}_2, \,...\,\mathbf{R}_{N'}\}$. Freely Diffusing cross-linkers are depicted in black. Cross-linkers networked to an $\mathbf{r}_n$ and an $\mathbf{R}_{n'}$ are depicted in green and are assigned an advantage of $\mathrm{e}^\epsilon$.}
\end{figure*}

 Introducing the collective variables defined in eqs.~\eqref{eq:CLrhodef} and \eqref{eq:CAandCBdef}, one can move the products into the exponents to become sums and write this in a continuous description as follows:

\begin{equation}
    \mathbb{Q}= \int [ \mathrm{d}\phi] [\mathrm{d} \phi^*][ \mathrm{d}\psi] [\mathrm{d} \psi^*]\,\, 
\mathrm{e}^{\mathbb{F}[\phi, \phi^*, \psi, \psi^*]}
\label{eq:Qcl-tracks-repeat1}
\end{equation}
with 
\begin{multline}
    \mathbb{F}[\phi, \phi^*, \psi, \psi^*]= \frac{1}{\tau}\int_{\mathbf{r},t} C_A(\mathbf{r},t)\ln(1 +\phi^*(\mathbf{r},t)) \,\\
    + \frac{1}{\tau}\int_{\mathbf{r},t}C_B(\mathbf{r},t)\ln(1+\psi^*(\mathbf{r},t))\\
+\frac{1}{\tau}\int_{\mathbf{Y},t}\rho(\mathbf{Y}, \mathbf{0},t) \ln \left(1+\phi(\mathbf{Y},t)\psi(\mathbf{Y} ,t)\mathrm{e}^\epsilon \right)\\
-  \alpha\int_{\mathbf{r},t} \, \phi(\mathbf{r},t)\,\phi^*(\mathbf{r},t)-  \alpha\int_{\mathbf{r},t} \, \psi(\mathbf{r},t)\,\psi^*(\mathbf{r},t) \, .
\label{eq:Fcl-t-repeat1}
\end{multline}
Note that the cross-linker extension length has been set to zero here, such that this serves as a lowest order approximation, neglecting the second order term in the pre-averaging expansions discussed in Sec.~\ref{sec:CL-Bnetworking} and \ref{sec:intra}. The second order term can be included, but leads to the requirement of an artificially large minimum repulsive potential to maintain stability of the system and is therefore omitted here.

As usual, to deal with the integrals over the fields $\phi, \phi^*, \psi$ and $\psi^*$, a saddle point approximation is implemented. In order to find a suitable set of saddle point solutions $\{\bar{\phi}, \bar{\phi}^*,\bar{\psi}, \bar{\psi}^*\}$, one needs to solve the following set of simultaneous equations
\begin{subequations}
    \label{eq:SPidentities_cl-A-B}
    \begin{equation}
       0=\left.\frac{\delta\mathbb{F}}{\delta\phi(\mathbf{r},t)} \right|_{\bar{\phi},\bar{\phi}^*,\bar{\psi},\bar{\psi}^* }  = - \alpha \bar{\phi}(\mathbf{r},t) + \frac{1}{\tau}\frac{C_A(\mathbf{r},t)}{1 + \bar{\phi}^*(\mathbf{r},t)} 
    \end{equation}
    \begin{equation}
        0=\left.\frac{\delta\mathbb{F}}{\delta\psi^*(\mathbf{r},t)} \right|_{\bar{\phi},\bar{\phi}^*,\bar{\psi},\bar{\psi}^* } = - \alpha \bar{\psi}(\mathbf{r},t) + \frac{1}{\tau}\frac{C_B(\mathbf{r},t)}{ 1+\bar{\psi}^*(\mathbf{r},t)}\, ,
    \end{equation}
    \begin{equation}
        0=\left.\frac{\delta\mathbb{F}}{\delta\psi(\mathbf{r},t)} \right|_{\bar{\phi},\bar{\phi}^*,\bar{\psi},\bar{\psi}^* }  = - \alpha \bar{\phi}^*(\mathbf{r},t) + \tfrac{\rho(\mathbf{r},\mathbf{0},t)\bar{\psi}(\mathbf{r},t)\mathrm{e}^\epsilon}{\tau \left(1+\bar{\phi}(\mathbf{r},t)\bar{\psi}(\mathbf{r},t)\mathrm{e}^\epsilon\right)}\,, 
    \end{equation}
    \begin{equation}
        0=\left.\frac{\delta\mathbb{F}}{\delta\psi^*(\mathbf{r},t)} \right|_{\bar{\phi},\bar{\phi}^*,\bar{\psi},\bar{\psi}^* }  = - \alpha \bar{\psi}^*(\mathbf{r},t) + \tfrac{\rho(\mathbf{r},\mathbf{0},t)\bar{\phi}(\mathbf{r},t)\mathrm{e}^\epsilon}{\tau \left(1+\bar{\phi}(\mathbf{r},t)\bar{\psi}(\mathbf{r},t)\mathrm{e}^\epsilon\right)}\,. 
    \end{equation}
\end{subequations}
 Assuming that $\mathrm{e}^\epsilon$ is large, $\bar{\phi}(\mathbf{r},t)\bar{\psi}(\mathbf{r},t)\mathrm{e}^\epsilon \gg 1$ such that we may approximate $1+\bar{\phi}(\mathbf{r},t)\bar{\psi}(\mathbf{r},t)\mathrm{e}^\epsilon \approx \bar{\phi}(\mathbf{r},t)\bar{\psi}(\mathbf{r},t)\mathrm{e}^\epsilon$ leading to the following saddle point equations:
\begin{subequations}
    \label{eq:SPidentities_cl-strong}
    \begin{equation}
       \bar{\phi}(\mathbf{r},t) = \frac{1}{\alpha \tau}\frac{C_A(\mathbf{r},t)}{1 + \bar{\phi}^*(\mathbf{r},t)} 
    \end{equation}
    \begin{equation}
         \bar{\psi}(\mathbf{r},t) = \frac{1}{\alpha\tau}\frac{C_B(\mathbf{r},t)}{1+\bar{\psi}^*(\mathbf{r},t)}\, ,
    \end{equation}
    \begin{equation}
        \bar{\phi}^*(\mathbf{r},t) =\frac{\rho(\mathbf{r},\mathbf{0},t)}{\alpha \tau \left(\bar{\phi}(\mathbf{r},t)\right)}\,, 
    \end{equation}
    \begin{equation}
       \bar{\psi}^*(\mathbf{r},t) = \frac{\rho(\mathbf{r},\mathbf{0},t)}{ \alpha \tau \left(\bar{\psi}(\mathbf{r},t)\right)}\,. 
    \end{equation}
\end{subequations}

Solving eqs.~\eqref{eq:SPidentities_cl-strong} leads to 
\begin{subequations}
        \label{eq:SPsols_cl-strong}
    \begin{equation}
       \bar{\phi}(\mathbf{r},t) = \frac{C_A(\mathbf{r},t)-\rho(\mathbf{r},\mathbf{0},t)}{\alpha \tau} 
    \end{equation}
    \begin{equation}
         \bar{\psi}(\mathbf{r},t) =  \frac{C_B(\mathbf{r},t)-\rho(\mathbf{r},\mathbf{0},t)}{\alpha \tau} \, ,
    \end{equation}
    \begin{equation}
        \bar{\phi}^*(\mathbf{r},t) =\frac{\rho(\mathbf{r},\mathbf{0},t)}{C_A(\mathbf{r},t)-\rho(\mathbf{r},\mathbf{0},t)}\,, 
    \end{equation}
    \begin{equation}
       \bar{\psi}^*(\mathbf{r},t) = \frac{\rho(\mathbf{r},\mathbf{0},t)}{C_B(\mathbf{r},t)-\rho(\mathbf{r},\mathbf{0},t)}\,. 
    \end{equation}
\end{subequations}

Recalling eq.~\eqref{eq:Fcl-t}, these solutions can be substituted into
\begin{equation}
     \mathbb{Q}= \mathrm{e}^{\mathbb{F}[\bar{\phi}, \bar{\phi}^*, \bar{\psi}, \bar{\psi}^*]}
     \label{eq:QSP-strong}
\end{equation}
 such that
 \begin{widetext}
 \begin{multline}
    \mathbb{F}[\bar{\phi}, \bar{\phi}^*, \bar{\psi}, \bar{\psi}^*]= \frac{1}{\tau}\int_{\mathbf{r},t} C_A(\mathbf{r},t)\ln\left(\tfrac{C_A(\mathbf{r},t)}{C_A(\mathbf{r},t)-\rho(\mathbf{r},\mathbf{0},t)})\right) 
    + \frac{1}{\tau}\int_{\mathbf{r},t}C_B(\mathbf{r},t)\ln\left(\tfrac{C_B(\mathbf{r},t)}{C_B(\mathbf{r},t)-\rho(\mathbf{r},\mathbf{0},t)}\right)\\
+\frac{1}{\tau}\int_{\mathbf{Y},t}\rho(\mathbf{Y}, \mathbf{0},t) \ln \left(1 + 
\tfrac{1}{\alpha^2 \tau^2}(C_A(\mathbf{r},t)-\rho(\mathbf{r},\mathbf{0},t))(C_B(\mathbf{r},t)-\rho(\mathbf{r},\mathbf{0},t))\mathrm{e}^\epsilon\right)
-  \tfrac{2}{\tau}\int_{\mathbf{r},t} \,\rho(\mathbf{r},\mathbf{0},t)  ,
\label{eq:Fcl-strong}
\end{multline}
 \end{widetext}
 which is the saddle point approximation. Again,  let $C_A(\mathbf{r},t) = \bar{C}_A + \Delta C_A(\mathbf{r},t)  $, $C_B(\mathbf{r},t) = \bar{C}_B + \Delta C_B(\mathbf{r},t)  $ and $\rho (\mathbf{r},\mathbf{0},t) = \bar{\rho} + \Delta \rho(\mathbf{r},\mathbf{0},t)  $ and expand $\mathbb{F}$ up to second order in the fluctuations $\Delta C_A(\mathbf{r},t),\Delta C_B(\mathbf{r},t)$ and $\Delta \rho(\mathbf{r},\mathbf{0},t)  $ to obtain the values of the following potentials:
 \begin{widetext}
\begin{subequations}
    \begin{equation}
        w_A =\frac{\bar{\rho} \left(-\alpha ^4 \bar{\rho} \tau ^4+2 \alpha ^2 \tau ^2 \bar{\rho} \mathrm{e}^{\epsilon } (\bar{C}_A-\bar{\rho}) (\bar{\rho}-\bar{C}_B)+ \mathrm{e}^{2 \epsilon } (\bar{C}_A-\bar{\rho})^3 (\bar{C}_B-\bar{\rho})^2\right)}{\bar{C}_A \tau  (\bar{C}_A-\bar{\rho})^2 \left(\alpha ^2 \tau ^2+ \mathrm{e}^{\epsilon } (\bar{C}_A-\bar{\rho}) (\bar{C}_B-\bar{\rho})\right)^2}
    \end{equation}
    \begin{equation}
        w_B =\frac{\bar{\rho} \left(-\alpha ^4 \bar{\rho} \tau ^4+2 \alpha ^2 \tau ^2 \bar{\rho} \mathrm{e}^{\epsilon } (\bar{C}_A-\bar{\rho}) (\bar{\rho}-\bar{C}_B)+\mathrm{e}^{2 \epsilon } (\bar{C}_A-\bar{\rho})^2 (\bar{C}_B-\bar{\rho})^3\right)}{\bar{C}_B \tau  (\bar{C}_B-\bar{\rho})^2 \left(\alpha ^2 \tau ^2+ \mathrm{e}^{\epsilon } (\bar{C}_A-\bar{\rho}) (\bar{C}_B-\bar{\rho})\right)^2}
    \end{equation}
    \begin{multline}
        w_\rho =\tfrac{2 \alpha ^2  \tau ^2 \bar{\rho} \mathrm{e}^{\epsilon } (\bar{C}_A-\bar{\rho}) (\bar{\rho}-\bar{C}_B) \left(\bar{C}_A^2-3 \bar{\rho} (\bar{C}_A+\bar{C}_B)+\bar{C}_A \bar{C}_B+\bar{C}_B^2+3 \bar{\rho}^2\right)-\alpha ^4 \tau ^4 \left(\bar{\rho}^2 (\bar{C}_A+\bar{C}_B)-4 \bar{C}_A \bar{C}_B \bar{\rho}+\bar{C}_A \bar{C}_B (\bar{C}_A+\bar{C}_B)\right)}{\tau  (\bar{C}_A-\bar{\rho})^2 (\bar{C}_B-\bar{\rho})^2 \left(\alpha ^2 \tau ^2+ \mathrm{e}^{\epsilon } (\bar{C}_A-\bar{\rho}) (\bar{C}_B-\bar{\rho})\right)^2}\\
        + \tfrac{\mathrm{e}^{2 \epsilon } (\bar{C}_A-\bar{\rho})^3 (\bar{C}_B-\bar{\rho})^3 (\bar{C}_A+\bar{C}_B-2 \bar{\rho})}{\tau  (\bar{C}_A-\bar{\rho})^2 (\bar{C}_B-\bar{\rho})^2 \left(\alpha ^2 \tau ^2+ \mathrm{e}^{\epsilon } (\bar{C}_A-\bar{\rho}) (\bar{C}_B-\bar{\rho})\right)^2}
    \end{multline}
    \begin{equation}
        v_{A,\rho} =\tfrac{\alpha ^4 \bar{\rho} \tau ^4+\alpha ^2 \tau ^2 \mathrm{e}^{\epsilon } (\bar{\rho}-\bar{C}_A) (\bar{C}_A (\bar{C}_B-2 \bar{\rho})+\bar{\rho} (4 \bar{\rho}-3 \bar{C}_B))+\mathrm{e}^{2 \epsilon } (\bar{\rho}-\bar{C}_A)^3 (\bar{C}_B-\bar{\rho})^2}{\tau  (\bar{C}_A-\bar{\rho})^2 \left(\alpha ^2 \tau ^2+\mathrm{e}^{\epsilon } (\bar{C}_A-\bar{\rho}) (\bar{C}_B-\bar{\rho})\right)^2}
    \end{equation}
    \begin{equation}
        v_{B,\rho} = \tfrac{\alpha ^4 \bar{\rho} \tau ^4+ \alpha ^2 \tau ^2 \mathrm{e}^{\epsilon } (\bar{\rho}-\bar{C}_B) (\bar{C}_A (\bar{C}_B-3 \bar{\rho})-2 \bar{\rho} (\bar{C}_B-2 \bar{\rho}))+\mathrm{e}^{2 \epsilon } (\bar{C}_A-\bar{\rho})^2 (\bar{\rho}-\bar{C}_B)^3}{\tau  (\bar{C}_B-\bar{\rho})^2 \left(\alpha ^2 \tau ^2+\mathrm{e}^{\epsilon } (\bar{C}_A-\bar{\rho}) (\bar{C}_B-\bar{\rho})\right)^2}
    \end{equation}
    \begin{equation}
        v_{A,B} =-\frac{\alpha ^2 \tau  \bar{\rho}  \mathrm{e}^{\epsilon }}{\left(\alpha ^2 \tau ^2+ \mathrm{e}^{\epsilon } (\bar{C}_A-\bar{\rho}) (\bar{C}_B-\bar{\rho})\right)^2}\, .
    \end{equation}
\end{subequations}
\end{widetext}
To obtain a measure of the number of cross-linking instances, we can take the partial derivative of \eqref{eq:Fcl-t}, \textit{i.e.}
\begin{equation}
    \frac{\partial \mathrm{ln}{\mathbb{Q}}}{\partial \epsilon} =  \frac{\partial \mathbb{F}}{\partial \epsilon} = \frac{1}{\tau} \int_ {\mathbf{r},t}\frac{\rho(\mathbf{r},\mathbf{0},t) \phi (\mathbf{r},t) \psi(\mathbf{r},t) \mathrm{e}^\epsilon}{ 1+\phi (\mathbf{r},t) \psi(\mathbf{r},t) \mathrm{e}^\epsilon }\,.
    \label{eq:numberDensity}
\end{equation}
Once again assuming large values of $\epsilon$ the approximation  $1+\bar{\phi}(\mathbf{r},t)\bar{\psi}(\mathbf{r},t)\mathrm{e}^\epsilon \approx \bar{\phi}(\mathbf{r},t)\bar{\psi}(\mathbf{r},t)\mathrm{e}^\epsilon$ can be applied to \eqref{eq:numberDensity} to obtain
\begin{equation}
     \frac{\partial \mathbb{F}_s}{\partial \epsilon} = \frac{1}{\tau} \int_ {\mathbf{r},t}\rho(\mathbf{r},\mathbf{0},t) \, = \frac{1}{\tau} \int_ {t}M\,.
\end{equation}
Thus, in the strong cross-linking limit, the number of inter-species cross-linking instances depends on the number of cross-linkers $M$ in the system.

\subsection{Dynamical Implementation}
\label{sec:CLpolyCorrs}

Returning to the full system as depicted in Fig.~\ref{fig:crosslinker&tracks}, once both the saddle point approximations and small density fluctuation expansions have been implemented, the networking functional, first given by eq.~\eqref{eq:Qfullintrainter} should have the following form:
\begin{multline}
\mathbb{Q}[\Delta C_\mathrm{A},\Delta C_\mathrm{B}, \Delta  \rho] = \mathcal{N} \mathrm{e}^{-
\int_{\mathbf{k},\omega}  \Delta C_\mathrm{A}(\mathbf{k},\omega)v_{\mathrm{A}, \rho} \Delta \rho_{-\mathbf{k}} (-\omega) }
 \\
 \times \mathrm{e}^{ - \int_{\mathbf{k},\omega} \Delta C_\mathrm{B}(\mathbf{k},\omega) v_{\mathrm{B}, \rho} \Delta \rho_{-\mathbf{k}} (-\omega) -\int_{\mathbf{k},\omega} \Delta C_\mathrm{A}(\mathbf{k},\omega) v_{\mathrm{A}, \mathrm{B}} \Delta C_\mathrm{B}(\mathbf{k},\omega)}
 \\
 \times \mathrm{e}^{ 
 -\frac{1}{2}\int_{\mathbf{k},\omega} \Delta \rho_{\mathbf{k}} (\omega) w_\rho^2 \Delta \rho_{-\mathbf{k}} (-\omega)- \frac{1}{2} \int_{\mathbf{k},\omega} \Delta C_\mathrm{A}(\mathbf{k},\omega) w_{\mathrm{A}}^2 \Delta C_\mathrm{A}(-\mathbf{k},-\omega) }\\
 \times \mathrm{e}^{ 
- \frac{1}{2} \int_{\mathbf{k},\omega}  \Delta C_\mathrm{B}(\mathbf{k},\omega) w_\mathrm{B}^2 \Delta C_\mathrm{B}(-\mathbf{k},-\omega)}\,.
\, 
\label{eq:QCLwAdv_itoVs-main}
\end{multline} 
 Thus, the networking potentials required are $v_\mathrm{AB},v_{\mathrm{A}\rho},v_{\mathrm{B}\rho}, w_\mathrm{A},w_\mathrm{B}$ and $ w_\rho$. To differentiate between the intra-species cross-linking networking potentials derived in Sec.~\ref{sec:intra} and the inter-species cross-linking networking potentials derived in Sec.~\ref{sec:inter}, superscripts will be used here, such that the former are denoted by $v_{C\rho}^{\text{intra}},w_C^{\text{intra}}$ and $w_\rho^{\text{intra}}$ and the latter by $v_\mathrm{AB}^\text{inter},v_{\mathrm{A}\rho}^\text{inter},v_{\mathrm{B}\rho}^\text{inter}, w_\mathrm{A}^\text{inter},w_\mathrm{B}^\text{inter}$ and $ w_\rho^\text{inter}$. Recall that in  Sec.~\ref{sec:intra}, the intra-species cross-linking was considered between polymers with background density $\bar{C}$ , while the full system, considers intra-species cross-linking between polymers of type $\mathrm{A}$ as well as polymers of type $\mathrm{B}$ along with the inter-species cross-linking between polymers of type $\mathrm{A}$ and $\mathrm{B}$. Thus to obtain the appropriate potentials for intra-species cross-linking between polymers of type $\mathrm{A}$ and $\mathrm{B}$, respectively, those derived for polymers with density $\bar{C}$ can be used with a substitution of the relevant density variables in each case, \textit{i.e.} $\bar{C} \rightarrow \bar{C}_\mathrm{A}$ to obtain $v_{\mathrm{A}\rho}^{\text{intra-}\mathrm{A}},w_\mathrm{A}^{\text{intra-}\mathrm{A}}$ and $w_\rho^{\text{intra-}\mathrm{A}}$ or  $\bar{C} \rightarrow \bar{C}_\mathrm{B}$ to obtain $v_{\mathrm{B}\rho}^{\text{intra-}\mathrm{B}},w_\mathrm{B}^{\text{intra-}\mathrm{B}}$ and $w_\rho^{\text{intra-}\mathrm{B}}$. Since, networking potentials have been considered separately in their derivations in Sec.~\ref{sec:derivationPotentials}, but now need to be applied in the same system to match the networking functional in eq.~\eqref{eq:QCLwAdv_itoVs-main}, we approximate this at the consecutive application of networking functionals for each type of cross-linking, \textit{i.e.}
\begin{multline}
    \mathbb{Q}[\Delta C_\mathrm{A},\Delta C_\mathrm{B}, \Delta  \rho] \approx \mathbb{Q}^\text{inter}[\Delta C_\mathrm{A},\Delta C_\mathrm{B}, \Delta  \rho] \\\times \mathbb{Q}^{\text{intra-}\mathrm{A}}[\Delta C_\mathrm{A},\Delta C_\mathrm{B}, \Delta  \rho] \\\times \mathbb{Q}^{\text{intra-}\mathrm{B}}[\Delta C_\mathrm{A},\Delta C_\mathrm{B}, \Delta  \rho]\, ,
\end{multline}
such that each of the networking potentials of the full system can be obtained by summing the contributions from each type of cross-linking, as follows:
\begin{subequations}
    \begin{equation}
        v_\mathrm{AB}=v_\mathrm{AB}^\text{inter} (0)
    \end{equation}
    \begin{equation}
        v_{\mathrm{A}\rho}=v_{\mathrm{A}\rho}^\text{inter}(0)+v_{\mathrm{A}\rho}^{\text{intra-}\mathrm{A}}(0)
    \end{equation}
    \begin{equation}
        v_{\mathrm{B}\rho}=v_{\mathrm{B}\rho}^\text{inter}(0)+v_{\mathrm{B}\rho}^{\text{intra-}\mathrm{B}}(0)
    \end{equation}
    \begin{equation}
        w_\mathrm{A}=w_\mathrm{A}^\text{inter}(0)+w_\mathrm{A}^{\text{intra-}\mathrm{A}}(0)
    \end{equation}
    \begin{equation}
        w_\mathrm{B}=w_\mathrm{B}^\text{inter}(0)+w_\mathrm{B}^{\text{intra-}\mathrm{B}}(0)
    \end{equation}
    \begin{equation}
        w_\rho=w_\rho^\text{inter}(0)+w_\rho^{\text{intra-}\mathrm{A}}(0)+w_\rho^{\text{intra-}\mathrm{B}}(0)
    \end{equation}
\end{subequations}
Note that, in the above expressions, the intra- and inter-species potentials are all evaluated at $k=0$. This choice has been made simply because the $k$- contribution corresponding to the cross-linker extension approximation is not available in all cases, so setting $k=0$ considers the approximation where cross-linkers are  regarded as point-particles, for consistency across comparisons.
 
 This expression will have the following form:
\begin{widetext}
       \begin{multline}
    \mathbb{Z}_\mathrm{full}[J_{\mathrm{A}}, J_{\mathrm{B}}, J_\rho] = \mathcal{N} \int [\mathrm{d} \Delta \rho ] \,[\mathrm{d} \Delta C_\mathrm{A}]\,[\mathrm{d} \Delta C_\mathrm{B}]\, \mathrm{e}^{-\frac{1}{2}\int_{\mathbf{k},\omega} \Delta \rho(\mathbf{k}, \omega) \mathbb{B}_{\rho}(\mathbf{k},\omega) \Delta \rho(-\mathbf{k}, -\omega)}\\
\times \mathrm{e}^{ - \frac{1}{2} \int_{\mathbf{k},\omega} \Delta C_\mathrm{A}(\mathbf{k},\omega) \mathbb{B}_{\mathrm{A} }(\mathbf{k},\omega)  \Delta C_\mathrm{A}(-\mathbf{k},-\omega) 
- \frac{1}{2} \int_{\mathbf{k},\omega}  \Delta C_\mathrm{B}(\mathbf{k},\omega) \mathbb{B}_{\mathrm{B}}(\mathbf{k},\omega) \Delta C_\mathrm{B}(-\mathbf{k},-\omega)}\\
\times \mathrm{e}^{+\int_{\mathbf{k},\omega} \Delta C_\mathrm{A}(\mathbf{k},\omega)v_{\mathrm{A},\rho} \Delta \rho({-\mathbf{k}},-\omega) +\int_{\mathbf{k},\omega}{\Delta C_\mathrm{B}(\mathbf{k},\omega)v_{\mathrm{B},\rho} \Delta \rho(-\mathbf{k},-\omega)+  \int_{\mathbf{k},\omega} \Delta C_\mathrm{A}(\mathbf{k},\omega) v_{\mathrm{A},\mathrm{B}} \Delta C_\mathrm{B}(-\mathbf{k},-\omega)}}\\
\times \mathrm{e}^{+\int_{\mathbf{k},\omega} J_{\mathrm{A}}(\mathbf{k},\omega)\Delta C_\mathrm{A}(-\mathbf{k},-\omega)+\int_{\mathbf{k},\omega} J_{\mathrm{B}}(\mathbf{k},\omega)\Delta C_\mathrm{B}(-\mathbf{k},-\omega)+\int_{\mathbf{k},\omega} J_{\rho}(\mathbf{k},\omega) \Delta \rho( -\mathbf{k},-\omega)}\, ,
       \end{multline}    
\end{widetext}
        where 
        \begin{subequations}
        \begin{equation}
            \mathbb{B}_{\rho}(\mathbf{k},\omega) = S_{0,\rho }^{-1}(\mathbf{k},\omega) +w_\rho\, + v, \,
        \end{equation}
        \begin{equation} 
        \mathbb{B}_{\mathrm{A}}(\mathbf{k},\omega) = S_{0,\mathrm{A} }^{-1}(\mathbf{k},\omega)   +w_\mathrm{A}+v
        \end{equation}
         and 
         \begin{equation}
             \mathbb{B}_{\mathrm{B}}(\mathbf{k},\omega) = S_{0,\mathrm{B} }^{-1}(\mathbf{k},\omega)  +w_\mathrm{B}+v\, .
         \end{equation}
         \label{eq:BexpressionsABrho}
        \end{subequations}
Here, $v$ is an additional repulsive potential added between particles of the same type, in order to avoid a collapse of the system, as discussed throughout Ref.~\cite{dutoitDynamicalNetworkingUsing2025}. 

After implementing the Gaussian functional integrals, one can use functional derivatives with respect to $J_{\mathrm{A}}$, $J_{\mathrm{B}}$ and $ J_\rho$ to obtain the following expressions for the correlation and cross-correlation functions:
\begin{widetext}
\begin{subequations}
\begin{multline}
 { \langle\!\langle}{\Delta C_\mathrm{A}(\mathbf{k},\omega)} {\Delta C_\mathrm{A}(-\mathbf{k},-\omega)}{ \rangle\! \rangle }= 
 \tfrac{2 \mathbb{B}_{\mathrm{B}}(\mathbf{k},\omega)  \mathbb{B}_{\rho}(\mathbf{k},\omega) -2 v_{\mathrm{B},\rho}^2}{2 \mathbb{B}_{\mathrm{A}}(\mathbf{k},\omega)  \left(\mathbb{B}_{\mathrm{B}}(\mathbf{k},\omega)  \mathbb{B}_{\rho}(\mathbf{k},\omega) -v_{\mathrm{B},\rho}^2\right)-2 \left( \mathbb{B}_{\mathrm{B}}(\mathbf{k},\omega)v_{\mathrm{A},\rho}^2 -2 v_{\mathrm{A},\mathrm{B}}v_{\mathrm{A},\rho} v_{\mathrm{B},\rho}+\mathbb{B}_{\rho}(\mathbf{k},\omega) v_{\mathrm{A},\mathrm{B}}^2\right)}
\, ,    
\label{eq:AcorrCL}
\end{multline}
\begin{multline}
 { \langle\!\langle}{\Delta C_\mathrm{B}(\mathbf{k},\omega)} {\Delta C_\mathrm{B}(-\mathbf{k},-\omega)}{ \rangle\! \rangle }=
 \tfrac{2 \mathbb{B}_{\mathrm{A}}(\mathbf{k},\omega)  \mathbb{B}_{\rho}(\mathbf{k},\omega) -2 v_{\mathrm{A},\rho}^2}{2 \mathbb{B}_{\mathrm{A}}(\mathbf{k},\omega)  \left(\mathbb{B}_{\mathrm{B}}(\mathbf{k},\omega)  \mathbb{B}_{\rho}(\mathbf{k},\omega) -v_{\mathrm{B},\rho}^2\right)-2 \left( \mathbb{B}_{\mathrm{B}}(\mathbf{k},\omega)v_{\mathrm{A},\rho}^2 -2 v_{\mathrm{A},\mathrm{B}}v_{\mathrm{A},\rho} v_{\mathrm{B},\rho}+\mathbb{B}_{\rho}(\mathbf{k},\omega) v_{\mathrm{A},\mathrm{B}}^2\right)}
\, ,    
\end{multline}
\begin{multline}
 { \langle\!\langle}{\Delta \rho({\mathbf{k}},\omega)} {\Delta \rho({-\mathbf{k}},(-\omega)}{ \rangle\! \rangle }= 
 \tfrac{2 \mathbb{B}_{\mathrm{A}}(\mathbf{k},\omega)  \mathbb{B}_{\mathrm{B}}(\mathbf{k},\omega) -2 v_{\mathrm{A},\mathrm{B}}^2}{2 \mathbb{B}_{\mathrm{A}}(\mathbf{k},\omega)  \left(\mathbb{B}_{\mathrm{B}}(\mathbf{k},\omega)  \mathbb{B}_{\rho}(\mathbf{k},\omega) -v_{\mathrm{B},\rho}^2\right)-2 \left( \mathbb{B}_{\mathrm{B}}(\mathbf{k},\omega)v_{\mathrm{A},\rho}^2 -2 v_{\mathrm{A},\mathrm{B}}v_{\mathrm{A},\rho} v_{\mathrm{B},\rho}+\mathbb{B}_{\rho}(\mathbf{k},\omega) v_{\mathrm{A},\mathrm{B}}^2\right)}
\, ,    
\end{multline}
\begin{multline}
 { \langle\!\langle}{\Delta C_\mathrm{A}(\mathbf{k},\omega)} {\Delta \rho(-\mathbf{k},-\omega)}{ \rangle\! \rangle }=
 \tfrac{-2 \mathbb{B}_{\mathrm{B}}(\mathbf{k},\omega)  v_{\mathrm{A},\rho}+2 v_{\mathrm{A},\mathrm{B}} v_{\mathrm{B},\rho}}{2 \mathbb{B}_{\mathrm{A}}(\mathbf{k},\omega)  \left(\mathbb{B}_{\mathrm{B}}(\mathbf{k},\omega)  \mathbb{B}_{\rho}(\mathbf{k},\omega) -v_{\mathrm{B},\rho}^2\right)-2 \left( \mathbb{B}_{\mathrm{B}}(\mathbf{k},\omega)v_{\mathrm{A},\rho}^2 -2 v_{\mathrm{A},\mathrm{B}}v_{\mathrm{A},\rho} v_{\mathrm{B},\rho}+\mathbb{B}_{\rho}(\mathbf{k},\omega) v_{\mathrm{A},\mathrm{B}}^2\right)}
\, ,    
\end{multline}
\begin{multline}
 { \langle\!\langle}{\Delta C_\mathrm{B}(\mathbf{k},\omega)} {\Delta \rho,({-\mathbf{k}},-\omega)}{ \rangle\! \rangle }=
 \tfrac{-2 \mathbb{B}_{\mathrm{A}}(\mathbf{k},\omega)  v_{\mathrm{A},\rho}+2 v_{\mathrm{A},\mathrm{B}} v_{\mathrm{B},\rho}}{2 \mathbb{B}_{\mathrm{A}}(\mathbf{k},\omega)  \left(\mathbb{B}_{\mathrm{B}}(\mathbf{k},\omega)  \mathbb{B}_{\rho}(\mathbf{k},\omega) -v_{\mathrm{B},\rho}^2\right)-2 \left( \mathbb{B}_{\mathrm{B}}(\mathbf{k},\omega)v_{\mathrm{A},\rho}^2 -2 v_{\mathrm{A},\mathrm{B}}v_{\mathrm{A},\rho} v_{\mathrm{B},\rho}+\mathbb{B}_{\rho}(\mathbf{k},\omega) v_{\mathrm{A},\mathrm{B}}^2\right)}
\, .    
\end{multline}
and
\begin{multline}
 { \langle\!\langle}{\Delta C_\mathrm{A}(\mathbf{k},\omega)} {\Delta C_\mathrm{B}(-\mathbf{k},-\omega)}{ \rangle\! \rangle }=
 \tfrac{-2 \mathbb{B}_{\rho}(\mathbf{k},\omega) v_{\mathrm{A},\mathrm{B}}+2 v_{\mathrm{A},\rho} v_{\mathrm{B},\rho}}{2 \mathbb{B}_{\mathrm{A}}(\mathbf{k},\omega)  \left(\mathbb{B}_{\mathrm{B}}(\mathbf{k},\omega)  \mathbb{B}_{\rho}(\mathbf{k},\omega) -v_{\mathrm{B},\rho}^2\right)-2 \left( \mathbb{B}_{\mathrm{B}}(\mathbf{k},\omega)v_{\mathrm{A},\rho}^2 -2 v_{\mathrm{A},\mathrm{B}}v_{\mathrm{A},\rho} v_{\mathrm{B},\rho}+\mathbb{B}_{\rho}(\mathbf{k},\omega) v_{\mathrm{A},\mathrm{B}}^2\right)}\, .    
 \label{eq:ABcorrCL}
\end{multline}
\label{eq:corrsCL}
\end{subequations}
\end{widetext}
The minimum repulsive potential $v_\mathrm{min}$ required for the system not to collapse, can now be determined by investigating where the denominator of the above correlation goes to $0$, in this case this occurs where
\begin{multline}
    -2 v_{\mathrm{A}, \mathrm{B}}^2 (v_\mathrm{min}+w_\rho)-2 v_{\mathrm{A}, \rho}^2 (v_\mathrm{min}+w_\mathrm{B})-2 v_{\mathrm{B}, \rho}^2 (v_\mathrm{min}+w_\mathrm{A})\\
    +2 (v_\mathrm{min}+w_\mathrm{A}) (v_\mathrm{min}+w_\mathrm{B}) (v_\mathrm{min}+w_\rho)+4 v_{\mathrm{A}, \mathrm{B}} v_{\mathrm{A}, \rho} v_{\mathrm{B}, \rho} =0  \, .  
    \label{eq:vminfull}
\end{multline}
There is only one real solution for $v_\mathrm{min}$ to eq.~\eqref{eq:vminfull}, which gives the minimum value of the repulsive potential, \textit{i.e.} $v\geq v_\mathrm{min}$ in eqs.~\eqref{eq:BexpressionsABrho}. The full expression, although obtainable analytically, is not given here due to its length.

\begin{figure}[!htbp]
    \centering

    \begin{subfigure}[t]{0.45\textwidth}
        \centering
        \includegraphics[width=\linewidth]{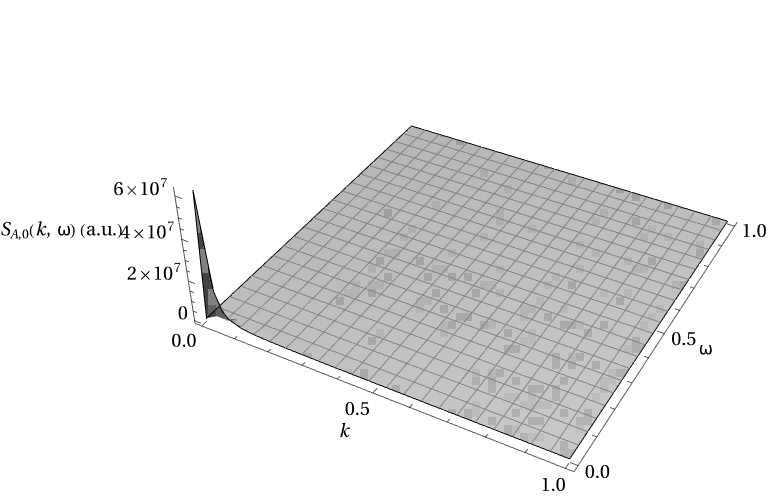}
        \caption{}
        \label{fig:SA0-polyCL}
    \end{subfigure}
    \begin{subfigure}[t]{0.45\textwidth}
        \centering
        \includegraphics[width=\linewidth]{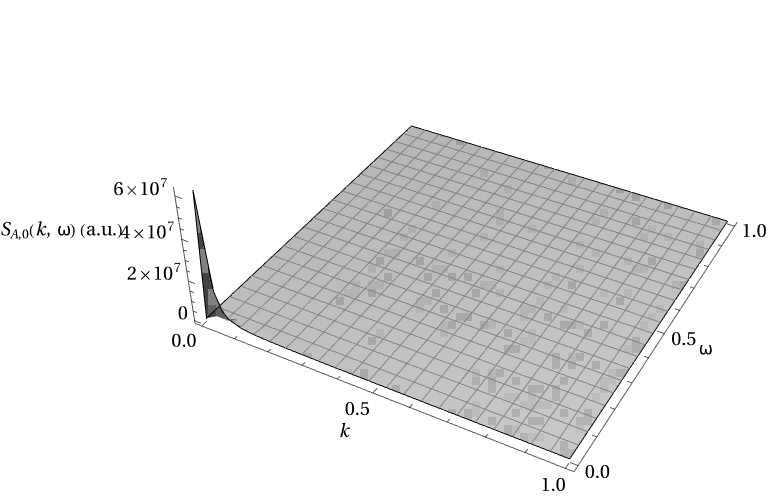}
        \caption{}
        \label{fig:SB0-polyCL}
    \end{subfigure}
    \par\bigskip
    \begin{subfigure}[t]{0.45\textwidth}
        \centering
        \includegraphics[width=\linewidth]{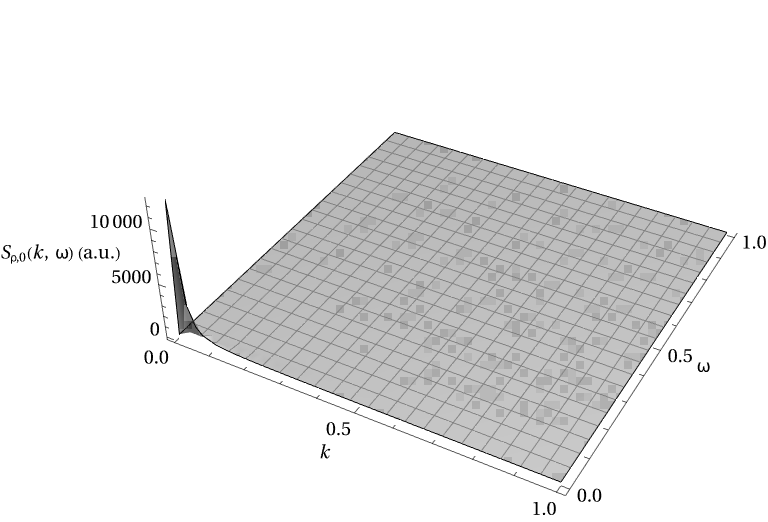}
        \caption{}
        \label{fig:Srho0-polyCL}
    \end{subfigure}

    \caption[Dynamic structure factors of polymers and cross-linkers without networking]{Dynamic structure factors for each dynamical system before cross-linking: (a) Polymer solution A, (b)Polymer solution B, and (c) cross-linker particles. Here $\gamma = 1$, $\lambda = 1$, $\gamma_A = 1$, $\gamma_B = 1$, $L_A = 100$ and $L_B = 100$.}
    \label{fig:S0plots-polyCL}
\end{figure}

In order to interpret this system, the correlation functions and networking potentials should be combined with dynamical structure factors for the polymers and cross-linker particles. For the polymers, we use the polymer solution dynamics from Ref.~\cite{fredricksonCollectiveDynamicsPolymer1990} also used in our previous work in Ref.~\cite{dutoitDynamicalNetworkingUsing2025}. For the cross-linkers, the dynamic structure factor for Brownian particles can be used, since  this serves as a lowest order approximation for the crosslinkers' extension as discussed in the pre-averaging approximation developed in Sec.~\ref{sec:CL-Bnetworking}.  So, the dynamic structure factors, before implementing networking or cross-linking, are given by
\begin{subequations}
    \begin{equation}
        S_{0,\mathrm{A}}(k, \omega) =  \frac{2 \gamma_\mathrm{A}  k^2}{\gamma_\mathrm{A}^2 \omega ^2+L_\mathrm{A}^ {-2} k^4 } ,
        \label{eq:SpolymerA}
    \end{equation}
    \begin{equation}
        S_{0,\mathrm{B}}(k, \omega) = \frac{2 \gamma_\mathrm{B}  k^2}{\gamma_\mathrm{B}^2 \omega ^2+L_\mathrm{B}^ {-2} k^4 },
         \label{eq:SpolymerB}
    \end{equation}
        \begin{equation}
            S_{0,\rho}(k, \omega) = \frac{D_\rho \,k^2}{D_\rho^2\,k^4 +\omega^2}\, ,\label{eq:Scrosslinkers}
        \end{equation}
\end{subequations}
where $D_\rho = \tfrac{\lambda}{4 \gamma^2}$ is the diffusion coefficient of the cross-linker particles. These dynamic structure factors are shown in Fig.~\ref{fig:S0plots-polyCL}. All three dynamic structure factors depict the sharp diffusive peaks at small $k$ and $\omega$, indicating strong density fluctuations on large length and time scales that decay rapidly. The amplitude of this peak, is however orders of magnitude larger for the polymers in Fig.~\ref{fig:SA0-polyCL} and \ref{fig:SB0-polyCL} than for the cross-linkers in Fig.~\ref{fig:Srho0-polyCL}, indicating much stronger fluctuations in the slow, large length scale density of the polymers.

\subsection{Results: Strong intra- and inter-species cross-linking}
\label{subsec:strongintrastronginter}

\begin{figure}[t]
    \centering
    \begin{minipage}{\columnwidth}
        \begin{subfigure}[t]{0.48\linewidth}
            \centering
            \includegraphics[width=\linewidth]{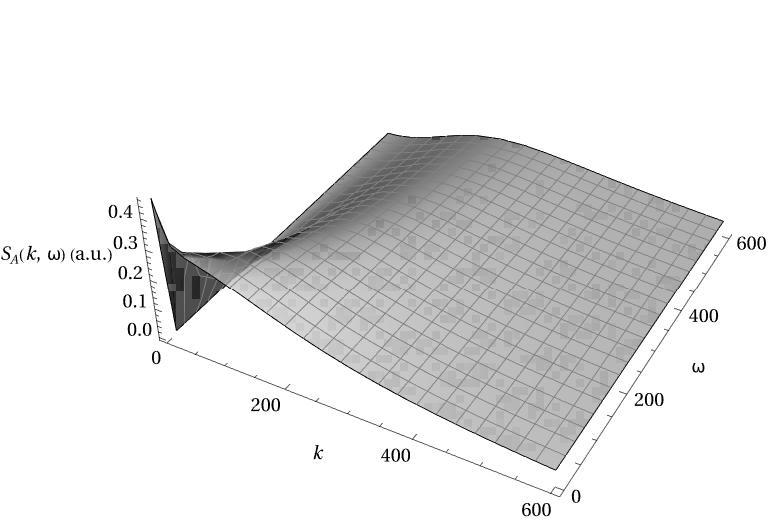}
            \caption{$S_\mathrm{A}(k, \omega)$}
            \label{fig:SAfull_intraAandB-s0_inter_s0}
        \end{subfigure}\hfill
        \begin{subfigure}[t]{0.48\linewidth}
            \centering
            \includegraphics[width=\linewidth]{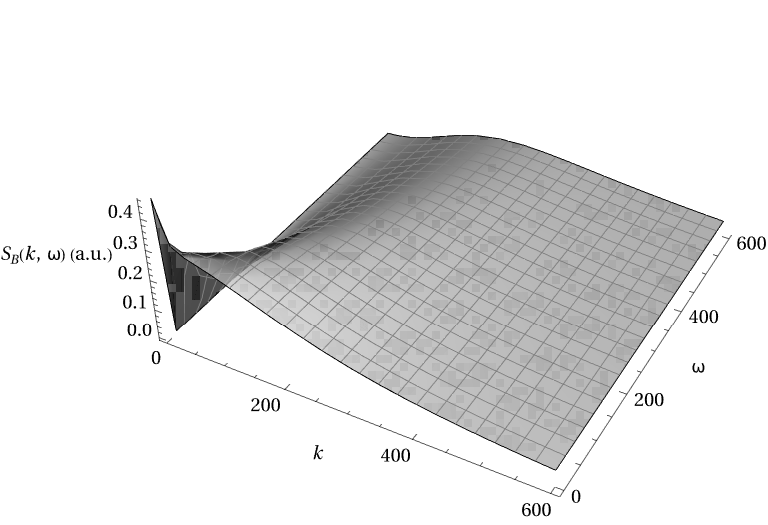}
            \caption{$S_\mathrm{B}(k, \omega)$}
            \label{fig:SBfull_intraAandB-s0_inter_s0}
        \end{subfigure}

        \begin{subfigure}[t]{0.48\linewidth}
            \centering
            \includegraphics[width=\linewidth]{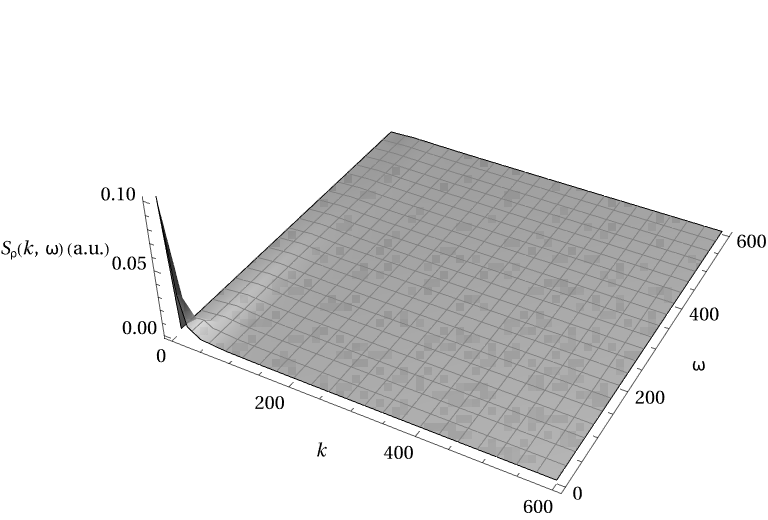}
            \caption{$S_\rho(k, \omega)$}
            \label{fig:Srhofull_intraAandB-s0_inter_s0}
        \end{subfigure}\hfill
        \begin{subfigure}[t]{0.48\linewidth}
            \centering
            \includegraphics[width=\linewidth]{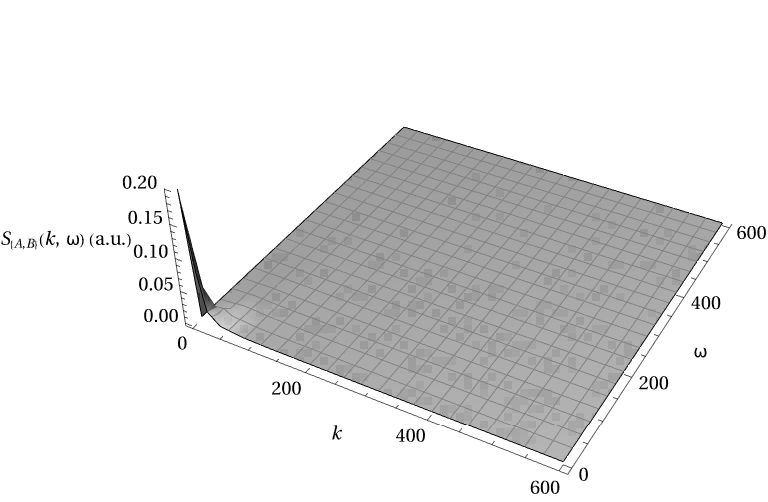}
            \caption{$S_\mathrm{AB}(k, \omega)$}
            \label{fig:SABfull_intraAandB-s0_inter_s0}
        \end{subfigure}

        \begin{subfigure}[t]{0.48\linewidth}
            \centering
            \includegraphics[width=\linewidth]{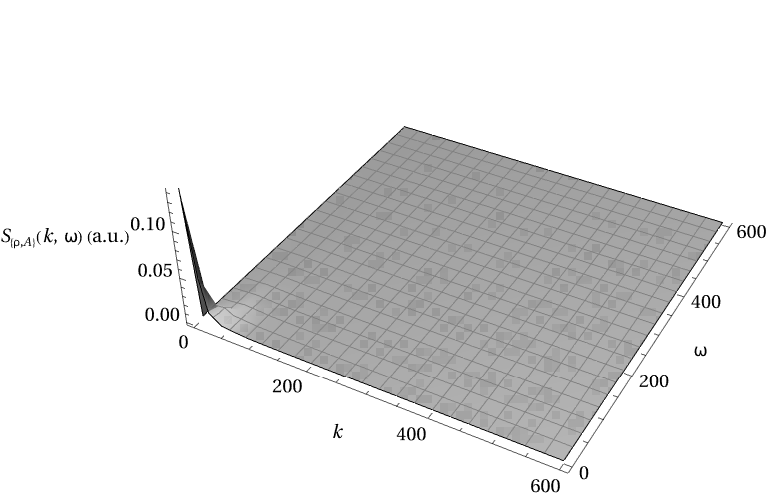}
            \caption{$S_{\rho_\mathrm{A}}(k, \omega)$}
            \label{fig:SrhoAfull_intraAandB-s0_inter_s0}
        \end{subfigure}\hfill
        \begin{subfigure}[t]{0.48\linewidth}
            \centering
            \includegraphics[width=\linewidth]{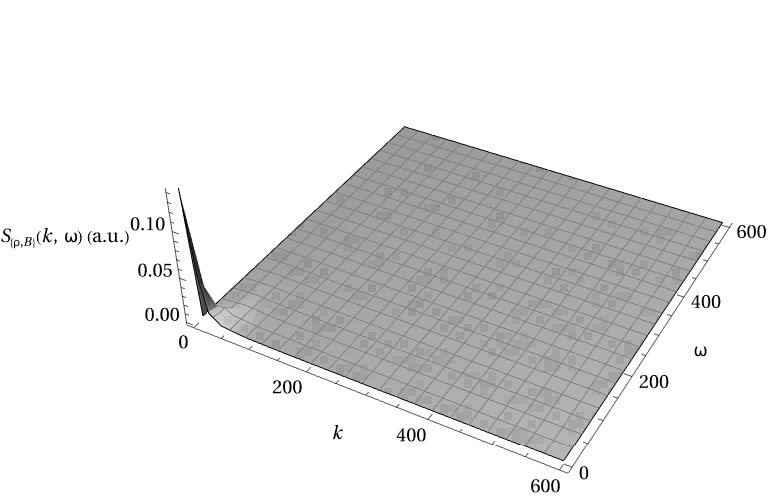}
            \caption{$S_{\rho_\mathrm{B}}(k, \omega)$}
            \label{fig:SrhoBfull_intraAandB-s0_inter_s0}
        \end{subfigure}
    \end{minipage}

    \caption[Dynamic structure factors for strong inter- and intra-species cross-linking]{Dynamic structure factors for cross-linked polymer mixture with strong inter- and intra-species cross-linking, approximating cross-linkers as point particles. Parameters: $\epsilon = 50$, $\mu = 50$, $\bar{\rho}_0 = 0.25$, $\bar{C}_A = 1$, $\bar{C}_B = 1$, $\gamma = 1$, $\kappa = 1$, $\lambda = 1$, $\gamma_A = 1$, $\gamma_B = 1$, $\alpha = 1$, $\tau = 1$, $L_A = 100$, $L_B = 100$, $v=2$.}
    \label{fig:full_intraAandB-s0_inter_s0}
\end{figure}

The dynamic structure factors for the combination of strong intra-species cross-linking and strong inter-species cross-linking is shown in Fig.~\ref{fig:full_intraAandB-s0_inter_s0}. Here the dynamic structure factors of the polymers, shown in Figs.~\ref{fig:SAfull_intraAandB-s0_inter_s0} and \ref{fig:SBfull_intraAandB-s0_inter_s0}, show the diffusive peaks at the origin that broaden as $k$ and $\omega$ increase. The sharp, narrow peaks at small $k$ and $\omega$ are very prominent, indicating that on long length and time scales the polymers exhibit typical diffusive behaviour with slow relaxation. As $k$ and $\omega$ increase, the peaks broaden and flatten in $\omega$ indicating that density fluctuations have no distinct relaxation frequency. At $\omega=0$ the broadening of the peak is also present as $k$ increases, with a gradual decay towards zero indicating that density fluctuations remain correlated at smaller length scales with a gradual decay towards zero at very large $k$.

The dynamic structure factor of the cross-linkers, shown in Fig.~\ref{fig:Srhofull_intraAandB-s0_inter_s0}, displays a typical narrow diffusive peak at the origin, with the cross-correlations of both polymers and cross-linkers in Figs.~\ref{fig:SABfull_intraAandB-s0_inter_s0}, \ref{fig:SrhoAfull_intraAandB-s0_inter_s0} and \ref{fig:SrhoBfull_intraAandB-s0_inter_s0} displaying the same shape. This indicates that density fluctuations are  correlated between both polymers as well as polymers and cross-linkers only at large length and time scales. 

\section{Molecular dynamics simulations}
\label{sec:MD}
\subsection{The model in LAMMPS}
\label{sec:LAMMPSmodel}
LAMMPS (Large-scale Atomic/Molecular Massively Parallel Simulator) \cite{thompsonLAMMPSFlexibleSimulation2022}, an open source code for classical molecular dynamics simulations widely used in the research community for both solid-state materials and soft matter systems, was utilised to perform molecular dynamics simulations of bead-spring polymers with cross-linking. All simulations were performed using reduced Lennard-Jones (LJ) units, in which the fundamental units of length $\sigma$, energy $\epsilon$, and mass $m$ were set to unity ($\sigma = \epsilon = m = 1$). Temperature is therefore measured in units of $\epsilon/k_B$, and time in units of $\tau = \sigma \sqrt{m/\epsilon}$. The discussion to follow outlines the system set-up and relevant parameters used.\\

Periodic boundary conditions were applied in all three spatial directions. Particles' trajectories were integrated using a velocity-Verlet scheme via the `\textit{fix nve}' command in LAMMPS, combined with a Langevin thermostat implemented via the `\textit{fix langevin}` command, to maintain the system temperature at roughly $T=1.0$. The damping parameter for the thermostat was set to $1.0$, and random seeds were provided to generate the stochastic thermal noise. A timestep of $\Delta t = 0.005$ was used throughout the simulations.\\

Polymer chains were represented as bead-spring chains, with beads connected using FENE (finite extensible nonlinear elastic) potentials given by

\begin{equation}
    U_\mathrm{FENE}(r) = -\tfrac{1}{2}k R_0^2 \ln{\left(1 - \left(\tfrac{r}{R_0}\right)^2\right) + 4 \epsilon \left[\left(\tfrac{\sigma}{r}\right)^{12} - \left(\tfrac{\sigma}{r}\right)^6\right] + \epsilon } 
    \label{eq:U_FENE} 
\end{equation}
for $r>R_0$. The first term describes the attractive bond whilst the remaining terms give a repulsive Lennard-Jones potential to avoid overlapping of bonded particles.  Here $k = 30.0$ is the bond strength and  $R_0 = 1.6$ is the maximum bond extension, as typically chosen for polymers, see e.g. \cite{huntComparisonModelLinear2009} .

In the analytical work in Sec.~\ref{sec:polymerCL}, the polymers were modelled as flexible chains, but biological filaments are more accurately modelled as semiflexible chains. Semiflexibility was introduced in the molecular dynamics simulations using a cosine angle potential 
\begin{equation}
    U_\theta = k_\theta (1-\cos\theta)
\end{equation}
applied between consecutive triplets of beads along each chain. The angular stiffness coefficient $k_\theta$ was specified as $k_\theta=20.0$.\\

Cross-linker particles  were modelled as two spherical beads bonded to one another.  FENE potentials as defined in \eqref{eq:U_FENE} were used for cross-linkers, with the same parameters as polymer bonds  with bond strength $k = 30.0$ and maximum bond extension $R_0 = 1.6$. \\

Non-bonded interactions were modelled using a Weeks-Chandler-Anderson (WCA) potential in LAMMPS via a truncated and shifted Lennard-Jones potential where the potential energy between two particles separated by a distance $r$ is given by
\begin{equation}
    U_{\mathrm{LJ-shifted}}(r) = 4 \epsilon \left[\left(\tfrac{\sigma}{r} \right)^{12}-\left(\tfrac{\sigma}{r} \right)^{6} \right] - 4 \epsilon \left[\left(\tfrac{\sigma}{r_\mathrm{c}} \right)^{12}-\left(\tfrac{\sigma}{r_\mathrm{c}} \right)^{6} \right]
\end{equation}
for $r < r_\mathrm{c}$, where $r_\mathrm{c}$ is the cutoff. At distances $r> r_\mathrm{c}$, the potential is zero such that there are no long-range interactions. Here the cutoff has been chosen at the minimum $r_\mathrm{c} =2^\frac{1}{6} \sigma$ for each particle pair to include only the repulsive part of the Lennerd-Jones potential, thereby modelling only excluded volume interactions. This is done so as not to incorporate additional effects such as clustering or clumping in order to investigate the effects of explicit bond formation due to cross-linking. 
The parameters for each type of particle pair were chosen as $\epsilon = 1.0$, $\sigma = 1.0$ and $r_\mathrm{c} =1.12$  for interactions between monomers of either polymer, $\epsilon = 1.0$, $\sigma = 2.0$ and $r_\mathrm{c}=2.24$ for interactions between cross-linker particles and $\epsilon = 1.0$, $\sigma = 1.5$,  $r_\mathrm{c}=1.68$ for interactions between cross-linker particles and monomers of either polymer.

Reversible bonding between polymer beads and cross-linkers were implemented using REACTOR \cite{gissingerREACTERHeuristicMethod2020a} via the LAMMPS command `\textit{fix bond/react}'. A two-step reaction process was identified:
\begin{itemize}
  \item Step 1: Binding of one bead of a cross-linker to a monomer in a polymer chain;
  \item Step 2: Formation of a second bond between the other bead of the cross-linker to a monomer, completing the cross-link, either on the same polymer chain (intra-chain) or between two polymer chains (inter-chain).
\end{itemize}

For more details regarding the implementation and parameters of this reversible bonding mechanism, please see the Supplementary Information.

The simulation workflow consisted of an equilibration run, a run with cross-linking implemented via `\textit{fix bond/react}' and a comparison run without cross-linking. In this system, $600$ polymers, $300 $ of each type of polymer, are included in a simulation box of  dimensions $100 \times 100 \times 100 $. Each of the $600$ polymers has a length of $40$ monomers. Also included in the simulation box are $2500$ cross-linkers with $2$ beads each to give a total of $29\,000$ particles in the box. 

\subsection{Results}
\label{sec:MDresults}
\begin{figure}[htbp]
    \centering

    \begin{subfigure}[b]{0.3\textwidth}
        \centering
        \includegraphics[width=\linewidth]{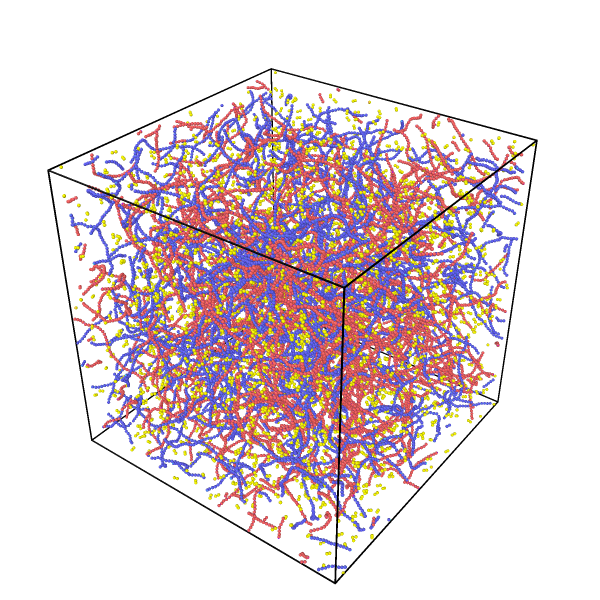}
        \caption{Initial configuration after equilibration run}
        \label{fig:network_initialconfig}
    \end{subfigure}

    \vspace{1em} 
    \begin{subfigure}[b]{0.3\textwidth}
        \centering
        \includegraphics[width=\linewidth]{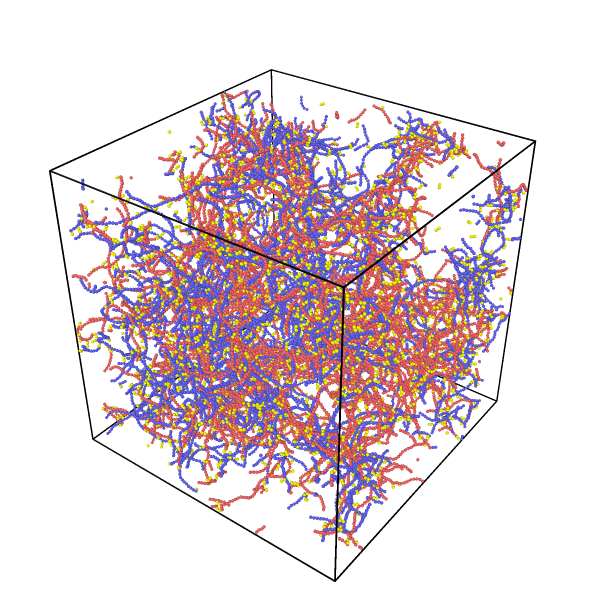}
        \caption{After cross-linking}
        \label{fig:network_aftercrosslinking}
    \end{subfigure}
    \hfill
    \begin{subfigure}[b]{0.3\textwidth}
        \centering
        \includegraphics[width=\linewidth]{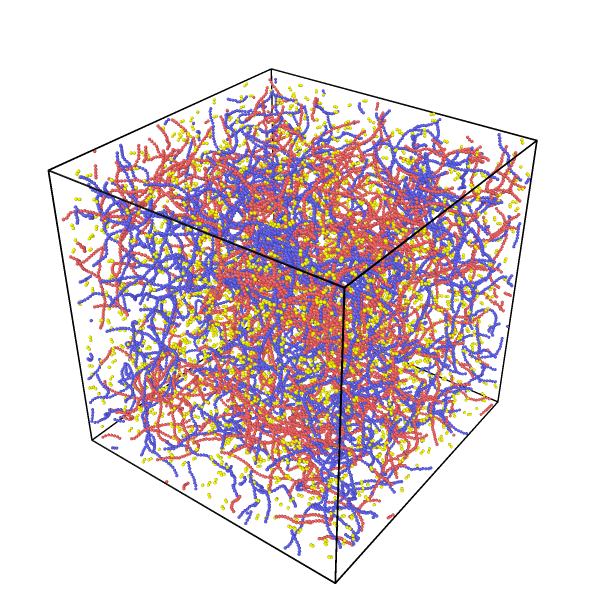}
        \caption{After run with cross-linking disabled}
        \label{fig:network_afternolinking}
    \end{subfigure}

    \caption[Snapshots from simulations of a cross-linked polymer network]{Snapshots from simulations showing polymer and cross-linker configurations at various points in the workflow. This run included $300$ polymers of type $\mathrm{A}$ (in red) and $300$ polymers of type $\mathrm{B}$ (in blue), with $40$ monomers each. There are also $2500$ cross-linkers (in yellow) in the simulation box of $100 \times 100 \times 100$ with periodic boundary conditions.}
    \label{fig:network_configurations}
\end{figure}

Figure~\ref{fig:network_configurations} shows snapshots of the simulation at the final timestep of the equilibration run in Fig. \ref{fig:network_initialconfig} and the final snapshots of the runs with and without cross-linking in Figs.~\ref{fig:network_aftercrosslinking} and \ref{fig:network_afternolinking}, respectively. The configuration shown in Fig.~\ref{fig:network_initialconfig}, is used as the initial configuration of the polymers and cross-linkers in both the runs with and without cross-linking.

\begin{figure}[htbp]
    \centering
    \includegraphics[width=0.8\linewidth]{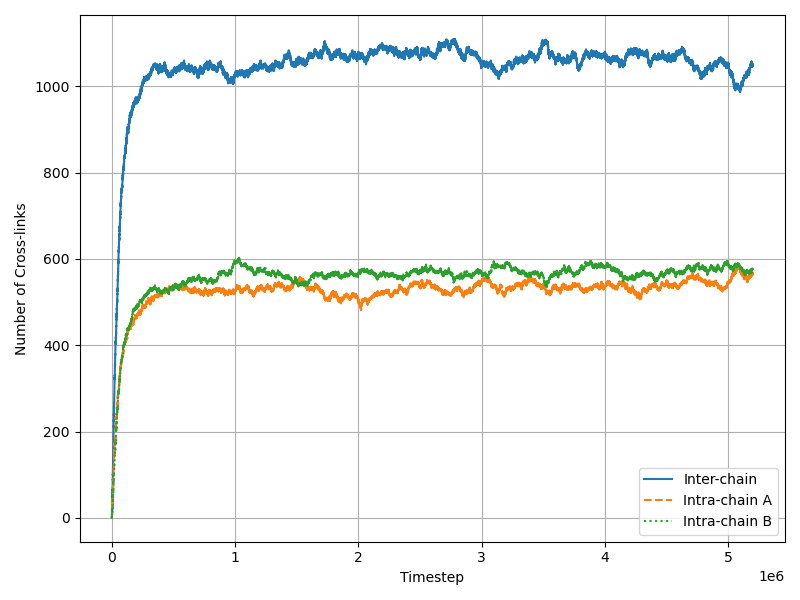}
    \caption[Time evolution of cross-link numbers in the polymer network]{The number of intra- and inter-species cross-links as a function of timestep throughout the simulation corresponding to the snapshot in Fig.~\ref{fig:network_aftercrosslinking}.}
    \label{fig:intraandintervst}
\end{figure}

In Figs.~\ref{fig:network_initialconfig} and \ref{fig:network_afternolinking} the polymers appear to be spread out throughout the simulation box, whilst in Fig.~\ref{fig:network_aftercrosslinking} cross-linking seems to result in clumping of polymers with the consequence of leaving empty spaces. Cross-linkers, on the other hand, seem to be mixed in between the polymers in all three cases. In the cross-liked example, the cross-linkers are less visible, likely due to bonding with polymers, but some can still be seen diffusing freely in empty spaces where no polymers are present. To confirm this, the number of cross-links that form throughout the simulations can be plotted as a function of timestep. This is done in Fig.~\ref{fig:intraandintervst}.

In Fig.~\ref{fig:intraandintervst} there is an initial sharp rise in both intra- and inter-species cross-link numbers, followed by fluctuations around a steady state as cross-links continue to break and form. This steady state seems to be reached at around $500\, 000$ timesteps. There are far more inter-species cross-links than either of the intra-chain $\mathrm{A}$ and intra-chain $\mathrm{B}$, with the number of inter-species cross-links fluctuating between $1000 -1200$ and both  intra-species cross-links fluctuating between $500-600$. Thus the total number of cross-linkers fluctuates between $2000 - 2400$ leaving at least $100$ cross-linkers either bonded only at one end, or not bound to a polymer at all. This corresponds to the visual observation of free cross-linkers in the snapshot in Fig.~\ref{fig:network_aftercrosslinking}.

To further quantify the effects of cross-linking, some of the polymer properties can be investigated and compared for the simulations with and without cross-linking. One such property is the persistence length, which gives a measure of the length scale on which a polymer is statistically straight \cite{Rubinstein}. There are various ways in which the persistence length can be calculated \cite{cifraDifferencesLimitsEstimates2004}, here the persistence length $\ell_\mathrm{p}$ of each chain is calculated according to the preferred method motivated in Ref.~ \cite{devilliersSimulationConfinedTethered2024} as follows
\begin{equation}
\ell_\mathrm{p} = -\frac{\langle \ell_{\text{bond}} \rangle}{\mathrm{ln}(\langle \cos \theta \rangle)}
\end{equation}
where $\langle \ell_{\text{bond}} \rangle$ is the average bond length and $\langle \cos \theta \rangle$ is the average cosine of the angle $\theta$ between adjacent bond vectors. Averages are taken over all angles along a polymer chain and over all time steps. Persistence lengths were calculated from the simulation data for both the run with cross-linking and without and used to calculate a mean value and standard deviation for the persistence length of each polymer in both cases. These results are shown in Table~\ref{tab:persistence_lengths}.

\begin{table}[htbp]
\centering
\begin{tabular}{lcc}
\hline
\textbf{Run} & \textbf{Mean $\ell_\mathrm{p}$} & \textbf{Standard Deviation} \\
\hline
With cross-linking & $14.3281$ & $ 0.2257$ \\
No linking         & $18.9712$ & $ 0.0458$ \\
\hline
\end{tabular}
\caption{Mean persistence lengths with and without cross-linking.}
\label{tab:persistence_lengths}
\end{table}

As seen in Table~\ref{tab:persistence_lengths}, the value of the mean persistence length for the system without cross-linking is around what is expected for an angle potential with a coefficient of $k=20$. The mean persistence length for the polymers without cross-linking is significantly lower, even taking into account the larger standard deviation. Thus this suggests that cross-linking seems to reduce the length scale on which the polymers are oriented along a straight line. This could be due to the cross-links' forcing polymers into configurations in which they bend more than they naturally would. If polymers were cross-linked to one another with similar orientations such that the polymers create bundles of aligned cross-linked polymers, one might expect the persistence length to increase with cross-linking. This phenomenon is, however, not observed here and indicates that bundles of this type are not occurring, and that clusters of randomly oriented polymers are more frequently forming. 

To further investigate the alignment of the polymers, we  also utilise a local order parameter defined using the second Legendre polynomial (see \cite{chaikinPrinciplesCondensedMatter2013}) as
\begin{equation}
S(r) = \frac{1}{2N_r} \sum_{i \neq j}
\left[ 3 \left( \hat{\mathbf{b}}_i \cdot \hat{\mathbf{b}}_j \right)^2 - 1 \right] 
\cdot \Theta\left( r - \left| \mathbf{r}_i - \mathbf{r}_j \right| \right)
\label{eq:order_param}
\end{equation}
where $r$ is the cut-off radius, $\hat{\mathbf{b}}_i$ and $ \hat{\mathbf{b}}_j$ are the bond vectors corresponding to a pair of monomers with positions $ \mathbf{r}_i$ and $ \mathbf{r}_j$, which may be on the same or different polymer chains. $N_r$ is the number of pairs that fall within the cut-off radius from one another and contribute to the value of $S(r)$. A value of $S(r)=1$, indicates strong parallel local alignment of bond vectors and $S(r)=0$ corresponds to random orientation.

\begin{figure}[htbp]
    \centering
    \includegraphics[width=0.75\linewidth]{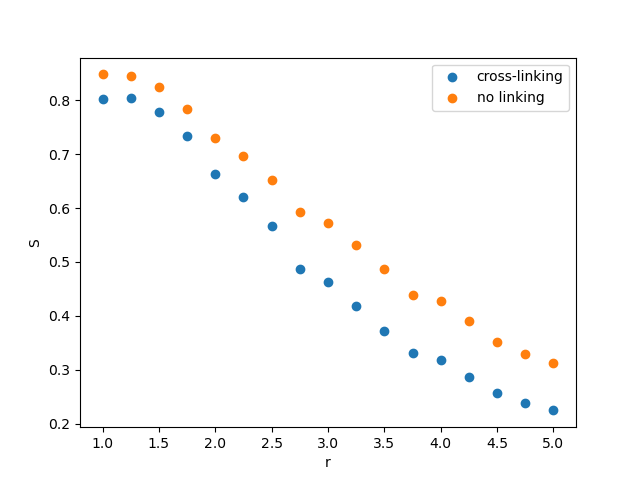}
    \caption[Local order parameter as a function of the cut-off radius]{Local order parameter as a function of the cut-off radius corresponding to eq.~\eqref{eq:order_param}, for simulations with cross-linking and without cross-linking, shown in the snapshots in Figs.~\ref{fig:network_aftercrosslinking} and \ref{fig:network_afternolinking}, respectively.}
    \label{fig:order_parameter}
\end{figure}

Figure \ref{fig:order_parameter} shows the order parameter $S(r)$ from eq.~\eqref{eq:order_param} calculated for the final configurations of the simulations with and without cross-linking, shown in Figs.~\ref{fig:network_aftercrosslinking} and \ref{fig:network_afternolinking}, as a function of the cutoff radius $r$. Both cases show an overall trend of a decrease in the order parameter as the cutoff radius is increased, with the order parameter being consistently lower in the cross-linking case than without cross-linking. In the cross-linking case the order parameter is at a maximum at around $0.8$ for a cutoff radius of $1$ indicating that bond vectors are largely aligned on this short scale, but become less and less aligned, the larger the region being considered becomes. The order parameter drops almost to $0.2$ for a radius of $5$ indicating that bonds are somewhat randomly aligned over this length scale. The same trend is observed in the scenario without cross-linking, but values remain consistently higher than in the cross-linking case. This indicates that the addition of cross-linking tends to add to the randomness of the orientation of the bond vectors. This is consistent with the finding of decreased mean persistence length of the polymers, which indicates that the cross-linkers are not contributing to the formation of parallel-aligned bundles of polymers, but rather introducing larger angles and orientations of polymers to one another. 

From the preceding discussion, it is clear that the addition of cross-linking significantly affects the configurations of the polymers in the system being considered. It has also been established that the cross-linkers, both form and break bonds throughout the simulation and may therefore affect the overall dynamics of the system. To investigate this further and to compare these simulations more directly to the analytical results derived hitherto, the dynamic structure factors of these simulated systems also are considered.

To obtain dynamic structure factors for the simulated polymer system, the open-source tool \textit{Dynasor} \cite{franssonDynasorToolExtracting2021} was used. Dynasor is a Python-based package designed to compute static and dynamic structure factors from molecular dynamics trajectories. This makes it well suited  not only to further investigation of the effects of cross-linking on the collective dynamics of the polymers in the simulated system, but also allow for comparing with the previously derived analytical results for the dynamic structure factors of a cross-linked polymer network in Sec.~\ref{sec:polymerCL}. The following discussion outlines the application of the tool to the simulation data and presents the resulting dynamic structure factors. 

To obtain suitable dump files of the trajectories in the molecular dynamics simulation from LAMMPS, the runs, with and without cross-linking are performed for $5\, 200\,000$ timesteps, with the trajectory written to a file every $10$ timesteps for the last $200\,000$ timesteps. This long run ensures that, both cross-link numbers and polymer dynamics have reached their equilibrium values, with the frequent sampling in the final timesteps allowing for better temporal resolution for the Dynasor analysis. Note that, from here onwards, the symbol $q$ is used for the spatial Fourier transform, where $k$ has been used in the theoretical work, for consistency with the Dynasor documentation.

\begin{figure}[htbp]
    \centering
    \begin{subfigure}[t]{0.45\linewidth}
    \centering
    \includegraphics[width=\linewidth]{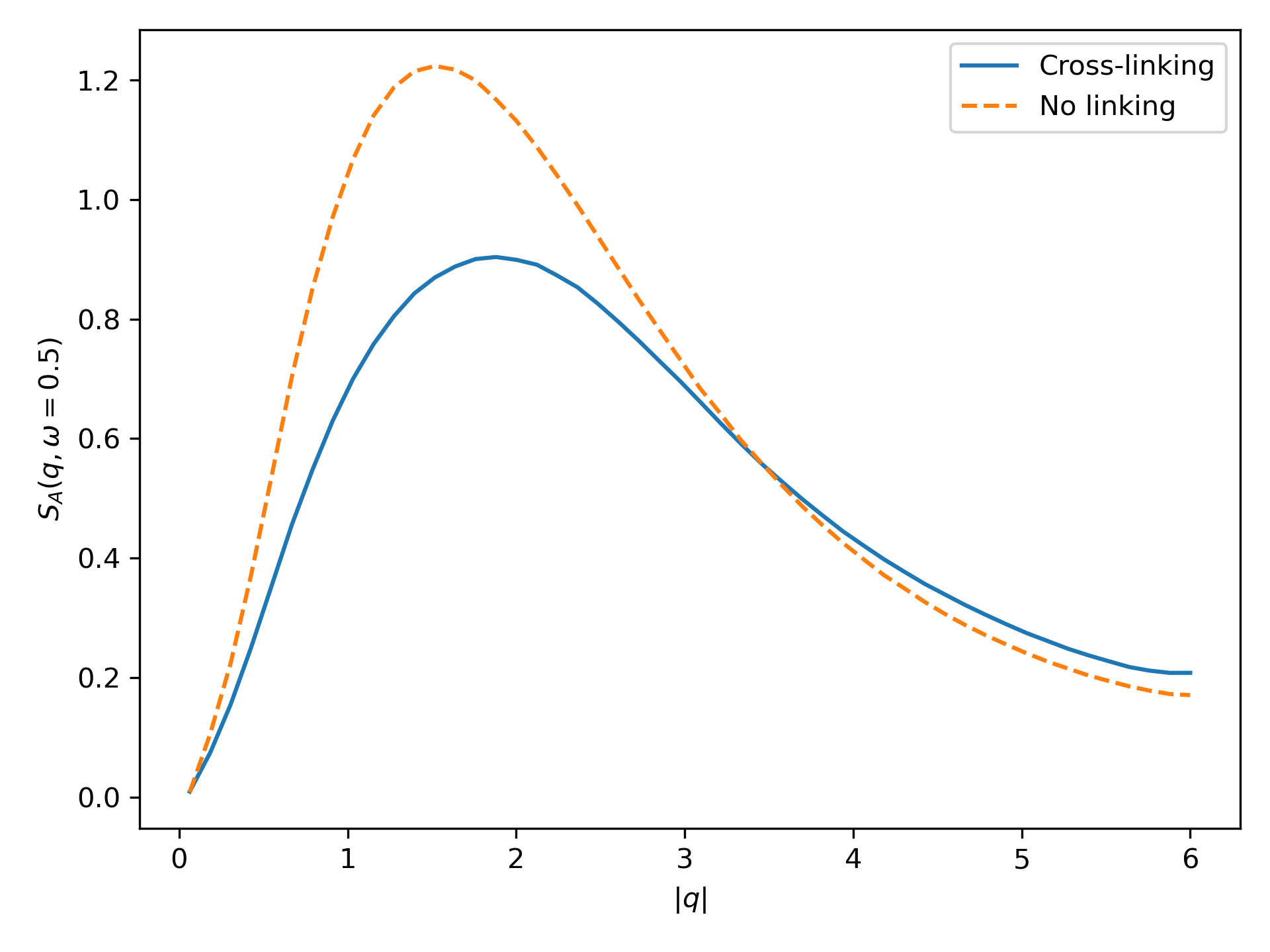}
    \caption{$S_\mathrm{A}(q, \omega=0.5)$}
    \label{fig:SAomega05} 
    \end{subfigure}
    \begin{subfigure}[t]{0.45\linewidth}
    \centering
    \includegraphics[width=\linewidth]{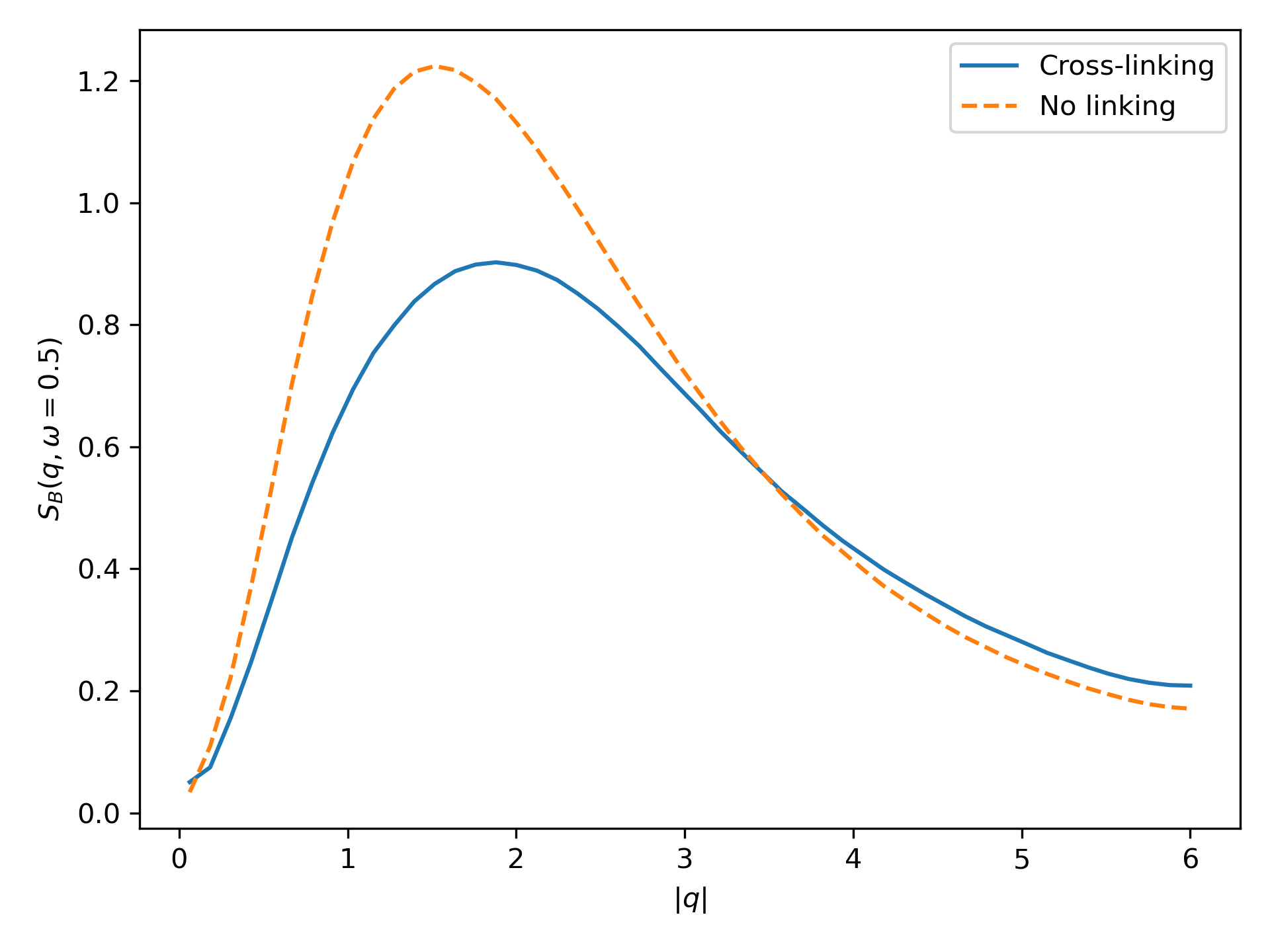}
    \caption{$S_\mathrm{B}(q, \omega=0.5)$}
    \label{fig:SBomega05} 
    \end{subfigure}

    \vspace{1em}
    
    \begin{subfigure}[t]{0.45\linewidth}
    \centering
    \includegraphics[width=\linewidth]{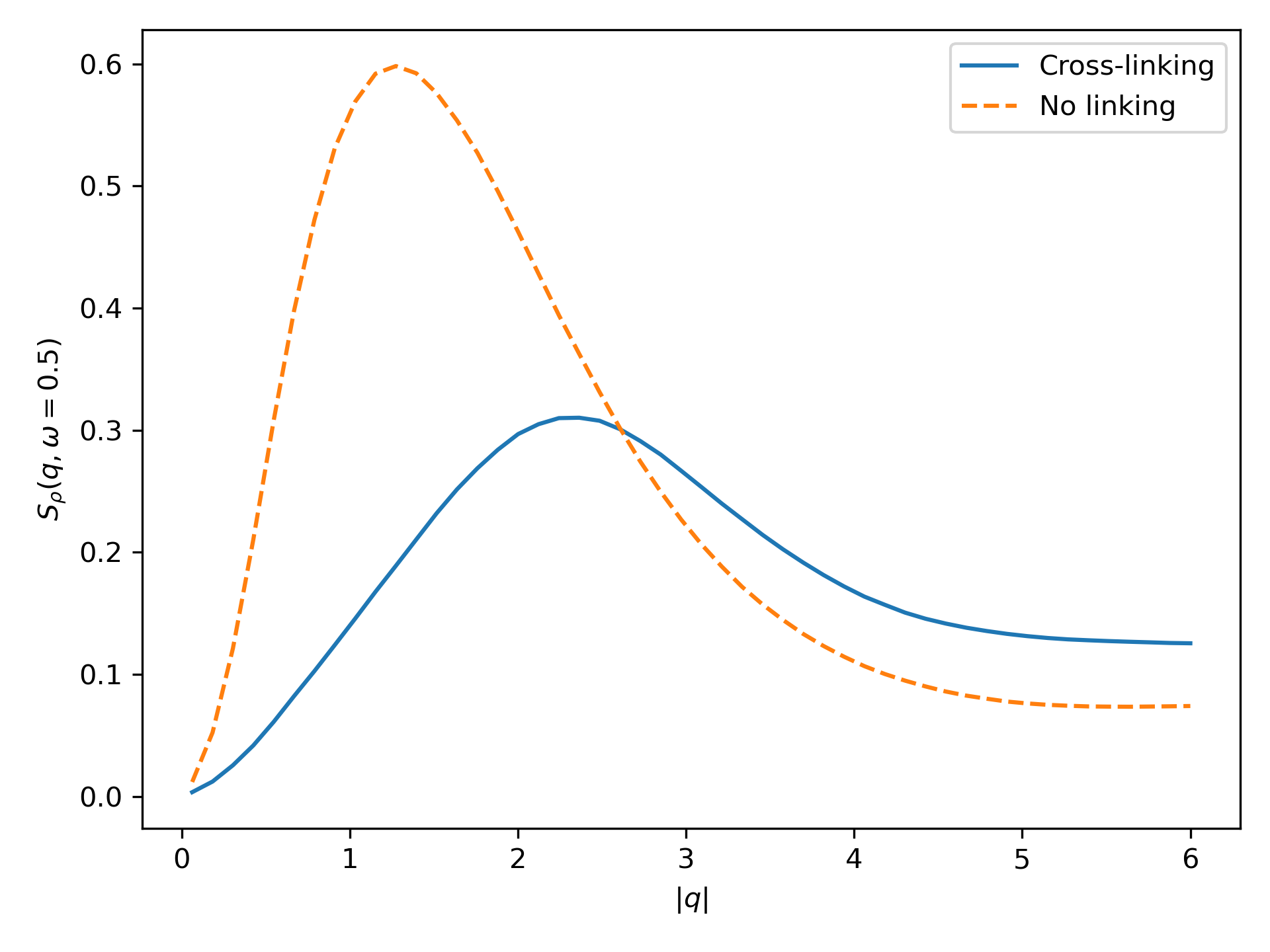}
    \caption{$S_\rho(q, \omega=0.5)$}
    \label{fig:Srhoomega05} 
    \end{subfigure}

    \caption{Comparison of the dynamic structure factors with and without cross-linking at $\omega = 0.5$.}
    \label{fig:Somegafixed}
\end{figure}

\begin{figure}[htbp]
    \centering
    \begin{subfigure}[t]{0.45\linewidth}
    \centering
    \includegraphics[width=\linewidth]{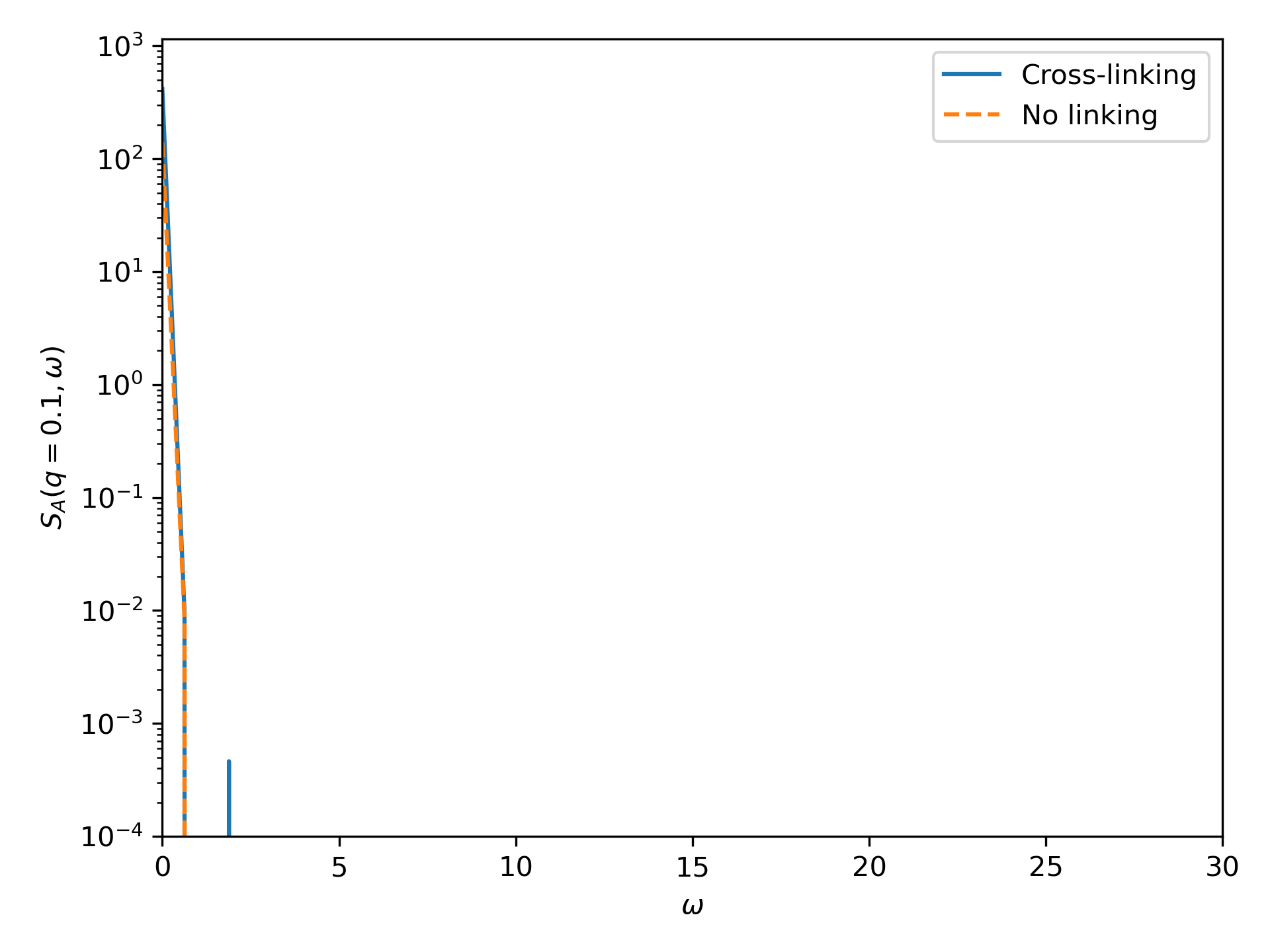}
    \caption{$S_\mathrm{A}(q=0.1, \omega)$}
    \label{fig:SAq01} 
    \end{subfigure}
    \begin{subfigure}[t]{0.45\linewidth}
    \centering
    \includegraphics[width=\linewidth]{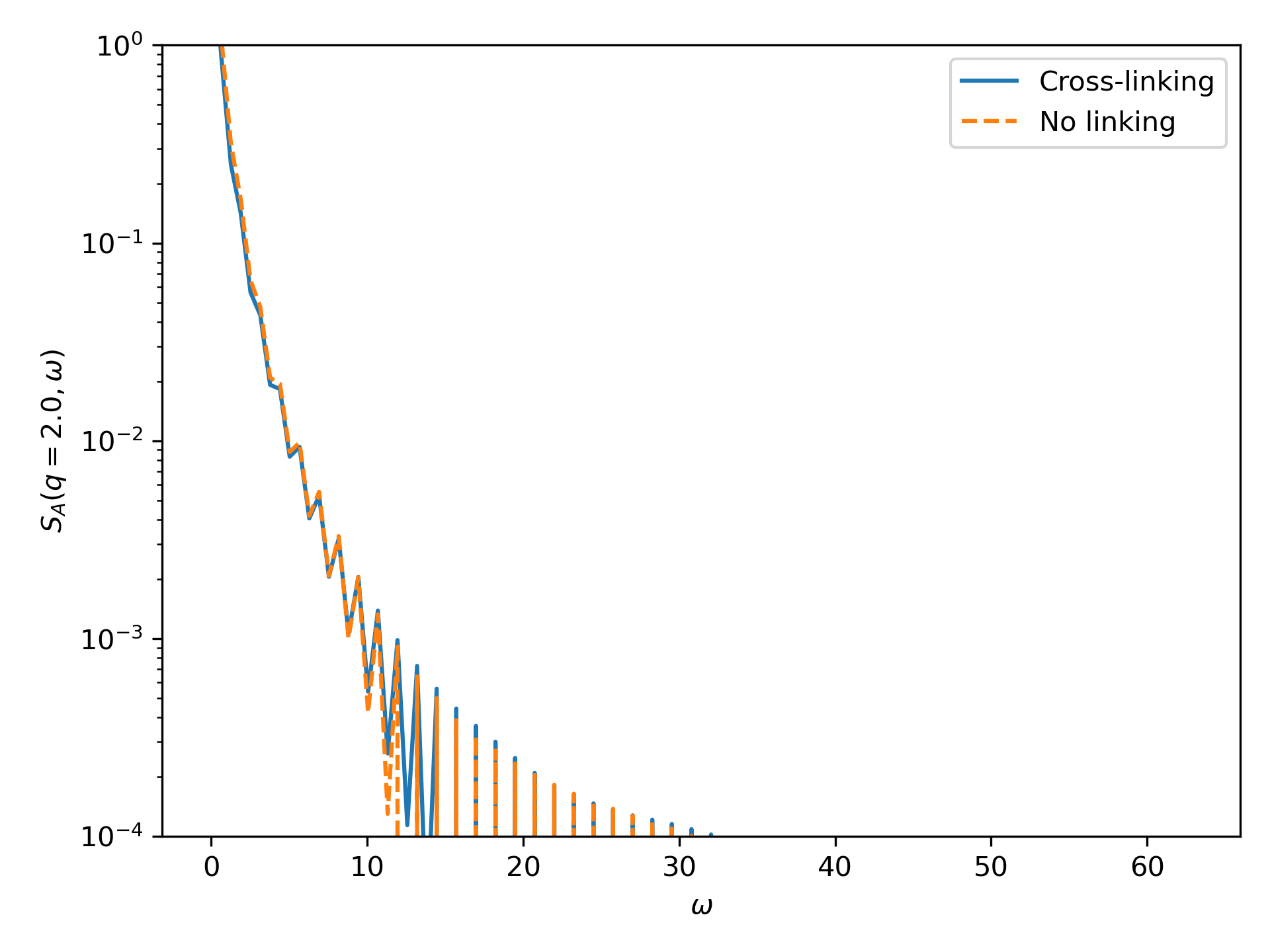}
    \caption{$S_\mathrm{A}(q=2.0, \omega)$}
    \label{fig:SAq2} 
    \end{subfigure}

    \vspace{1em}
    
    \begin{subfigure}[t]{0.45\linewidth}
    \centering
    \includegraphics[width=\linewidth]{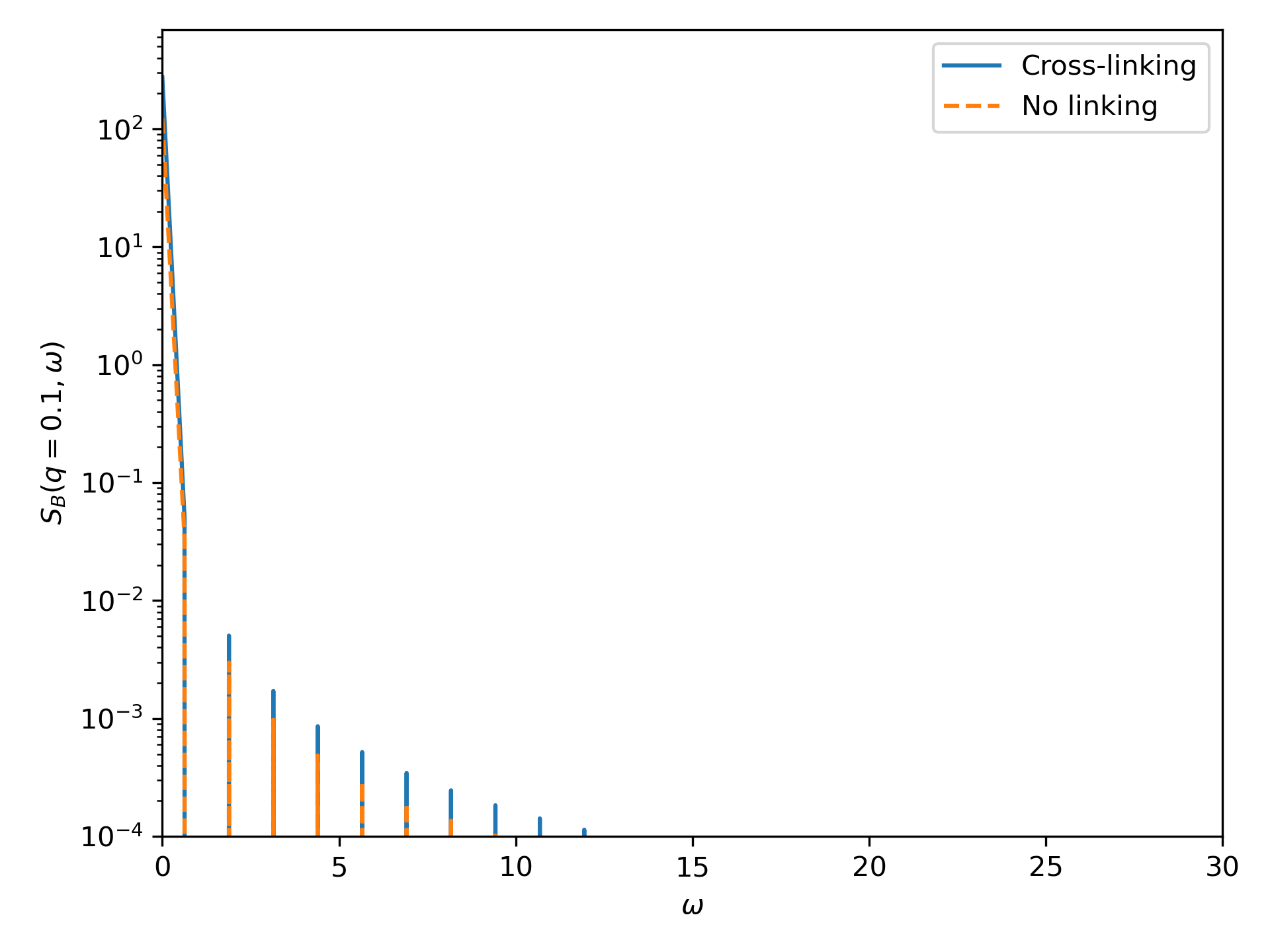}
    \caption{$S_\mathrm{B}(q=0.1, \omega)$}
    \label{fig:SBq01} 
    \end{subfigure}
    \begin{subfigure}[t]{0.45\linewidth}
    \centering
    \includegraphics[width=\linewidth]{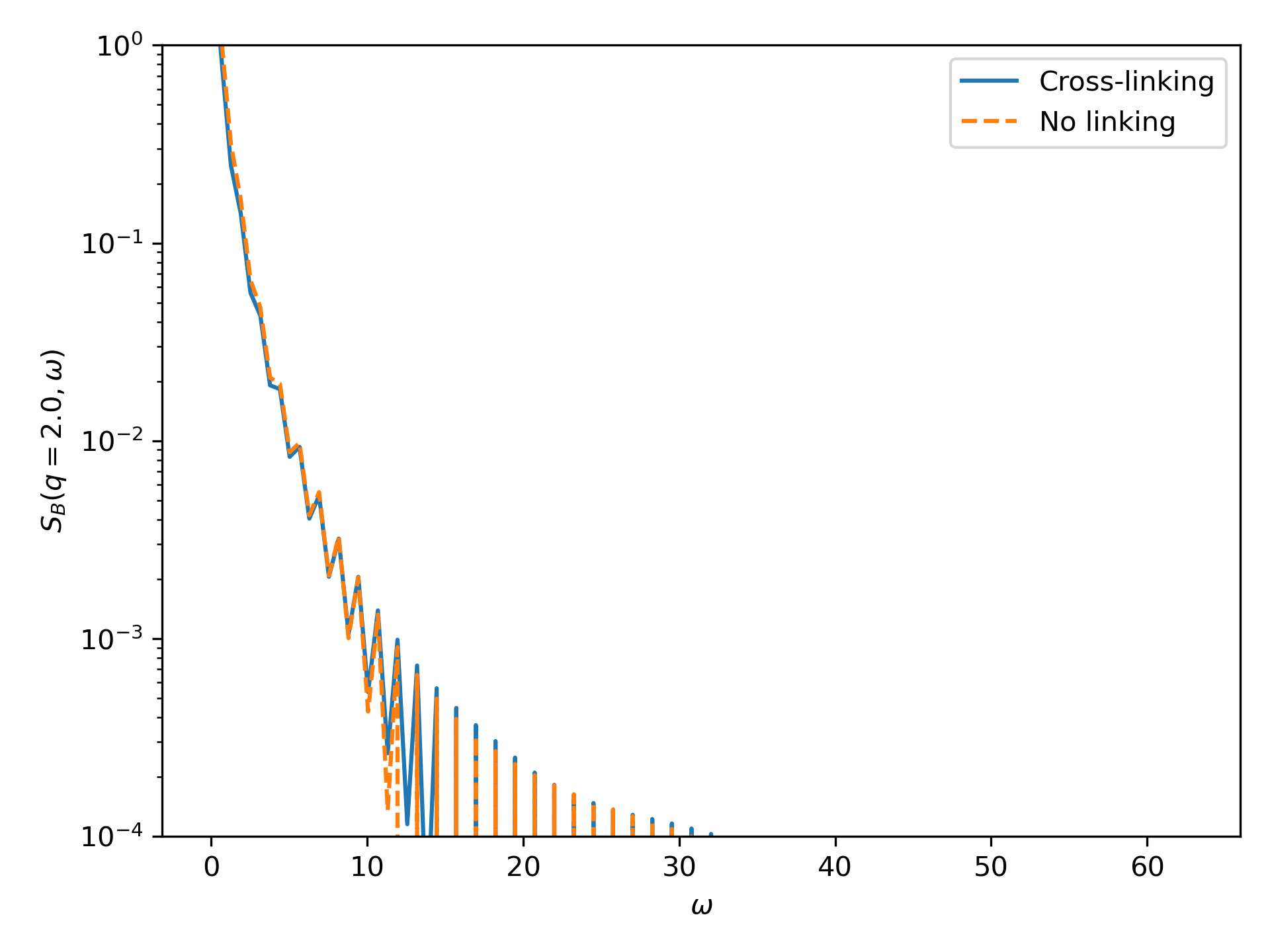}
    \caption{$S_\mathrm{B}(q=2.0, \omega)$}
    \label{fig:SBq2} 
    \end{subfigure}

    \vspace{1em}
    
   \begin{subfigure}[t]{0.45\linewidth}
    \centering
    \includegraphics[width=\linewidth]{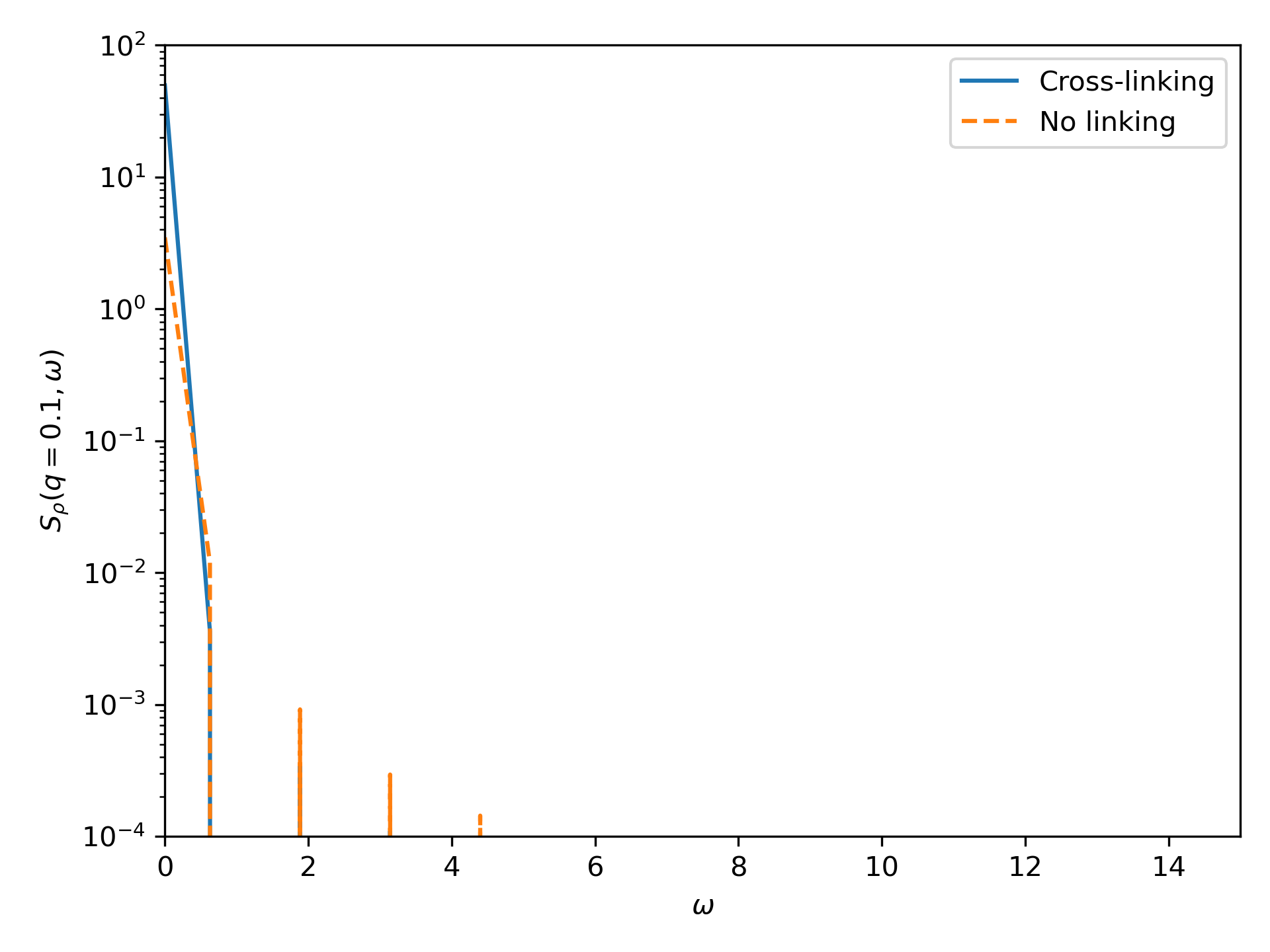}
    \caption{$S_\rho(q=0.1, \omega)$}
    \label{fig:Srhoq01} 
    \end{subfigure}
    \begin{subfigure}[t]{0.45\linewidth}
    \centering
    \includegraphics[width=\linewidth]{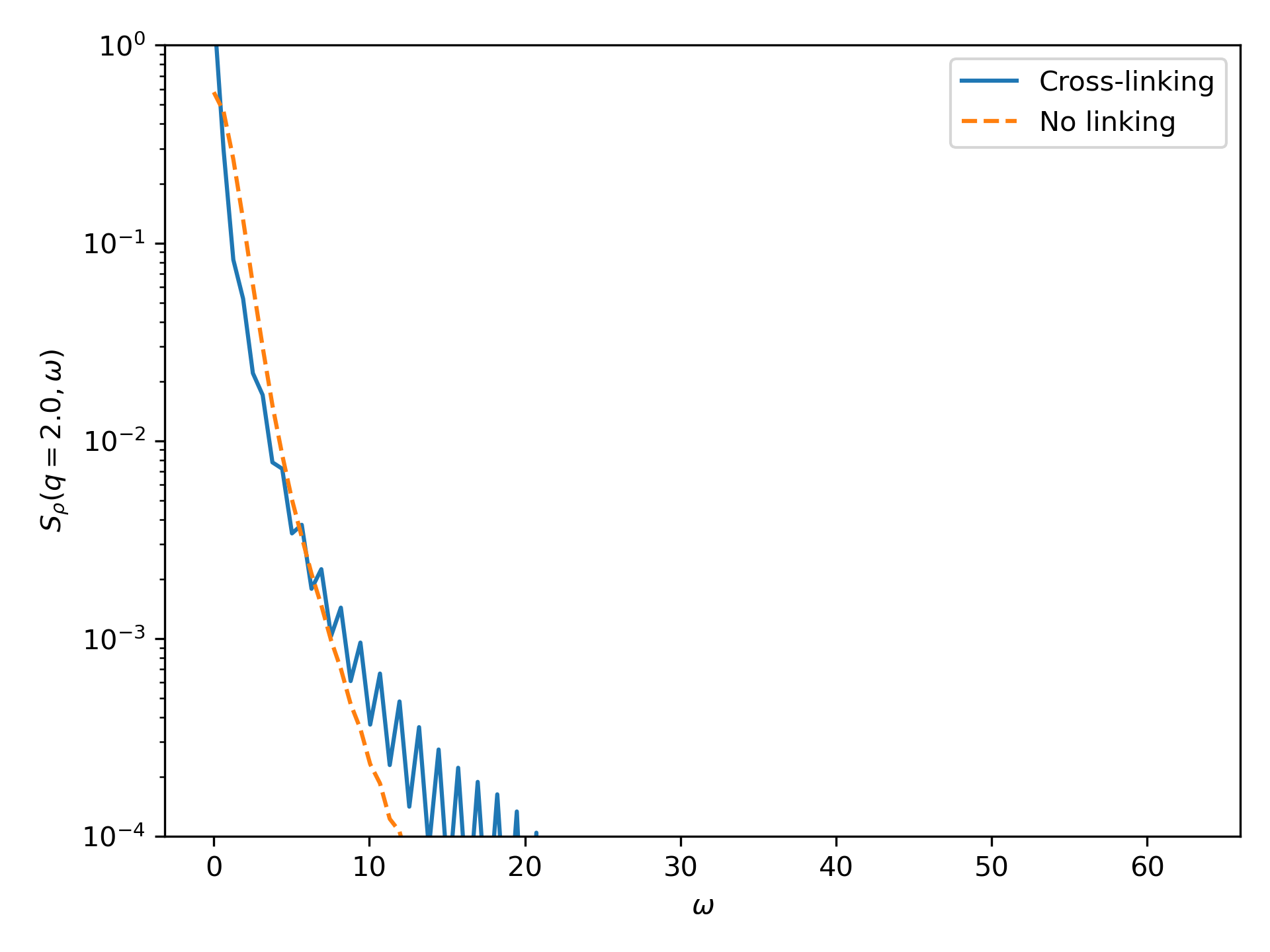}
    \caption{$S_\rho(q=2.0, \omega)$}
    \label{fig:Srhoq2} 
    \end{subfigure}

    \caption{Comparison of the dynamic structure factors with and without cross-linking at $q= 0.1$ and $q=2.0$.}
    \label{fig:Sqfixed}
\end{figure}

 Figure ~\ref{fig:Somegafixed} shows slices of the dynamic structure factors of the polymers in Figs.~\ref{fig:SAomega05}--\ref{fig:SBomega05} and that of the cross-linkers in Fig.~\ref{fig:Srhoomega05} for $\omega =0.5$. In all three cases, the peak for the simulation without cross-linking is higher and narrower than the peak for the simulation with cross-linking, but at higher $q$-values the dynamic structure factor for the cross-linking simulation has higher values than that of the simulation with no linking. This indicates that cross-linking tends to result in the correlation of density fluctuations becoming more prominent towards higher $q$ or smaller length scales.  In Fig.~\ref{fig:Srhoomega05}, the lower broader peak corresponding to the cross-linking simulation is also shifted such that the maximum occurs at a higher $q$. This indicates that the short length scale density fluctuations are even more prominent for the cross-linkers than for the polymers when cross-linking is implemented. This is also evident in the significantly higher long wavelength \textit{tail} of the peak for the cross-linking case in Fig.~\ref{fig:Srhoomega05}. In all three cases, the values plateau at large $q$ instead of falling off to zero. This is likely due to spectral leakage at low $\omega$ caused by finite-time windowing and statistical noise in the frequency dependence (as shown in Figs.~\ref{fig:Srhoq01} --\ref{fig:Srhoq2}) originating from the temporal binning and discrete Fourier transforms used in the analysis of the molecular dynamics trajectories. The three-dimensional plots of the dynamics structure factors provided in the Supplementary Information further show that these plateaus rapidly diminish with increasing $\omega$, thereby supporting that this arises from low-frequency leakage rather than a physical effect.

To further investigate the frequency dependence, the dynamic structure factors for the polymers are shown in Figs.~\ref{fig:SAq01} --\ref{fig:SBq2} and for the cross-linkers in Figs.~\ref{fig:Srhoq01} --\ref{fig:Srhoq2} for $q=0.1$ and $q=2.0$, comparing the values obtained from the simulations with and without cross-linking. These plots are noisy in comparison to those with fixed $\omega$-values in Fig.~\ref{fig:Somegafixed}. This is due to the way the dynamic structure factors were post-processed in Dynasor, with spherical averaging of the values over $q$, but no averaging in $\omega$. Since no built-in function for $\omega$-averaging was available in Dynasor at the time of writing, this step was not implemented, but could be considered in future work.

For the polymers, at both $q=0.1$ and $q=2.0$, in Figs. ~\ref{fig:SAq01}--\ref{fig:SBq2}, the peaks decay strongly from large values at small $\omega$ as evident in the logarithmic scale of the vertical axis, with the values corresponding to the cross-linking case in blue, slightly higher than the values in the no linking case. There appears to be some asymmetry between the $\mathrm{A}$ and $\mathrm{B}$ polymers in the $\omega$ behaviour at low $q$ in Figs.~\ref{fig:SAq01} and \ref{fig:SBq01}, but this is likely due to noise and different spatial configurations associated with the particular length scale considered here. Very slight differences in the dynamic structure factors near $q=0$ can also be seen in the plots for $\omega=0.5$ in Figs.~\ref{fig:SAomega05}--\ref{fig:SBomega05}.

For the cross-linkers, the dynamic structure factors in Figs.~\ref{fig:Srhoq01}--\ref{fig:Srhoq2}, also show the rapid decay of the peaks. The dynamic structure factor values corresponding to the no linking simulation being higher at long length scales where $q=0.1$ in Fig.~\ref{fig:Srhoq01}. At smaller length scales where $q=2.0$ in Fig.~\ref{fig:Srhoq2}, the dynamic structure factor seems to be higher in the cross-linking case near $\omega=0$ and at larger $\omega$-values. This corresponds to the shift in the peak of the dynamic structure factor to larger $q$-values seen in Fig.~\ref{fig:Srhoomega05}. Values of  the dynamic structure factor are higher for the no linking case on large length scales, or small $q$ where the peak of its dynamic structure factor occurs; when $q=2.0$, this is closer to the peak of the dynamic structure factor in the cross-linking case. The $\omega$ behaviour of the dynamic structure factor near this peak in the cross-linking case when $q=2.0$, shows an initial sharp peak, followed by a broadening in $\omega$ at larger $\omega$-values. This once again indicates, that density fluctuations of the cross-linkers become more strongly correlated on shorter time scales due to cross-linking.

\section{Conclusion}\label{sec:conclusion}

This paper extends the dynamical networking formalism presented in Ref.~\cite{dutoitDynamicalNetworkingUsing2025} to model cross-linked polymers, assigning different statistical advantages to intra- and inter-species cross-linking. Effective potentials due to networking were derived separately for strong intra- and inter-species cross-linking and combined to reconstruct the dynamic structure factors for the full system with both types of cross-linking. Cross-linking resulted in a broadening of the dynamic structure factors of both polymers and cross-linkers, with cross-correlations indicating that density fluctuations between polymers, as well as between polymers and cross-linkers, are correlated only at large length and time scales.

To validate and complement the analytical findings, Sec.~\ref{sec:MD} presents molecular dynamics simulations of two species of semi-flexible polymers with both intra- and inter-species cross-linking. Simulations showed that cross-linking is reversible and reduces polymer persistence length and local alignment, in comparison to systems without cross-linking. The resulting trajectories were further used to compute dynamic structure factors, revealing the same qualitative trends observed in theory, namely a broadening of the diffusive peaks and the emergence of higher tails. Larger-scale simulations and analysis with improved temporal resolution and windowing are needed to reduce noise in the frequency dependence of the dynamic structure factors and to quantify the effects of spectral leakage on the higher-$q$ tails at small $\omega$.

Taken together, the analytical framework and molecular dynamics simulations provide complementary perspectives, both showing the same qualitative trend of cross-linking leading to broadened diffusive peaks in the dynamic structure factor. The dynamic structure factors for polymers with and without cross-linking are expected to show only minor differences. In weak gelation, for example, cooperative diffusion coefficients are expected to be similar, with and without cross-linking, while more noticeable differences appear in quantities such as viscosity~\cite{gennesScalingConceptsPolymer1991}. This expectation is also supported by recent experimental evidence, where similar dynamic structure factors are observed using quasielastic neutron scattering in gel and solution states of weakly cross-linked polymer networks, with slightly higher tails corresponding to the gel state~\cite{aomuraQuasielasticNeutronScattering2023}. These findings suggest that the theoretical framework developed here captures the essential features of the system.

The experimental work in Ref.~\cite{aomuraQuasielasticNeutronScattering2023} also opens interesting avenues for future exploration within the theoretical framework presented here, particularly regarding the effects of cross-linking when polymer networks undergo deformation. Such scenarios are accessible within both the theoretical framework and the molecular dynamics approach developed in this work, providing a pathway for studying how cross-linking influences the mechanical response of polymer networks. In addition, the framework can be extended to incorporate active cross-linkers, enabling investigations of the stability and dynamics of active networks such as the cytoskeleton.

In conclusion, the theoretical framework for modelling the dynamics of cross-linked polymers at the mesoscopic scale captures qualitative trends observed in both molecular dynamics simulations and experimental work. It also lends itself to further development and application, offering a promising framework for understanding and predicting the dynamics of polymer networks in both synthetic and biological contexts.

\section*{Supplementary information}
Supplementary material (available as a PDF) provides extended simulation details, including descriptions of the LAMMPS setup for the two-polymer system with reversible cross-linking, implementation of reversible bonds using the REACTER package, and trajectory analyses performed with Dynasor. Additional figures supporting the main text are also included. The online version of this article contains the Supplementary Information file available at [DOI to be added by publisher].

\section*{Acknowledgements}

The authors would like to thank Mesfin Tsige for helpful discussions during the early stages of developing the molecular dynamics simulations presented in this work. This work is based on the research supported
	
	in part 
	by the National Research Foundation of South Africa (Grant Numbers 99116, MND210620613719 and PMDS240820261081). \\
	KKMN would like to thank the Isaac Newton Institute for Mathematical Sciences, Cambridge, for support and hospitality during the programme ``New statistical physics in living matter: non equilibrium states under adaptive control'' where some of the work on this paper was undertaken. This work was supported by EPSRC grant no EP/R014604/1.\\
    GP acknowledges support by the NICIS CHPC, South Africa, under grant allocation MATS0887.

\section*{Declarations}

\begin{itemize}
\item Funding: as per acknowledgements.
\item Competing interests: The authors have no financial or proprietary interests in any material discussed in this article. 
\item Ethics approval and consent to participate: Not applicable
\item Consent for publication: Not applicable
\item Data availability: All data supporting the findings of this study are described in the main text and Supplementary Information. The raw simulation data are not publicly available due to file size considerations, but processed data and analysis details are provided in the manuscript and Supplementary Information. Further information is available from the corresponding author upon reasonable request.
\item Materials availability: Not applicable
\item Code availability: The simulations were performed using the open-source molecular dynamics package LAMMPS, and the dynamical analysis was carried out using the Python package Dynasor. Custom analysis scripts are described in the Supplementary Information. No additional code was developed or deposited.
\item Author contribution: Conceptualisation of theoretical work (NdT and KKMN), Calculations (detailed calculations and numerical work: NdT; approximation approach, checks: NdT and KKMN), Conceptualisation of simulations (NdT, KKMN and GP), Simulations and analysis (NdT), , Interpretation (theoretical work and results: NdT and KKMN; simulations and results: NdT, KKMN and GP), Writing (NdT), Editing (NdT, KKMN and GP)
\end{itemize}

\appendix
\section{Random Phase Approximation for cross-linker particles}
The discussion below outlines the calculations done in order to implement a Random Phase Approximation (RPA) for $N$ cross-linkers. Utilising eq.~\eqref{eq:CL_MSR2}, \textit{i.e.} the generating functional for a single cross-linker, in a product, one can obtain the generating functional for $N$ cross-linkers, as follows:
\begin{multline}
     \mathbb{Z} = \mathcal{N}\int [\mathrm{d} \mathbf{y}_1][\mathrm{d} \mathbf{Y}_1]...[\mathrm{d} \mathbf{y}_N][\mathrm{d} \mathbf{Y}_N]\, [\mathrm{d} \hat{\mathbf{y}}_1][\mathrm{d} \hat{\mathbf{Y}}_1] ... [\mathrm{d} \hat{\mathbf{y}}_N][\mathrm{d} \hat{\mathbf{Y}}_N]\\
     \times\mathrm{e}^{\sum_{j=1}^N \left(- \mathrm{i} \int_t \hat{\mathbf{Y}}_j \cdot \left( -\gamma_y \dot{\mathbf{Y}}_j(t)  + \frac{\lambda_y}{4}\hat{\mathbf{Y}}_j(t)\right) \right) }\\ 
     \times \mathrm{e}^{\sum_{j=1}^N \left(-\mathrm{i} \int_t \hat{\mathbf{y}}_j \cdot \left( -\gamma_y \dot{\mathbf{y}}_j (t)  - 2\kappa \mathbf{y}_j(t) - \lambda_y \hat{\mathbf{y}}_j(t)\right) \right)}  \, .
     \label{eq:CL_MSR_N}
\end{multline}
Now, the following collective variable may be introduced for the coordinates of the $N$ cross-linkers, along with its corresponding auxiliary variable:
\begin{subequations}
    \begin{equation}
         \rho_{\mathbf{K}, \mathbf{k}} (t) = \sum_{j=1}^N \mathrm{e}^{\mathrm{i} \mathbf{K} \cdot\mathbf{Y}_j(t)+\mathrm{i} \mathbf{k} \cdot\mathbf{y}_j(t)}\, ,
    \end{equation}
    \begin{equation}
        \hat{\rho}_{\mathbf{K}, \mathbf{k}} (t) = \sum_{j=1}^N \left(\mathrm{i} \mathbf{K} \cdot \hat{\mathbf{Y}}_j(t)+\mathrm{i} \mathbf{k} \cdot\hat{\mathbf{y}}_j(t) \right)\mathrm{e}^{\mathrm{i} \mathbf{K} \cdot\mathbf{Y}_j(t)+\mathrm{i} \mathbf{k} \cdot\mathbf{y}_j(t)}\, .
    \end{equation}
\end{subequations}

Note that these have been written in terms of the Fourier transformed spatial variable $\mathbf{K}$  and $\mathbf{k}$, following \cite{fredricksonCollectiveDynamicsPolymer1990}, as a matter of mathematical convenience. To move towards the collective variables, we may introduce them into the generating functional in the form of a somewhat camouflaged one, \textit{i.e.}

 \begin{widetext}
\begin{multline}
     \mathbb{Z} = \mathcal{N}\int [\mathrm{d} \mathbf{y}_1][\mathrm{d} \mathbf{Y}_1]...[\mathrm{d} \mathbf{y}_N][\mathrm{d} \mathbf{Y}_N]\, [\mathrm{d} \hat{\mathbf{y}}_1][\mathrm{d} \hat{\mathbf{Y}}_1] ... [\mathrm{d} \hat{\mathbf{y}}_N][\mathrm{d} \hat{\mathbf{Y}}_N]\,  [ \mathrm{d} \rho_{\mathbf{K}, \mathbf{k}}] [\mathrm{d}\hat{\rho}_{\mathbf{K}, \mathbf{k}}] \, \\\delta  \left(\rho_{\mathbf{K}, \mathbf{k}} (t) - \sum_{j=1}^N \mathrm{e}^{\mathrm{i} \mathbf{K} \cdot\mathbf{Y}_j(t)+\mathrm{i} \mathbf{k} \cdot\mathbf{y}_j(t)}\right)  \delta \left(\hat{\rho}_{\mathbf{K}, \mathbf{k}} (t) - \sum_{j=1}^N \left(\mathrm{i} \mathbf{K} \cdot \hat{\mathbf{Y}}_j(t)+\mathrm{i} \mathbf{k} \cdot\hat{\mathbf{y}}_j(t) \right)\mathrm{e}^{\mathrm{i} \mathbf{K} \cdot\mathbf{Y}_j(t)+\mathrm{i} \mathbf{k} \cdot\mathbf{y}_j(t)} \right) \\
     \times\mathrm{e}^{\sum_{j=1}^N \left(-\mathrm{i} \int_t \hat{\mathbf{y}}_j \cdot \left( -\gamma_y \dot{\mathbf{y}}_j (t)  - 2\kappa \mathbf{y}_j(t) - \lambda_y \hat{\mathbf{y}}_j(t)\right) \right)+\sum_{j=1}^N \left(- \mathrm{i} \int_t \hat{\mathbf{Y}}_j \cdot \left( -\gamma_y \dot{\mathbf{Y}}_j(t)  + \frac{\lambda_y}{4}\hat{\mathbf{Y}}_j(t)\right) \right) } \, .
     \label{eq:CL_RPA1}
\end{multline}     
 \end{widetext}
To eventually be able to implement the integrals over $\rho_{\mathbf{K}, \mathbf{k}}$ and $\hat{\rho}_{\mathbf{K}, \mathbf{k}}$ the Dirac Delta functions may be rewritten and a second order expansion may be implemented for some of the exponential terms such that the generating functional becomes
 \begin{widetext}
\begin{multline}
     \mathbb{Z} = \mathcal{N}\int[\mathrm{d}\rho_{\mathbf{K}, \mathbf{k}}]  [\mathrm{d}\hat{\rho}_{\mathbf{K}, \mathbf{k}}]  [ \mathrm{d} \psi_{\mathbf{K}, \mathbf{k}}] [\mathrm{d}\hat{\psi}_{\mathbf{K}, \mathbf{k}}] \mathrm{e}^{\mathrm{i} \int_{\mathbf{K},\mathbf{k},t} \psi_{\mathbf{K}, \mathbf{k}}(t) \rho_{\mathbf{K}, \mathbf{k}} (t) + \mathrm{i} \int_{\mathbf{K},\mathbf{k},t} \hat{\psi}_{\mathbf{K}, \mathbf{k}} (t)  \hat{\rho}_{\mathbf{K}, \mathbf{k}} (t)} \\
     \left\{ \int [\mathrm{d} \mathbf{y}_1][\mathrm{d} \mathbf{Y}_1]...[\mathrm{d} \mathbf{y}_N][\mathrm{d} \mathbf{Y}_N]\, [\mathrm{d} \hat{\mathbf{y}}_1][\mathrm{d} \hat{\mathbf{Y}}_1] ... [\mathrm{d} \hat{\mathbf{y}}_N][\mathrm{d} \hat{\mathbf{Y}}_N]\, \mathrm{e}^{\sum_{j=1}^N \left(- \mathrm{i} \int_t \hat{\mathbf{Y}}_j \cdot \left( -\gamma_y \dot{\mathbf{Y}}_j(t)  + \frac{\lambda_y}{4}\hat{\mathbf{Y}}_j(t)\right) \right) }\right.\\
     \times \mathrm{e}^{\sum_{j=1}^N \left(-\mathrm{i} \int_t \hat{\mathbf{y}}_j \cdot \left( -\gamma_y \dot{\mathbf{y}}_j (t)  - 2\kappa \mathbf{y}_j(t) - \lambda_y \hat{\mathbf{y}}_j(t)\right) \right)} \left[1- \mathrm{i} \int_{\mathbf{K},\mathbf{k},t} \psi_{\mathbf{K}, \mathbf{k}}(t)\sum_{j=1}^N \mathrm{e}^{\mathrm{i} \mathbf{K} \cdot\mathbf{Y}_j(t)+\mathrm{i} \mathbf{k} \cdot\mathbf{y}_j(t)}\right.\\ 
     \left.+ \int_{\mathbf{K},\mathbf{k},t} \hat{\psi}_{\mathbf{K}, \mathbf{k}} (t)\sum_{j=1}^N \left(\mathrm{i} \mathbf{K} \cdot \hat{\mathbf{Y}}_j(t)+\mathrm{i} \mathbf{k} \cdot\hat{\mathbf{y}}_j(t) \right)
     \mathrm{e}^{\mathrm{i} \mathbf{K} \cdot\mathbf{Y}_j(t)+\mathrm{i} \mathbf{k} \cdot\mathbf{y}_j(t)} \right.\\
     \left. -\mathrm{i} \int_{\mathbf{K},\mathbf{k},t}\int_{\mathbf{K}',\mathbf{k}',t'} \psi_{\mathbf{K}, \mathbf{k}}(t) \hat{\psi}_{\mathbf{K}', \mathbf{k}'} (t')\sum_{j=1}^N \left(\mathrm{i} \mathbf{K}' \cdot \hat{\mathbf{Y}}_j(t')+\mathrm{i} \mathbf{k'} \cdot\hat{\mathbf{y}}_j(t') \right)\mathrm{e}^{\mathrm{i} (\mathbf{K} \cdot\mathbf{Y}_j(t)+\mathbf{K}' \cdot\mathbf{Y}_j(t'))+\mathrm{i} (\mathbf{k} \cdot\mathbf{y}_j(t) +\mathbf{k}' \cdot\mathbf{y}_j(t'))}  \right. \\
     \left. \left. + \frac{1}{2}  \int_{\mathbf{K},\mathbf{k},t} \int_{\mathbf{K}',\mathbf{k}',t'} \psi_{\mathbf{K}, \mathbf{k}}(t ) \psi_{\mathbf{K}', \mathbf{k}'}(t')\sum_{j=1}^N \mathrm{e}^{\mathrm{i} (\mathbf{K} \cdot\mathbf{Y}_j(t)+\mathbf{K}' \cdot\mathbf{Y}_j(t'))+\mathrm{i} (\mathbf{k} \cdot\mathbf{y}_j(t) +\mathbf{k}' \cdot\mathbf{y}_j(t'))} \right]\right\} .
     \label{eq:CL_RPA_allTerms}
\end{multline}
\end{widetext}
In the above expression, the term containing the product $ \hat{\psi}_{\mathbf{K}, \mathbf{k}} (t)\hat{\psi}_{\mathbf{K}, \mathbf{k}} (t')$ does not appear since the rules for hatted fields in the MSR formalism, as set out in \cite{Jensen1981a}, indicate that all averages containing the product of two hatted fields must be zero. Now, implementing the approximation, all terms that merely result in the average of the collective variable may be neglected and the remaining terms may be rewritten in terms of averages over the stochastic forces denoted by $\langle ... \rangle$ and detailed in Appendix \ref{app:RPAaves}  such that 
\begin{widetext}
\begin{multline}
     \mathbb{Z} = \mathcal{N}\int[\mathrm{d}\rho_{\mathbf{K}, \mathbf{k}}]  [\mathrm{d}\hat{\rho}_{\mathbf{K}, \mathbf{k}}]  [ \mathrm{d} \psi_{\mathbf{K}, \mathbf{k}}] [\mathrm{d}\hat{\psi}_{\mathbf{K}, \mathbf{k}}] \mathrm{e}^{\mathrm{i} \int_{\mathbf{K},\mathbf{k},t} \psi_{\mathbf{K}, \mathbf{k}}(t) \rho_{\mathbf{K}, \mathbf{k}} (t) + \mathrm{i} \int_{\mathbf{K},\mathbf{k},t} \hat{\psi}_{\mathbf{K}, \mathbf{k}} (t)  \hat{\rho}_{\mathbf{K}, \mathbf{k}} (t)} \\
     \left\{ \left[1 + \frac{1}{2}  \int_{\mathbf{K},\mathbf{k},t} \int_{\mathbf{K}',\mathbf{k}',t'} \psi_{\mathbf{K}, \mathbf{k}}(t ) \psi_{\mathbf{K}', \mathbf{k}'}(t')\sum_{j=1}^N \langle  \mathrm{e}^{\mathrm{i} \mathbf{k}(\mathbf{y}_j(t)-\mathbf{y}_j(t'))}\rangle  \langle  \mathrm{e}^{\mathrm{i}\mathbf{K} (\mathbf{Y}_j(t)-\mathbf{Y}_j(t'))}\rangle\right. \right. \\
     \left. \left.  -\mathrm{i} \int_{\mathbf{K},\mathbf{k},t}\int_{\mathbf{K}',\mathbf{k}',t'} \psi_{\mathbf{K}, \mathbf{k}}(t) \hat{\psi}_{\mathbf{K}', \mathbf{k}'} (t')\sum_{j=1}^N \left(\mathrm{i} \mathbf{K} \langle \mathbf{\hat{Y}}_j(t) \mathrm{e}^{\mathrm{i}\mathbf{K} (\mathbf{Y}_j(t)-\mathbf{Y}_j(t'))}\rangle  \langle  \mathrm{e}^{\mathrm{i} \mathbf{k}(\mathbf{y}_j(t)-\mathbf{y}_j(t'))}\rangle \right. \right. \right.\\
     \left. \left. \left.+\mathrm{i} \mathbf{k}\langle \mathbf{\hat{y}}_j(t) \mathrm{e}^{\mathrm{i} \mathbf{k}(\mathbf{y}_j(t)-\mathbf{y}_j(t'))}\rangle  \langle  \mathrm{e}^{\mathrm{i}\mathbf{K} (\mathbf{Y}_j(t)-\mathbf{Y}_j(t'))}\rangle \right)\right]\right\} .
     \label{eq:CL_RPA_withAverages}
\end{multline}
\end{widetext}
After evaluating the averages (see Appendix \ref{app:RPAaves}) and bringing everything back into the exponent, this yields
\begin{multline}
     \mathbb{Z} = \mathcal{N}\int[\mathrm{d}\rho_{\mathbf{K}, \mathbf{k}}]  [\mathrm{d}\hat{\rho}_{\mathbf{K}, \mathbf{k}}]  [ \mathrm{d} \psi_{\mathbf{K}, \mathbf{k}}] [\mathrm{d}\hat{\psi}_{\mathbf{K}, \mathbf{k}}] \mathrm{e}^{\mathrm{i} \int_{\mathbf{K},\mathbf{k},t} \psi_{\mathbf{K}, \mathbf{k}}(t) \rho_{\mathbf{K}, \mathbf{k}} (t)}\\
     \times \mathrm{e}^{ + \mathrm{i} \int_{\mathbf{K},\mathbf{k},t} \hat{\psi}_{\mathbf{K}, \mathbf{k}} (t)  \hat{\rho}_{\mathbf{K}, \mathbf{k}} (t)- \frac{1}{2}\int_{\mathbf{K},\mathbf{k},t} \int_{t'}\psi_{\mathbf{K}, \mathbf{k}}(t)\mathcal{A}_{\mathbf{K}, \mathbf{k}}(t,t')\psi_{\mathbf{-K}, \mathbf{-k}}(t')}\\
     \times \mathrm{e}^{+\mathrm{i}\int_{\mathbf{K},\mathbf{k},t} \int_{t'}\psi_{\mathbf{K}, \mathbf{k}}(t)\mathcal{A}_{\mathbf{K}, \mathbf{k}}(t,t')\mathcal{B}_{\mathbf{K}, \mathbf{k}}(t,t')\hat{\psi}_{\mathbf{-K}, \mathbf{-k}}(t')}
\end{multline}
where
\begin{subequations}
    \begin{equation}
    \mathcal{A}_{\mathbf{K}, \mathbf{k}}(t,t') = N \mathrm{e}^{-\frac{\lambda_y}{4 \gamma_y^2 }|\mathbf{K}|^2(t-t') -\frac{\lambda_y}{\kappa \gamma_y }|\mathbf{k}|^2\left(1 + \coth{\left( \tfrac{\kappa}{\gamma_y}(t-t') \right)}\right)}\, \, 
    \label{eq:CLRPA-A}
\end{equation}
and
\begin{equation}
    \mathcal{B}_{\mathbf{K}, \mathbf{k}}(t,t')= \frac{\mathbf{K} \, \mathbf{k} \left(\tanh \left(\frac{\kappa  (t-t')}{\gamma }\right)+2\right) }{2 \left(\coth \left(\frac{\kappa  (t-t')}{\gamma }\right)+1\right)}\mathcal{A}_{\mathbf{K}, \mathbf{k}}(t,t')
\end{equation}
\end{subequations}

Finally, evaluating the Gaussian integrals over $ \hat{\psi}_{\mathbf{K}, \mathbf{k}} (t)$ and $\hat{\psi}_{\mathbf{K}, \mathbf{k}} (t')$ the generating functional for the cross-linkers is given by
\begin{widetext}
\begin{multline}
     \mathbb{Z} = \mathcal{N}\int[\mathrm{d}\rho_{\mathbf{K}, \mathbf{k}}]  [\mathrm{d}\hat{\rho}_{\mathbf{K}, \mathbf{k}}] \mathrm{e}^{-\mathrm{i}\int_{\mathbf{K},\mathbf{k},t} \int_{t'}\hat{\rho}_{\mathbf{K}, \mathbf{k}}(t)(\mathcal{B}_{\mathbf{K}, \mathbf{k}}(t,t')\mathcal{A}_{\mathbf{-K}, \mathbf{-k}}(t,t'))^{-1}\rho_{\mathbf{-K}, \mathbf{-k}}(t')}\\
     \times\mathrm{e}^{- \frac{1}{2}\int_{\mathbf{K},\mathbf{k},t} \int_{t'}\hat{\rho}_{\mathbf{K}, \mathbf{k}}(t)(\mathcal{B}_{\mathbf{K}, \mathbf{k}}(t,t')\mathcal{A}_{\mathbf{K}, \mathbf{k}}(t,t')\mathcal{B}_{\mathbf{-K}, \mathbf{-k}}(t,t'))^{-1}\hat{\rho}_{\mathbf{-K}, \mathbf{-k}}(t')}
     \label{eq:CLRPAfinal}
\end{multline}
\end{widetext}
\section{Averages for the cross-linker RPA }
\label{app:RPAaves}

To calculate the averages involving the extension of the cross-linker particles, a propagator or kernel for a single harmonic oscillator will be used to calculate the values of the integrals over a fixed time interval. After evaluating the integrals over the stochastic force and the $\hat{\mathbf{y}}_\alpha(t)$, the first average is given by
\begin{multline}
    \langle  \mathrm{e}^{\mathrm{i} \mathbf{k}\cdot(\mathbf{y}_j(t)-\mathbf{y}_j(t'))}\rangle = \prod_{\alpha=1}^M \int [\mathrm{d} \mathbf{y}_\alpha]\,  \mathrm{e}^{\mathrm{i} \mathbf{k}\cdot(\mathbf{y}_j(t)-\mathbf{y}_j(t'))}\\
    \times \mathrm{e}^{- \frac{\gamma_y^2}{4 \lambda_y}\int_t \dot{\mathbf{y}}_\alpha^2(t) -\frac{\kappa \gamma_y}{\lambda_y}\int_t \mathbf{y}_\alpha(t)\, \dot{\mathbf{y}}_\alpha(t)-\frac{\kappa^2}{\lambda_y}\int_t \mathbf{y}_\alpha^2(t) }
\end{multline}
 dropping the subscripts for ease of notation  and assuming that the integration is over some time interval from $\tau_1$ to $\tau_2$ such that $\mathbf{y}(\tau_1) = \mathbf{y}_1$ and  $\mathbf{y}(\tau_2) = \mathbf{y}_2$ this becomes
\begin{equation}
    \langle  \mathrm{e}^{\mathrm{i} \mathbf{k}\cdot(\mathbf{y}_j(t)-\mathbf{y}_j(t'))}\rangle = \int \mathrm{d }\mathbf{y}_1 \, \mathrm{d}\mathbf{y}_2 \, \mathrm{e}^{\mathrm{i} \mathbf{k}\cdot(\mathbf{y}_1-\mathbf{y}_2)}\, \mathbb{K}(\mathbf{y}_1, t ; \mathbf{y}_2, t')
    \label{eq:RPAavey-itoK} 
\end{equation}
where
\begin{widetext}
\begin{equation}
    \mathbb{K}(\mathbf{y}_1, \tau_1  ; \mathbf{y}_2, \tau_2) = \int_{\{\mathbf{y}(\tau_1) = \mathbf{y}_1;\,\mathbf{y}(\tau_2)=\mathbf{y}_2\}} [\mathrm{d} \mathbf{y}]\,  \mathrm{e}^{- \frac{\gamma_y^2}{4 \lambda_y}\int_{\tau_1}^{\tau_2} \mathrm{d} t \, \dot{\mathbf{y}}^2(t) -\frac{\kappa \gamma_y}{\lambda_y}\int_{\tau_1}^{\tau_2} \mathrm{d} t \, \mathbf{y}(t)\, \dot{\mathbf{y}}(t)-\frac{\kappa^2}{\lambda_y}\int_{\tau_1}^{\tau_2} \mathrm{d} t \, \mathbf{y}^2(t) }
    \label{eq:Kdef}
\end{equation}
\end{widetext}
is the kernel, propagator or Green's function for a single harmonic oscillator undergoing Brownian motion. 
By adding in a source function $\hat{\mathbf{J}}(t)$ for $\hat{\mathbf{y}}(t)$ the remaining average can also be determined using the propagator as follows
\begin{equation}
    \langle \mathbf{\hat{y}}_j(t) \mathrm{e}^{\mathrm{i}\mathbf{k} \cdot(\mathbf{y}_j(t)-\mathbf{y}_j(t'))}\rangle = \int \mathrm{d }\mathbf{y}_1 \, \mathrm{d}\mathbf{y}_2 \, \mathrm{e}^{\mathrm{i} \mathbf{k}\cdot(\mathbf{y}_1-\mathbf{y}_2)}\, \tfrac{\delta \, \mathbb{K}(\mathbf{y}_1, t  ; \mathbf{y}_2, t')}{\delta\,  \hat{\mathbf{J}}(t)}\,.
    \label{eq:RPAaveyhat-itoK}
\end{equation}

One can derive the kernel for a simple harmonic oscillator undergoing Brownian motion as given in eq.~\eqref{eq:KernelSHO} by using the \textit{principle of least action} as detailed in \cite{feynmanQuantumMechanicsPath2012}, to obtain
\begin{widetext}
\begin{equation}
    \mathbb{K}(\mathbf{y}_1, t  ; \mathbf{y}_2, t') = \mathcal{N} \mathrm{e}^{ - \tfrac{\kappa \gamma_y}{\lambda_y}\left(1-\coth{\left(\frac{2 \kappa}{\gamma_y}(t-t')\right)}\right) \mathbf{y}_1  \mathbf{y}_2 -\frac{1}{2}\mathrm{csch}\left(\frac{2 \kappa}{\gamma_y}(t-t')\right)(\mathbf{y}^2_1 + \mathbf{y}^2_2) }
    \label{eq:KernelSHO}
\end{equation}
\end{widetext}
such that \eqref{eq:RPAavey-itoK} and \eqref{eq:RPAaveyhat-itoK} result in
\begin{equation}
        \langle  \mathrm{e}^{\mathrm{i} \mathbf{k}\cdot(\mathbf{y}_j(t)-\mathbf{y}_j(t'))}\rangle  = \mathrm{e}^{-\frac{\lambda_y}{\kappa \gamma_y }|\mathbf{k}|^2\left(1 + \coth{\left( \tfrac{\kappa}{\gamma_y}(t-t') \right)}\right)}
\end{equation}
and
\begin{multline}
     \langle \mathbf{\hat{y}}_j(t) \mathrm{e}^{\mathrm{i}\mathbf{k} \cdot(\mathbf{y}_j(t)-\mathbf{y}_j(t'))}\rangle = \frac{i \, \mathbf{k} \gamma_y \left(\tanh \left(\frac{\kappa  }{\gamma_y }(t-t')\right)+2\right) }{2 \left(\coth \left(\frac{\kappa  }{\gamma_y }(t-t')\right)+1\right)}\\
     \times\mathrm{e}^{-\frac{\lambda_y}{\kappa \gamma_y }|\mathbf{k}|^2\left(1 + \coth{\left( \tfrac{\kappa}{\gamma_y}(t-t') \right)}\right)}
\end{multline}

\bibliographystyle{apsrev4-2} 
\bibliography{references}
\end{document}


\title{Supplementary Information for:\\
\textit{Dynamical Networking of Polymer Networks with Dedicated Cross-linker Particles}}

\author{Nadine du Toit }
\email{Corresponding author: 24461989@sun.ac.za}
\affiliation{\SUPhys}
 
\author{Kristian K. M\"uller-Nedebock \orcidlink{0000-0002-1772-1504}}
 \email{kkmn@sun.ac.za}
\affiliation{\SUPhys}
\affiliation{\NITheCS}
\author{Giuseppe Pellicane \orcidlink{0000-0002-3805-830X}}
\email{gpellicane@unime.it}
\affiliation{\NITheCS}
\affiliation{\UKZN}
\affiliation{\Messina}
\date{\today}

\maketitle
\tableofcontents
\noindent\textbf{How to read this file.} This Supplementary Information (SI) provides
extended simulation details and additional figures. Interpretation of
results is given in the main text. References and notation are consistent with the
main paper unless otherwise stated.

\section{Molecular dynamics simulations}

The main text includes molecular dynamics simulations of a two polymer system with reversible cross-linking. Whilst most details of the simulations are discussed in the main text, some additional details regarding the details of additional packages such as REACTER\cite{gissingerREACTERHeuristicMethod2020a} 
and Dynasor \cite{franssonDynasorToolExtracting2021} are given here. Details of the reversible bonding reactions implemented in the molecular dynamics simulations between polymer and cross-linker particles are discussed in Sec.~\ref{sec:reversiblebonds}. The analysis of the trajectories obtained during these molecular dynamics are discussed in Sec.~\ref{sec:Dynasor}.
\subsection{Reversible bonds}
\label{sec:reversiblebonds}
\begin{figure}
    \centering
    \subfile{./diagrams/BondLegend}
    \caption[Legend for molecule diagrams]{Legend for molecule diagrams showing bond and atom type labels indicated by different coloured lines and filled circles.}
    \label{fig:legend}
\end{figure}
   \begin{figure}
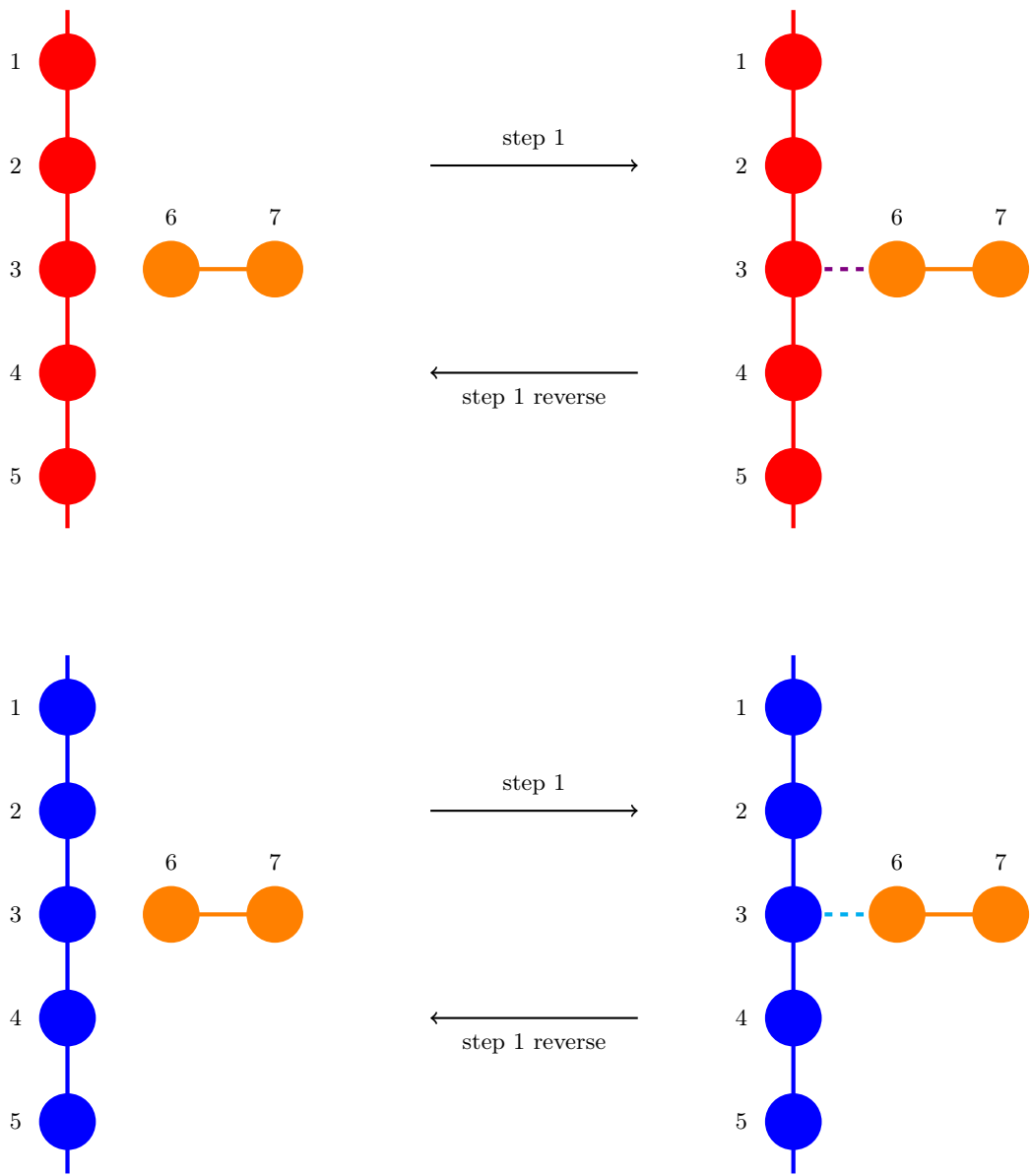

    \centering
    \subfile{./diagrams/step1A}

\vspace{5em}
    \subfile{./diagrams/step1B}
    \caption[Molecule reaction templates for step 1]{Molecule reaction templates and four possible reactions (indicated by arrows) for step 1 of '\textit{fix bond/react}'. Atom IDs 1-7 are indicated next to each atom. Bonds and atoms are colour coded according to the legend in Fig.~\ref{fig:legend}. For step 1, Atom IDs 1 and 5 are edgeIDs.}
    \label{fig:step1}
  \end{figure}
Reversible bonding between polymer beads and cross-linkers were implemented using REACTOR \cite{gissingerREACTERHeuristicMethod2020a} via the LAMMPS command `\textit{fix bond/react}`, in a two step process. This required specifying molecule templates both pre- and post-reaction, specifying the bonds and angles of the molecule such that bonds can be formed or broken accordingly. Along with the pre- and post-reaction molecule templates, a map file is required specifying the atom IDs of the initiator atoms, between which bonds will be formed or broken during the reaction, all atoms that are involved in the reaction and finally the atom IDs of edge atoms, which may be bonded to other atoms that are not part of the molecule template.  

\begin{figure}
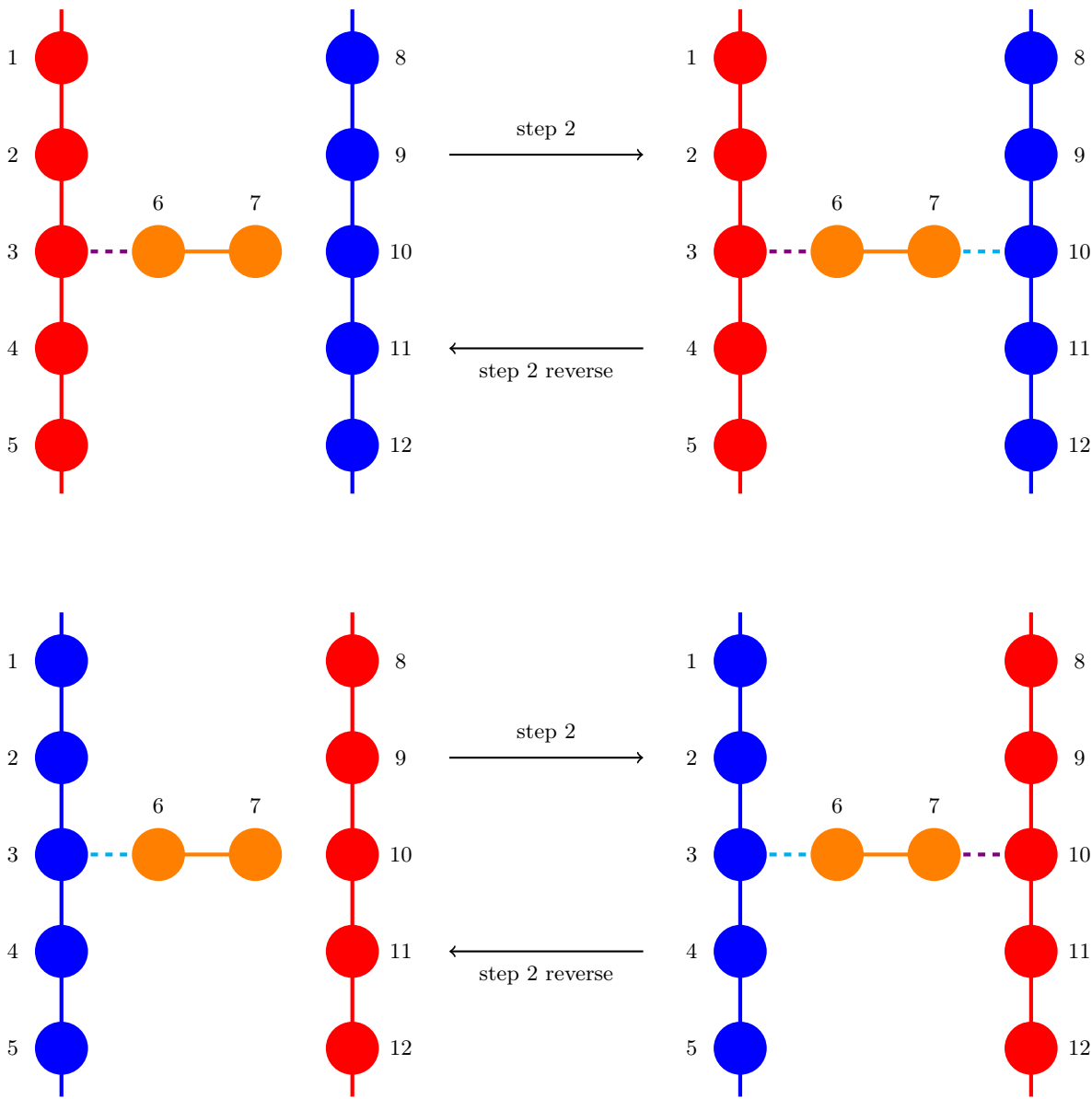

    \centering
    \subfile{./diagrams/step2interAB}
     
    \vspace{5em}
    \subfile{./diagrams/step2interBA}
    \caption[Molecule reaction templates for inter-species reactions of step 2]{Molecule reaction templates and four possible inter-chain reactions (indicated by arrows) for step 2 of \textit{fix bond/react}. Atom IDs 1 -12 are indicated next to each atom. Bonds and atoms are colour coded according to the legend in Fig.~\ref{fig:legend}. For step 2, Atom IDs 1, 5 8 and 12 are edgeIDs, whilst atom IDs 3 and 6 are the initiator atom IDs.}
    \label{fig:step2inter}
  \end{figure}

    \begin{figure}
    \centering
    \subfile{./diagrams/step2intraA}
    
    \vspace{5em}
    \subfile{./diagrams/step2intraB}
    \caption[Molecule reaction templates for intra-species reactions of step 2]{Molecule reaction templates and four possible intra-chain reactions (indicated by arrows) for step 2 of `\textit{fix bond/react}'. Atom IDS 1 -12 are indicated next to each atom. Bonds and atoms are colour coded according to the legend in Fig.~\ref{fig:legend}. For step 2, Atom IDs 1, 5 8 and 12 are edgeIDs. }
    \label{fig:step2intra}
  \end{figure}

The pre-and post-reaction molecule templates and steps are illustrated in Figs.~\ref{fig:step1}--\ref{fig:step2intra}. In order to be able to discern between intra- and inter-chain cross-links, the monomers of each of the polymers were assigned different atom type labels `\textit{beadPolyA}' and `\textit{beadPolyB}' and bond type labels `\textit{bondPolyA}' and `\textit{bonddPolyB}'  as shown in the colour coded legend in Fig.~\ref{fig:legend}, although they were given the same properties and treated with the same parameters. Differentiating between atom types in this way, there are two ways in which bonds can form according to step 1, either between atom type  `\textit{beadPolyA}'  and `\textit{beadCL}' to form a bond of type `\textit{bondPolyACL}'  or between atom type  `\textit{beadPolyB}'  and `\textit{beadCL}' to form a bond of type `\textit{bondPolyBCL}' as depicted in Fig.~\ref{fig:step1}. Switching the pre- and post-reaction templates, the reverse of each of these reactions is also specified such that cross-linkers that are bound at one end can detach and diffuse freely throughout the system. As for the second step, there are four possible bond creation reactions and their four reverse reactions as depicted in Figs.~\ref{fig:step2inter}--\ref{fig:step2intra}. Figure~\ref{fig:step2inter} shows the formation and breaking of inter-chain cross-links. Here a cross-linker already has a bead that has a bond with a monomer on either polymer A or polymer B. The second bead of the cross-linker bonds, during step 2, with a monomer on the other polymer type. The reverse of this reaction allows one of the bonds between a monomer and cross-linker that forms an inter-chain cross-link, to break such that the cross-linker remains bound only at one end. Similarly,  Fig.~\ref{fig:step2intra} depicts the two bond forming and two bond breaking reactions pertaining to intra-chain cross-links. In total there are thus 12 possible reactions that are implemented by the `\textit{fix bond/react}' command in the simulation---one reaction corresponding to each arrow in Figs.~\ref{fig:step1}--\ref{fig:step2intra}---with 4 possible reactions in step 1 and its reverse and 8 reactions in step 2 and its reverse. At each time step each of these reactions is attempted with a specified probability. For a reaction to be attempted, initiator atoms (given by atom IDs 3,6,7 and 10 in Figs. ~\ref{fig:step1}--\ref{fig:step2intra}) are identified according to the pre- and post-reaction templates and these atoms must be within a specified distance from one another. The reaction distances and probabilities for each step of the reaction are given in Table \ref{tab:bond/react_parameters}.
\begin{table}[ht]
    \centering
    \begin{tabular}{|c|c|c|c|}
        \hline
        Reaction & Minimum distance & Maximum distance & Reaction probability \\
        \hline
        Step 1 & 0.0 & 1.6 & 0.8 \\
        Step 1 reverse  & 1.1 & 2.5 & 0.05 \\
        Step 2  & 0.0 & 1.6 & 0.95 \\
        Step 2 reverse & 1.1 & 2.5 & 0.05 \\
        \hline
    \end{tabular}
    \caption{Reaction distances and probabilities for the steps of \textit{'fix bond/react'}.}
    \label{tab:bond/react_parameters}
    \end{table}
    After each successful reaction, \textit{i.e.} bond formation or breakage, a stabilisation period of 100 time steps is applied to avoid immediate rebonding or rebreaking. During this time, atoms involved in the reaction are integrated via an internal `\textit{fix nve/limit}', with a maximum displacement of 1.0 (in reduced units). This relatively large value allows atom motion to proceed without artificial restriction, with stabilisation serving primarily to prevent repeated reactions on short timescales rather than suppressing post-reaction forces. 
\subsection{Dynamic structure factors using Dynasor}
\label{sec:Dynasor}
\begin{figure}[htbp]
    \centering
    \includegraphics[width=0.75\linewidth]{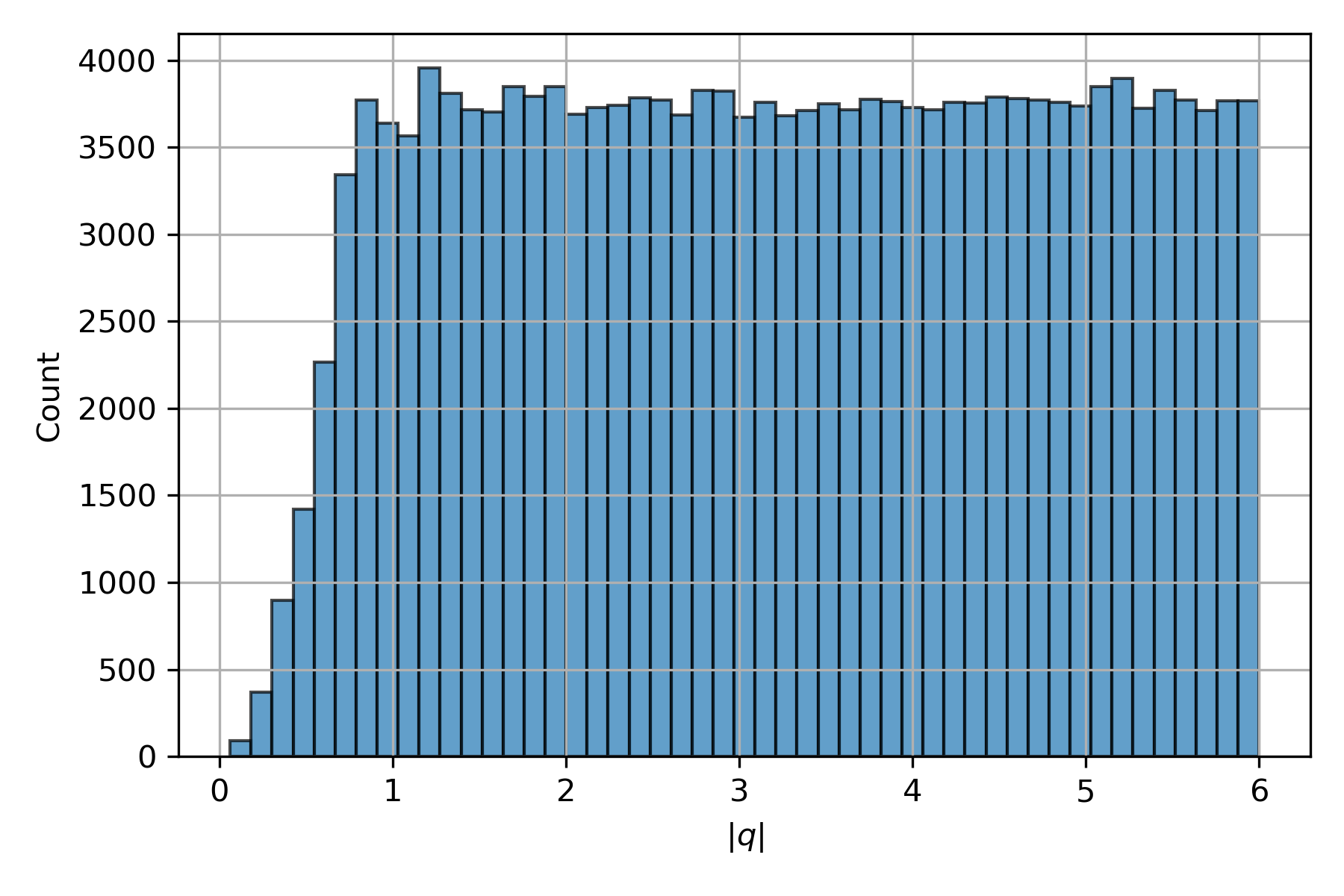}
    \caption[Distribution of $q$-points for Dynasor analysis]{Distribution of the $170\,011$ $q$-points used for calculating dynamic structure factors}
    \label{fig:qdistribution}
\end{figure}

After loading the LAMMPS trajectory files into Dynasor,  spherically distributed $q$-points were generated corresponding to the trajectory with a maximum $q$-value of $6$ and the maximum number of q-points set to $170\,000$. The $q$-points were filtered to remove values smaller than $q= \frac{2 \pi}{L}$, where $L=100$ is the length of the box, so as to avoid evaluations of the dynamic structure factor on length scales irrelevant to the simulation. This resulted in the sample of $q$-points shown in the histogram in Fig.~\ref{fig:qdistribution}, where the sample has been divided into $50$ bins.

\begin{figure}[t]
    \centering
    \begin{minipage}{\columnwidth}
        \begin{subfigure}[t]{0.48\linewidth}
            \centering
            \includegraphics[width=\linewidth]{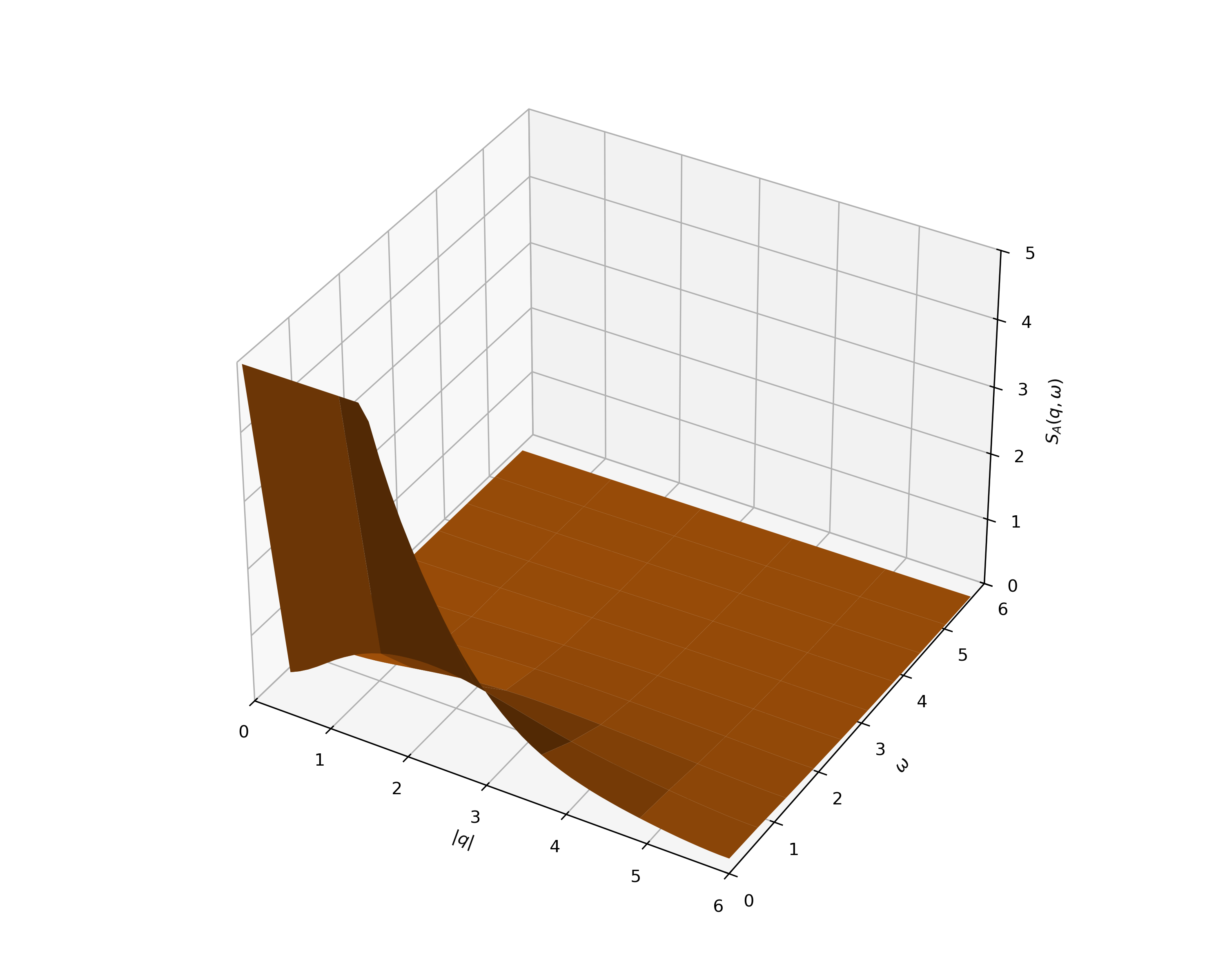}
            \caption{$S_\mathrm{A}(q, \omega)$ with cross-linking}
            \label{fig:SA_cl} 
        \end{subfigure}\hfill
        \begin{subfigure}[t]{0.48\linewidth}
            \centering
            \includegraphics[width=\linewidth]{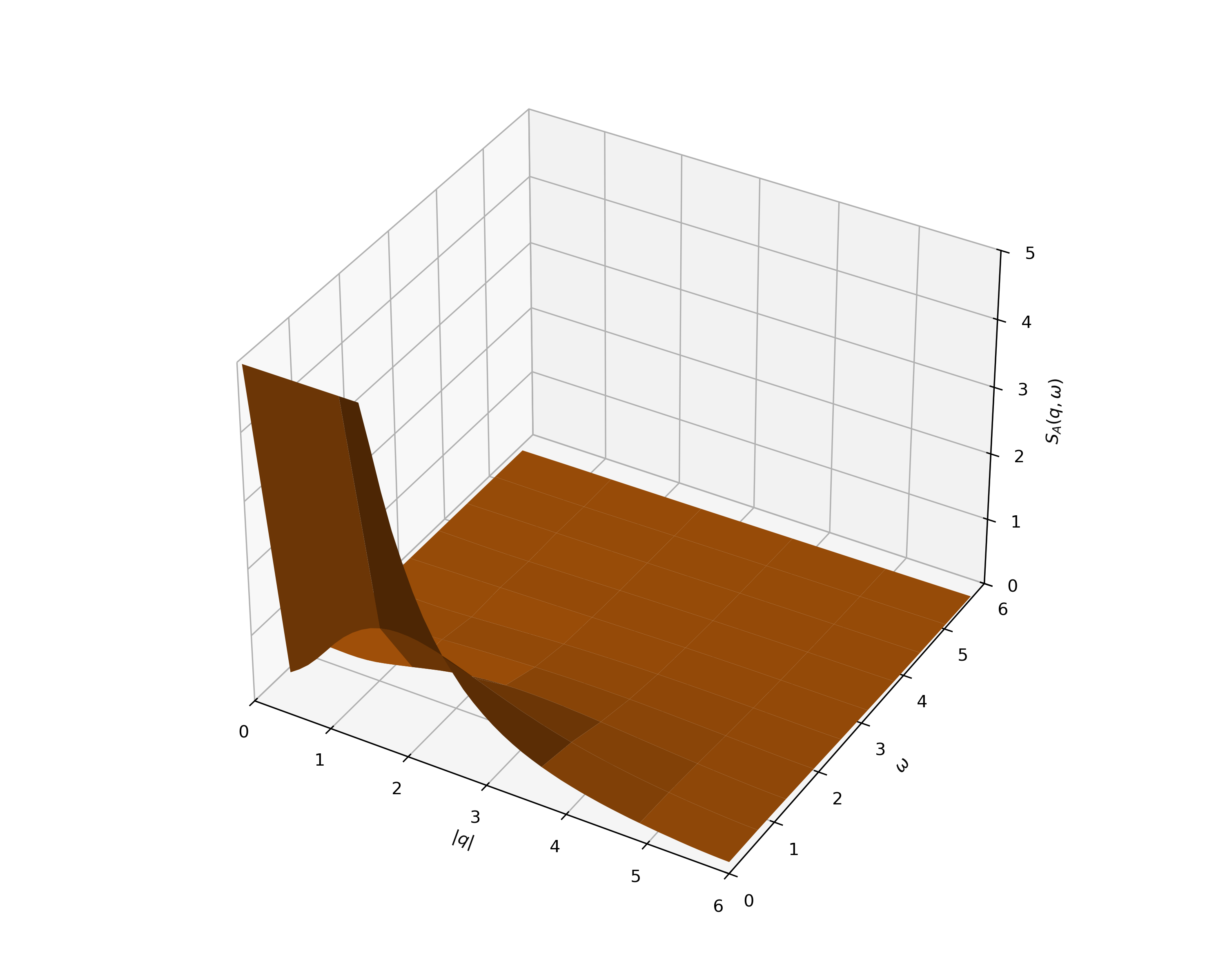}
            \caption{$S_\mathrm{A}(q, \omega)$ without cross-linking}
            \label{fig:SA_nl} 
        \end{subfigure}

        \begin{subfigure}[t]{0.48\linewidth}
            \centering
            \includegraphics[width=\linewidth]{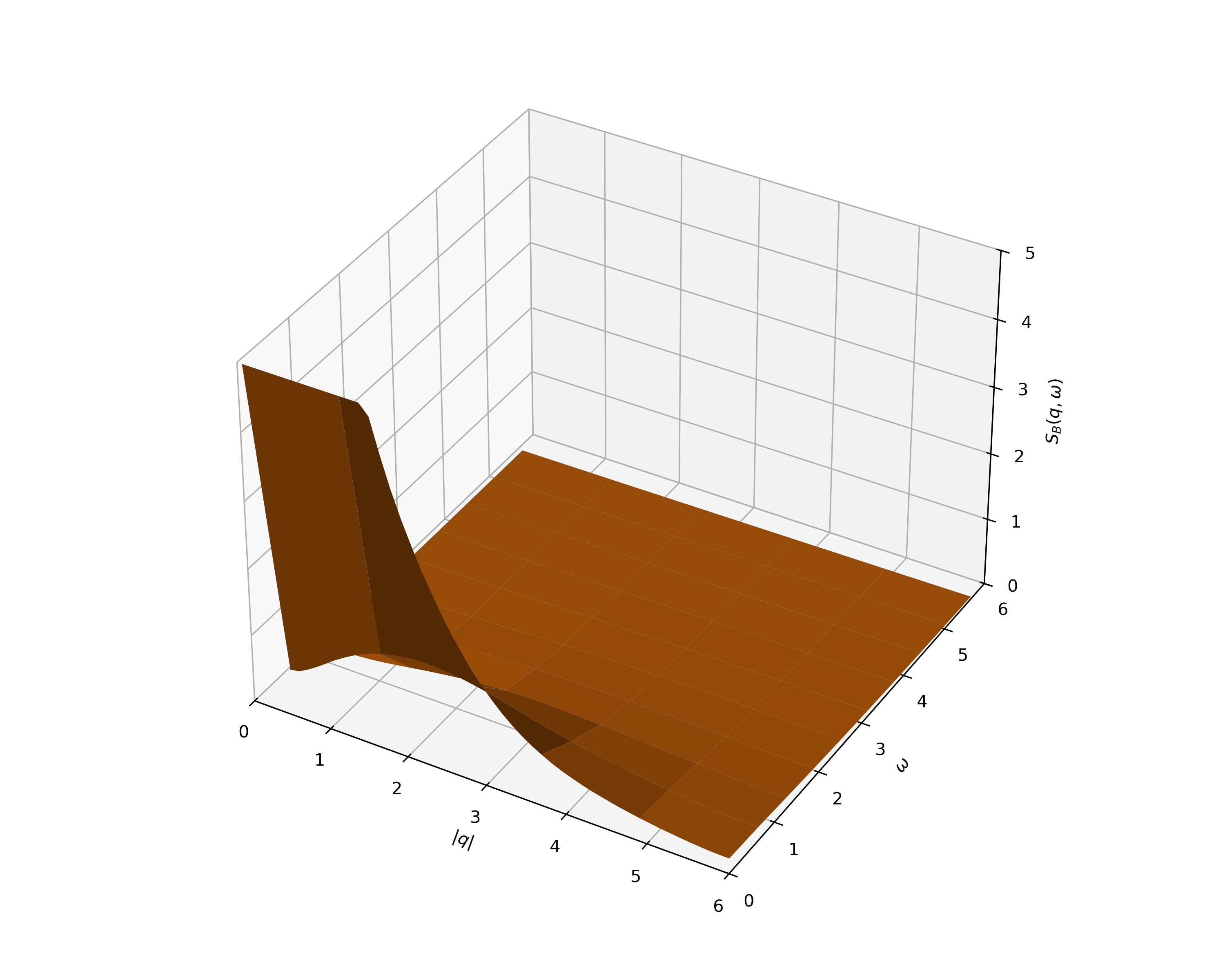}
            \caption{$S_\mathrm{B}(q, \omega)$ with cross-linking}
            \label{fig:SB_cl} 
        \end{subfigure}\hfill
        \begin{subfigure}[t]{0.48\linewidth}
            \centering
            \includegraphics[width=\linewidth]{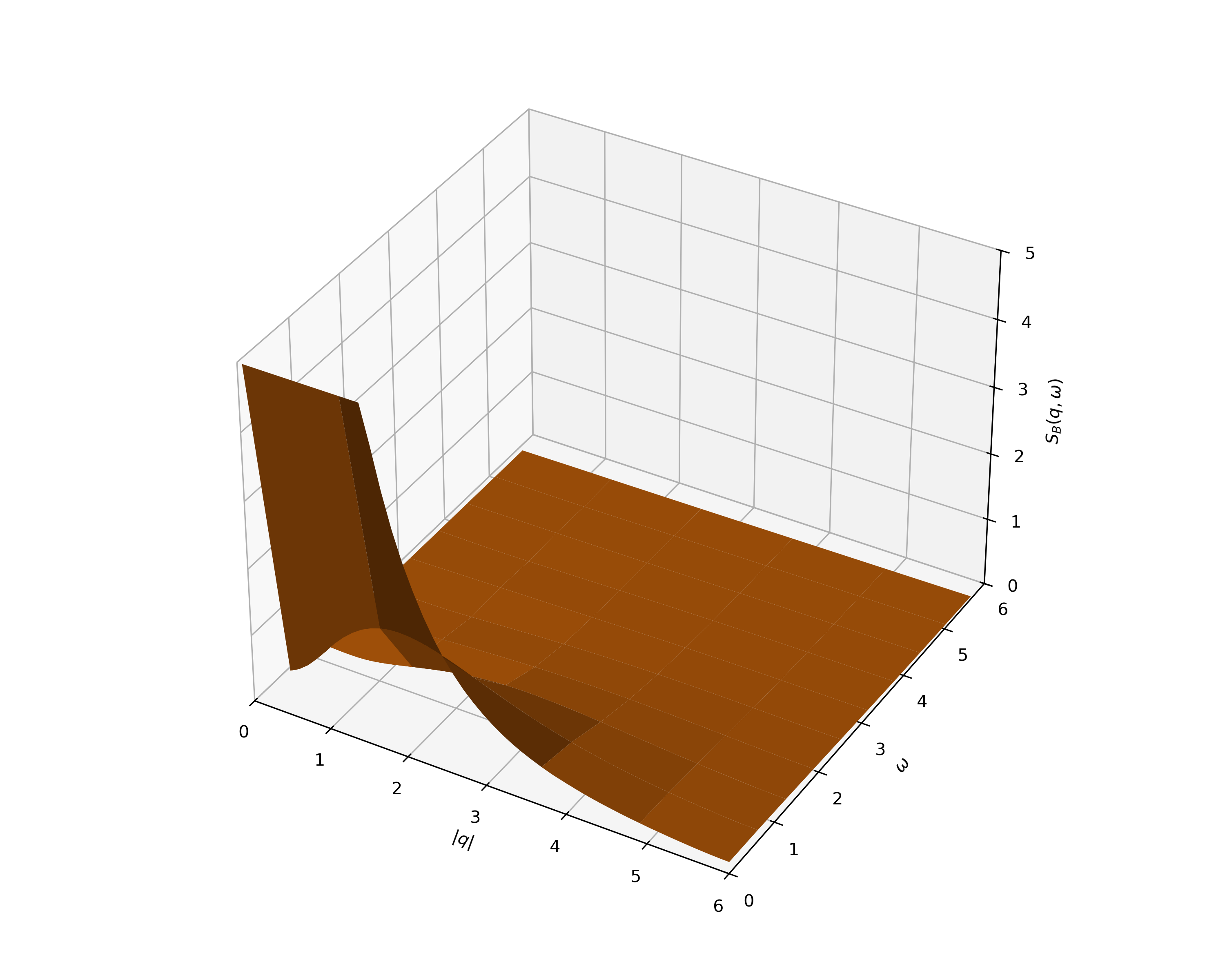}
            \caption{$S_\mathrm{B}(q, \omega)$ without cross-linking}
            \label{fig:SB_nl} 
        \end{subfigure}

        \begin{subfigure}[t]{0.48\linewidth}
            \centering
            \includegraphics[width=\linewidth]{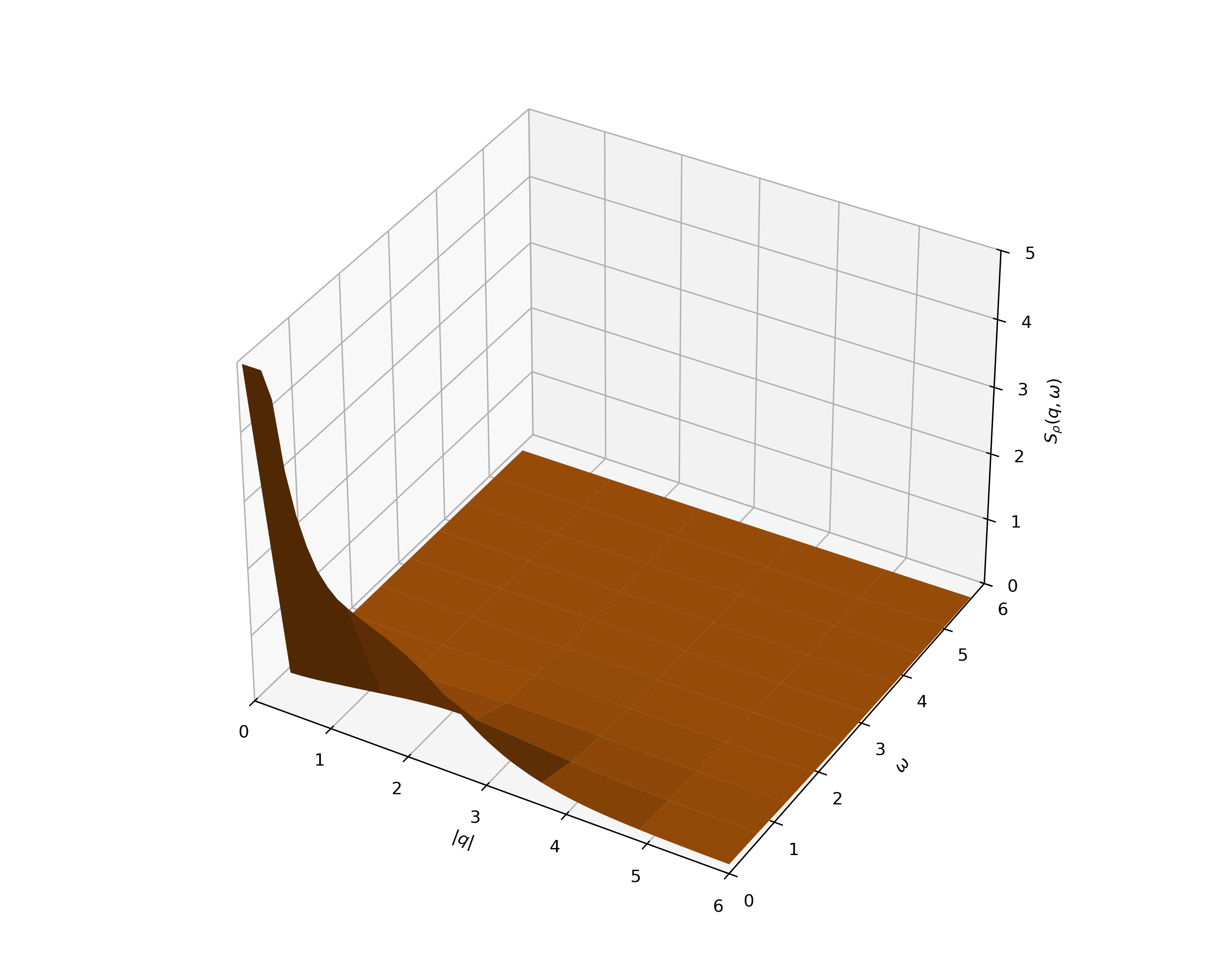}
            \caption{$S_\rho(q, \omega)$ with cross-linking}
            \label{fig:Srho_cl} 
        \end{subfigure}\hfill
        \begin{subfigure}[t]{0.48\linewidth}
            \centering
            \includegraphics[width=\linewidth]{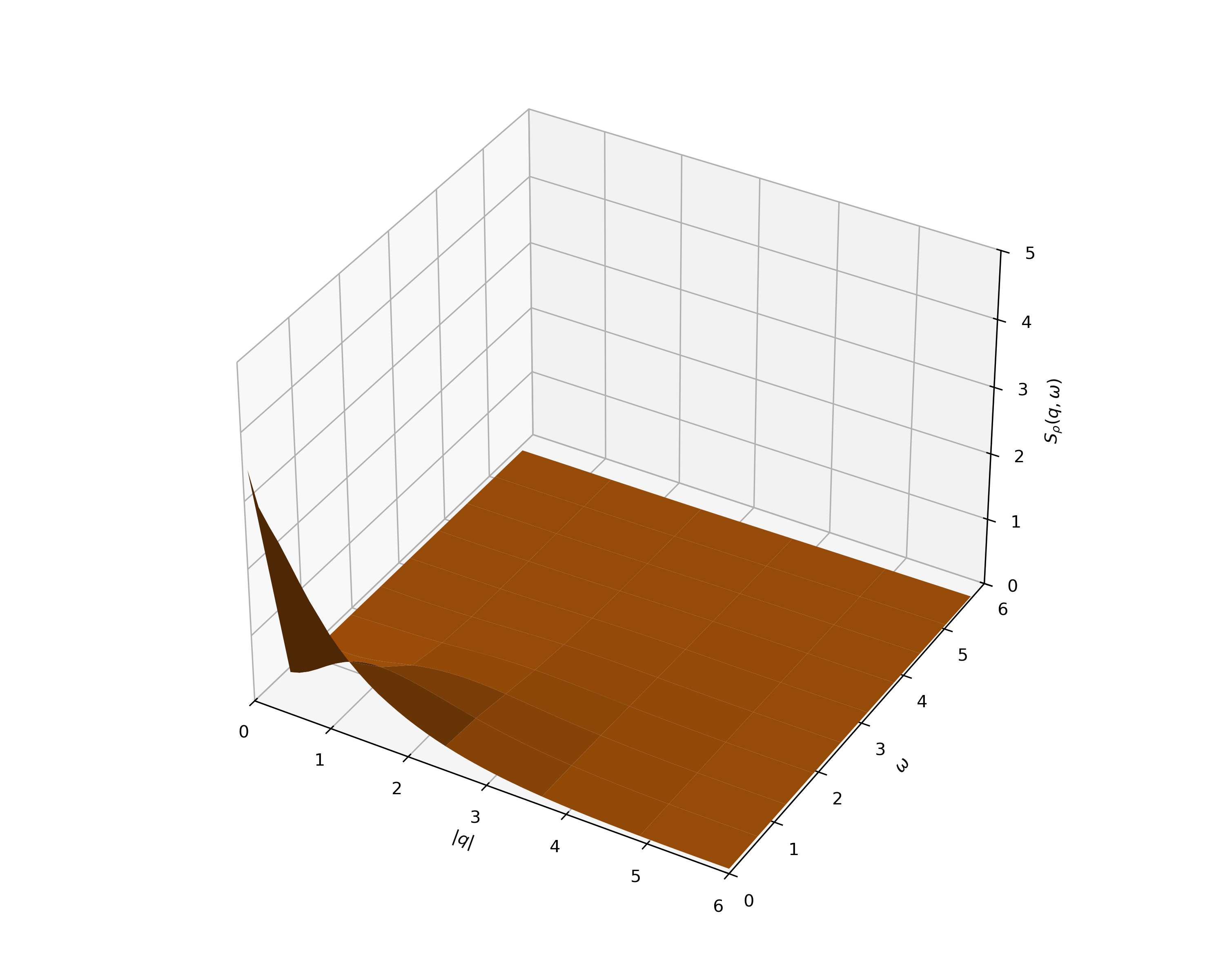}
            \caption{$S_\rho(q, \omega)$ without cross-linking}
             \label{fig:Srho_nl} 
        \end{subfigure}
    \end{minipage}

    \caption{Dynamic structure factors from polymer network simulations with and without cross-linking.}
    \label{fig:dynasorplots}
\end{figure}

The $q$-points in Fig.~\ref{fig:qdistribution} and the previously loaded trajectory, were passed to the Dynasor function to compute the dynamic structure factors, with a window size of $100$ and a timestep of $0.05$, corresponding to the output interval of the LAMMPS simulation. This resulted in a Dynasor sample containing the dynamic structure factors evaluated for the array of $q$-points and an associated frequency axis consisting of 101 $\omega$-values. As a final post-processing step,via a Dynasor function  the results were spherically averaged over $q$-points with $50$ bins to remove some of the noise. The dynamic structure factors obtained in this manner for both simulations with and without cross-linking are shown in Fig.~\ref{fig:dynasorplots}. These figures are the 3D plots corresponding to the 2D slices in the main text.

Figures \ref{fig:SA_cl} and \ref{fig:SA_nl}, show the dynamic structure factors for polymers of type $\mathrm{A}$ with and without cross-linking, respectively, whilst Figs.~\ref{fig:SB_cl} and \ref{fig:SB_nl} show the same for polymers of type $\mathrm{B}$.  All four plots show the usual shape of a peak at small $q$ and $\omega$-values that goes to zero at larger $q$ and $\omega$. The plots of the dynamic structure factors for the polymers with cross-linking, retain a similar shape but have noticeably broader peaks. The broadened peaks are especially prominent at $\omega=0$, indicating that fluctuations in polymer densities remain correlated on smaller length scales in the long time limit. This also corresponds to the overall qualitative trend of cross-linking resulting in broadening of the peaks of the dynamic structure factors, observed in the results of the analytical work presented in the main text. 

Figures \ref{fig:Srho_cl} and \ref{fig:Srho_nl} show the dynamic structure factors for the cross-linker, with and without cross-linking , respectively. Again, the peak at small $q$ and $\omega$ is observed, with a noticeable broadening in the peak, especially along the $q $ axis, with cross-linking. The peak of the dynamic structure factor also extends to significantly higher values at small $q$ and $\omega$ in the cross-linking case in Fig.~\ref{fig:Srho_cl}, indicating that a significant increase in the correlation of density fluctuations of the cross-linkers at large length and time scales. The observed feature makes sense, since majority of the cross-linkers are bound to polymers at any given time throughout the simulation, and this should result in an increase in collective motion of the cross-linkers.

\begin{figure}[t]
    \centering
    \begin{minipage}{\columnwidth}

        \begin{subfigure}[t]{0.48\linewidth}
            \centering
            \includegraphics[width=\linewidth]{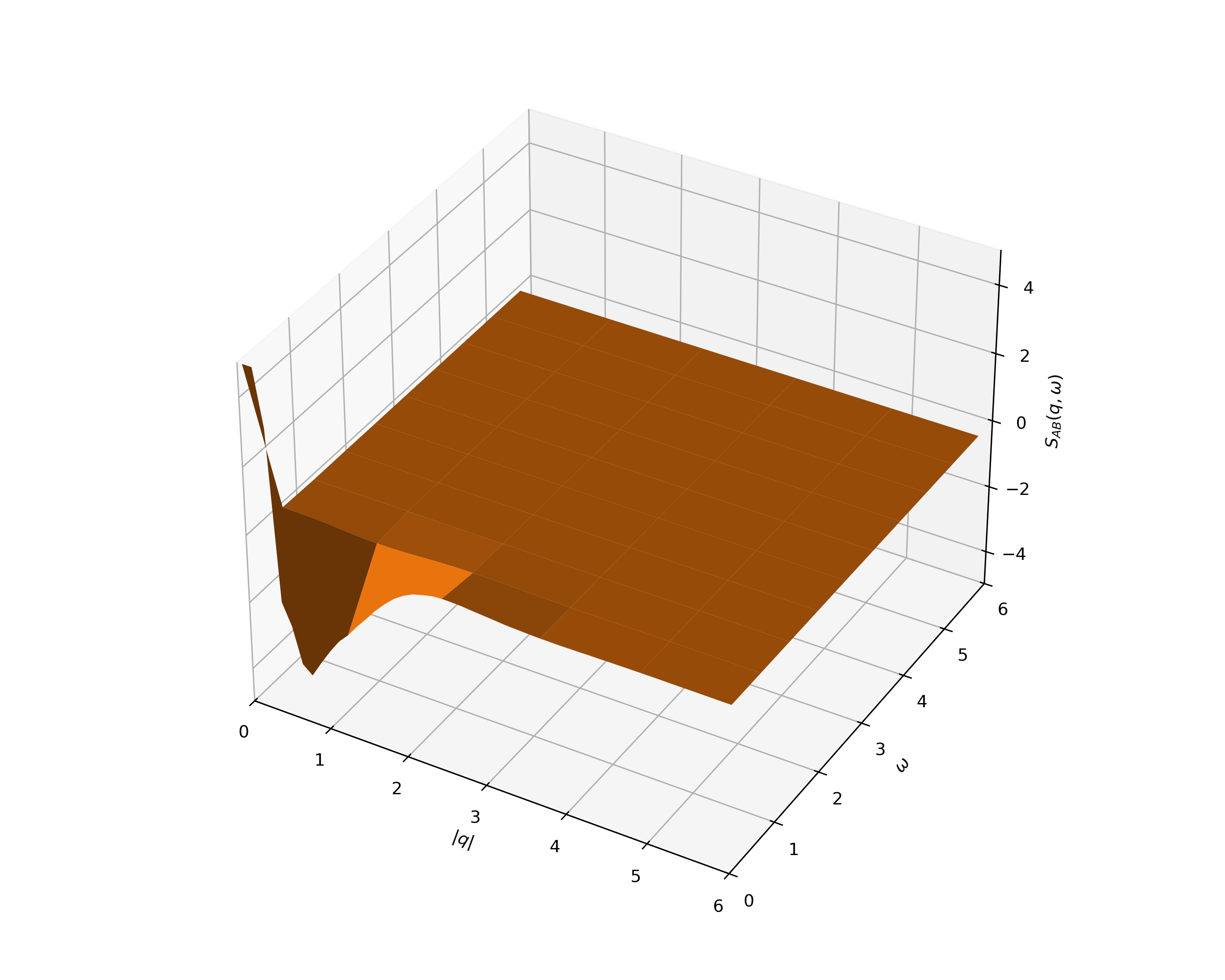}
            \caption{$S_\mathrm{AB}(q, \omega)$ with cross-linking}
            \label{fig:SAB_cl}
        \end{subfigure}\hfill
        \begin{subfigure}[t]{0.48\linewidth}
            \centering
            \includegraphics[width=\linewidth]{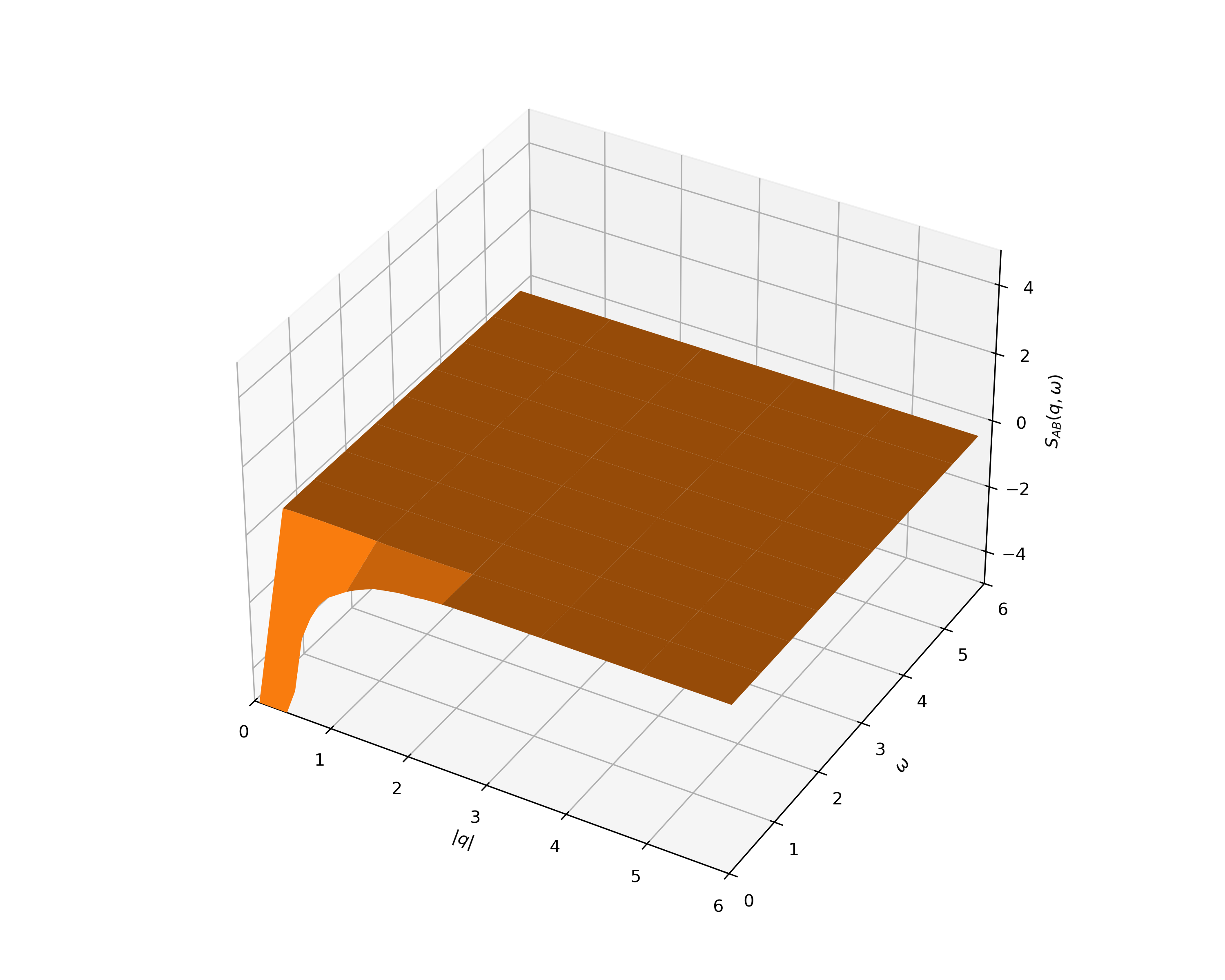}
            \caption{$S_\mathrm{AB}(q, \omega)$ without cross-linking}
            \label{fig:SAB_nl}
        \end{subfigure}

        \begin{subfigure}[t]{0.48\linewidth}
            \centering
            \includegraphics[width=\linewidth]{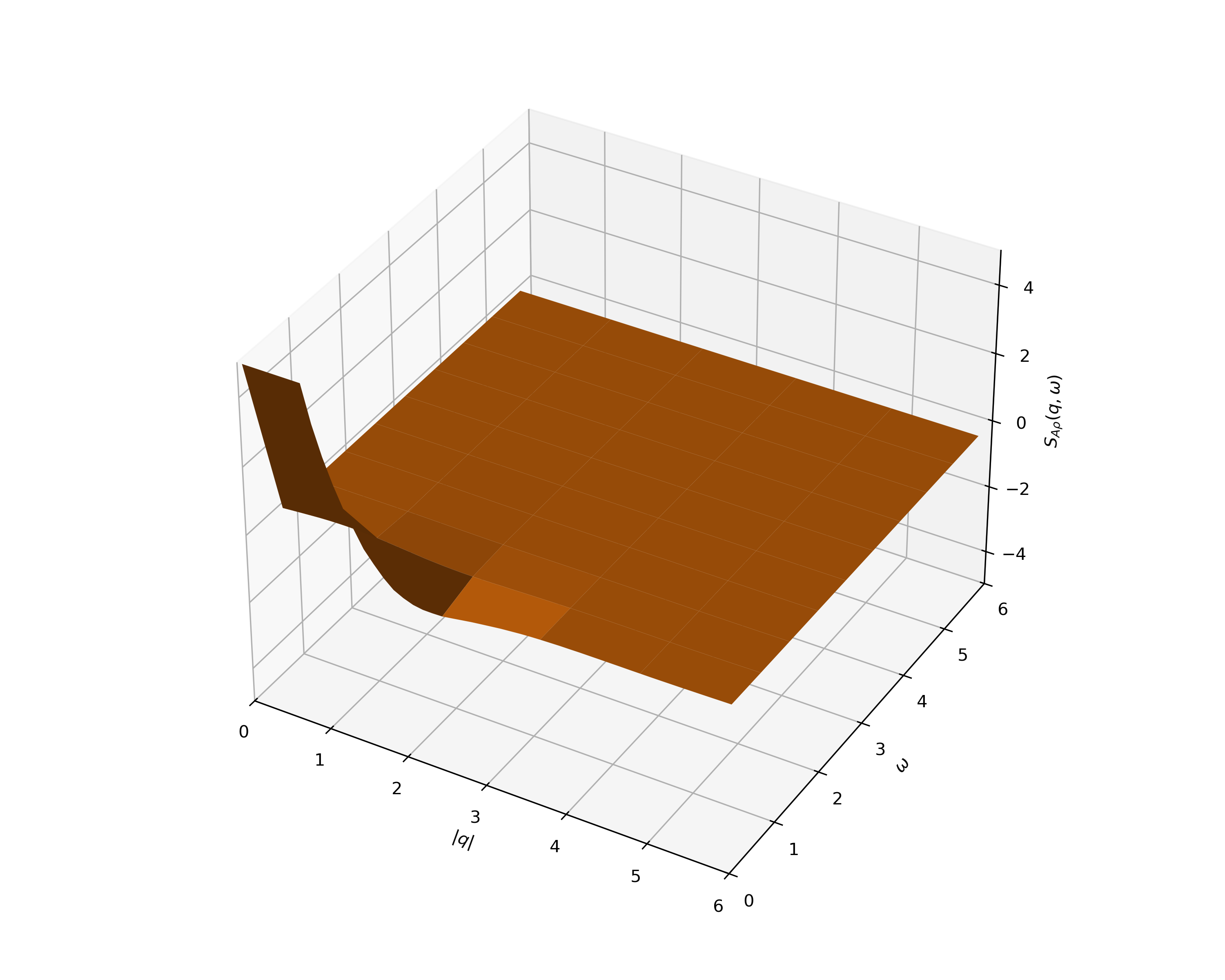}
            \caption{$S_{\mathrm{A}\rho}(q, \omega)$ with cross-linking}
            \label{fig:SrhoA_cl}
        \end{subfigure}\hfill
        \begin{subfigure}[t]{0.48\linewidth}
            \centering
            \includegraphics[width=\linewidth]{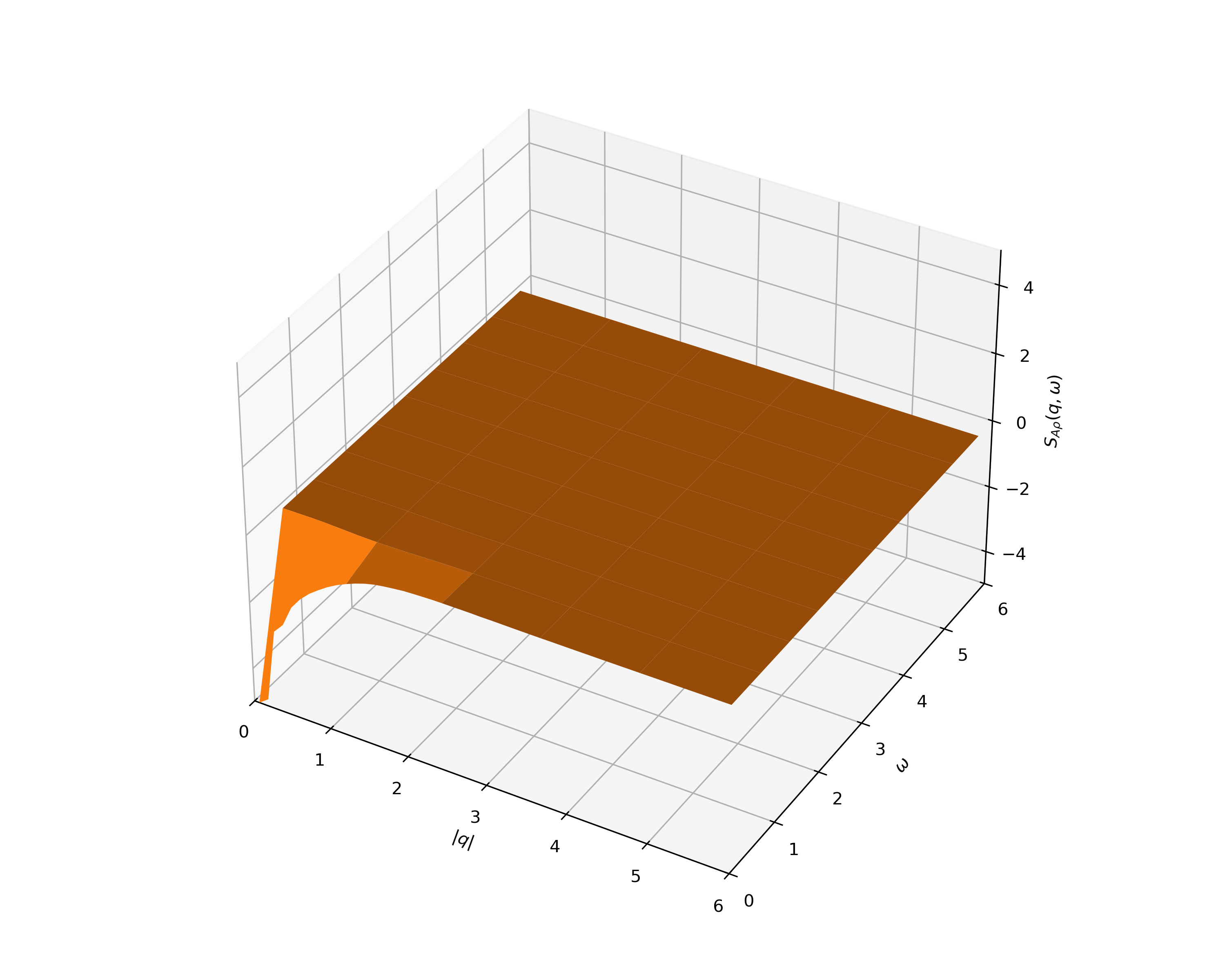}
            \caption{$S_{\mathrm{A}\rho}(q, \omega)$ without cross-linking}
            \label{fig:SrhoA_nl}
        \end{subfigure}

        \begin{subfigure}[t]{0.48\linewidth}
            \centering
            \includegraphics[width=\linewidth]{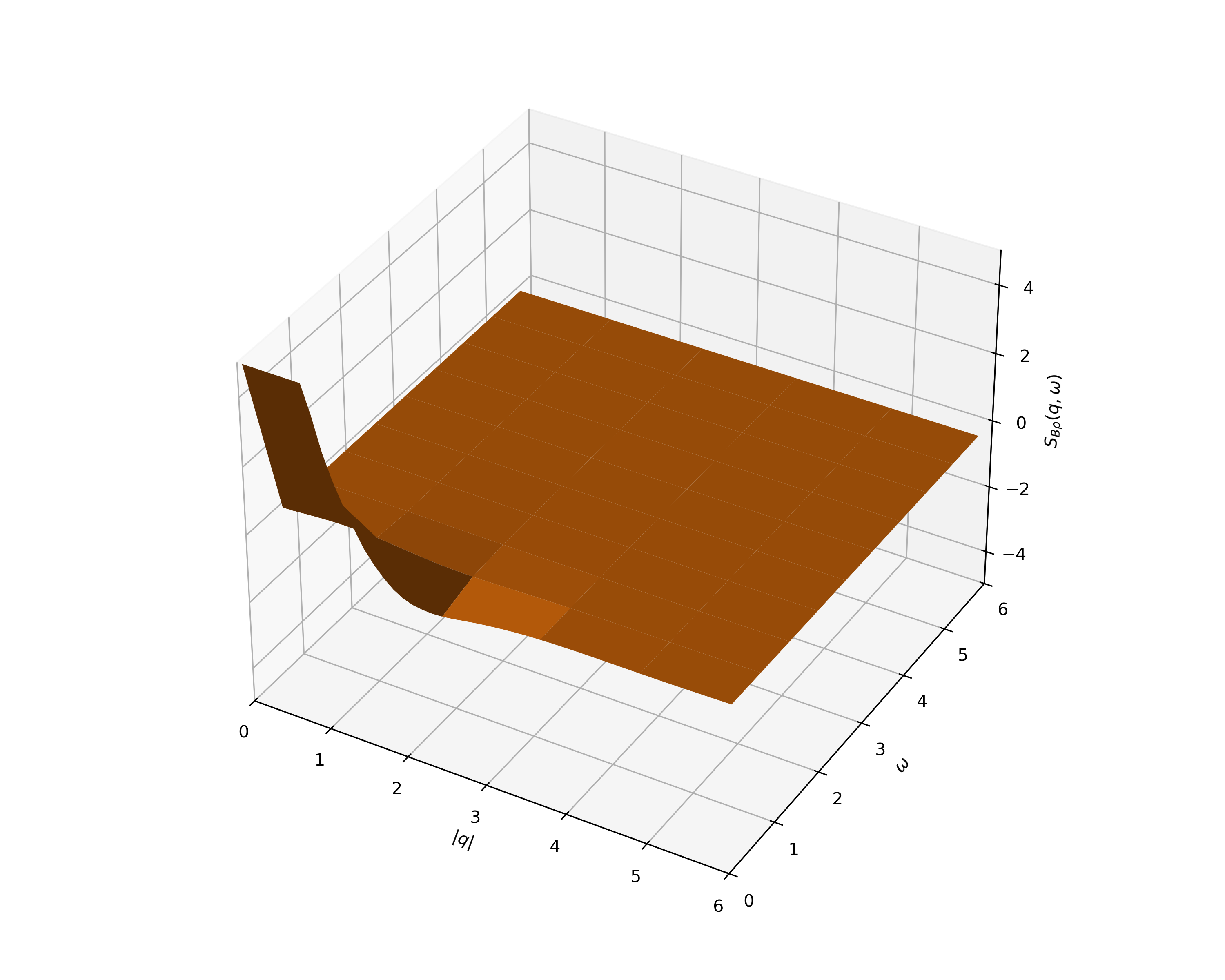}
            \caption{$S_{\mathrm{B}\rho}(q, \omega)$ with cross-linking}
            \label{fig:SrhoB_cl}
        \end{subfigure}\hfill
        \begin{subfigure}[t]{0.48\linewidth}
            \centering
            \includegraphics[width=\linewidth]{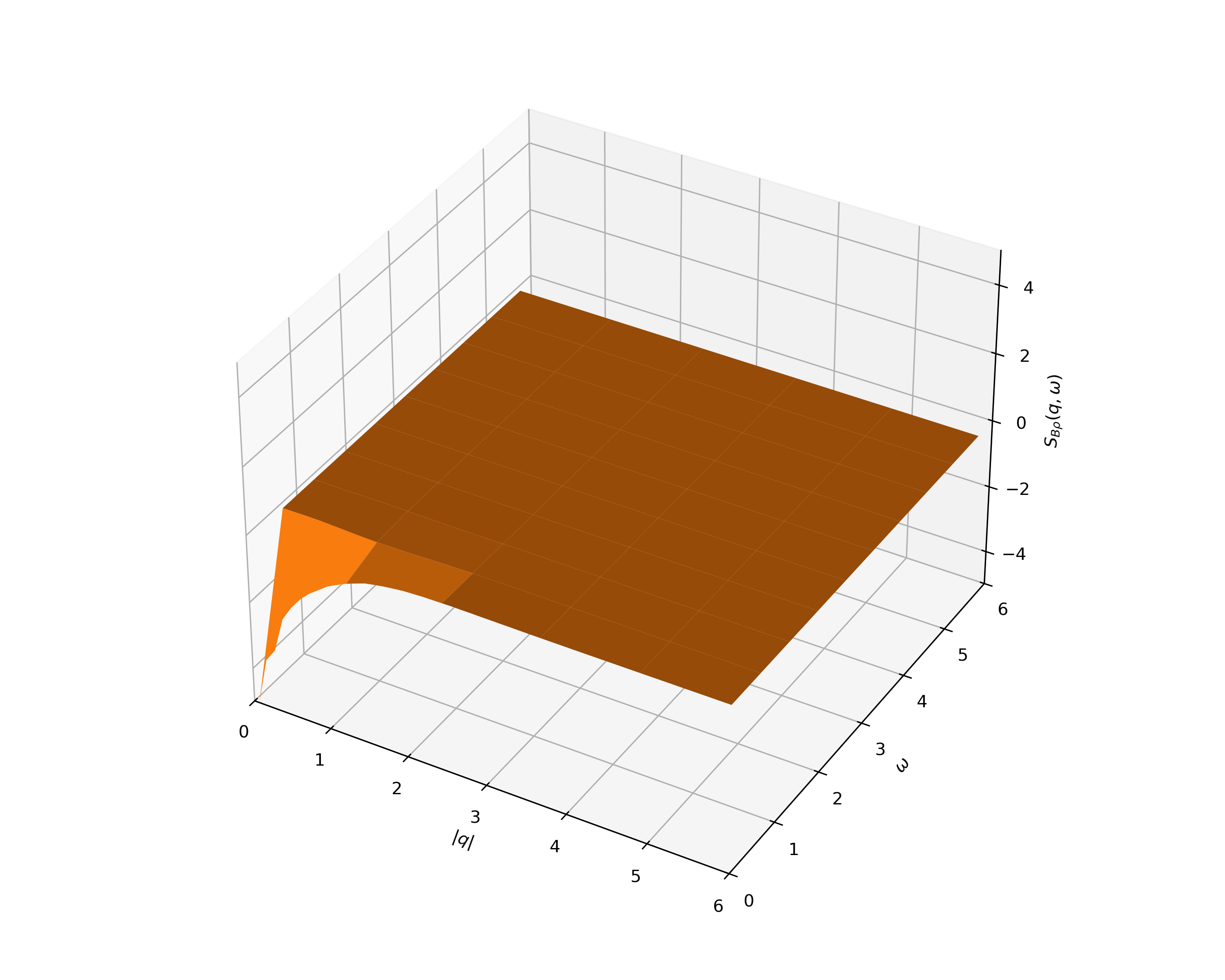}
            \caption{$S_{\mathrm{B}\rho}(q, \omega)$ without cross-linking}
            \label{fig:SrhoB_nl}
        \end{subfigure}

    \end{minipage}

    \caption{Dynamic structure factors for the cross-correlations from runs with and without cross-linking.}
    \label{fig:dynasorplots_cross}
\end{figure}

Finally, the dynamic structure factors can be investigated and compared for cross-correlations between polymers of different types and cross-linkers. This is shown in Fig.~\ref{fig:dynasorplots_cross}. The cross-correlations for the simulations without cross-linking are shown in Figs.~\ref{fig:SAB_nl}, \ref{fig:SrhoA_nl} and \ref{fig:SrhoB_nl}. These all show strong anti-correlations at large length and time scales that decay to zero at smaller length and time scales. The anti-peaks are narrower in $q$ for the cross-correlations between polymers and cross-linkers in Figs.\ref{fig:SrhoA_nl} and \ref{fig:SrhoB_nl} than between the two different types of polymers in Fig.~\ref{fig:SAB_nl}, indicating a stronger anti-correlation between polymers than between polymers and cross-linkers. This corresponds to what is expected in the simulations with no linking, since the only interaction between particles are the repulsive Lennard-Jones potentials discussed in the main text.

For the simulations with cross-linking, the cross-correlations are shown in Figs.~\ref{fig:SAB_cl}, \ref{fig:SrhoA_cl} and \ref{fig:SrhoB_cl}. These all show strong positive peaks at small $q$ and $\omega$, with varying magnitudes of anti-peaks indicating negative correlations at slightly larger $q$-values where $\omega=0$. For the cross-correlation between different polymer types in Fig.~\ref{fig:SAB_cl}, the anti-peak is the largest and occurs between $q=0$ and $q=1.0$. This indicates that, on long time scales, density fluctuations of polymers of different types are anti-correlated with one another on intermediate length scales, but become strongly correlated with one another on very large length scales near $q=0$, with no correlation on very short length scales. A similar pattern is observed in the cross-correlations between the cross-linkers and polymers in Figs.~\ref{fig:SrhoA_cl} and \ref{fig:SrhoB_cl}, but with a broader peak corresponding to strong correlations at long length scales followed by a very slight negative dip indicating anti-correlations around  $q=2$, with no correlations at smaller length scales. Thus density fluctuations of polymers and cross-linkers, at long time scales, are strongly correlated at large length scales and slightly anti-correlated on intermediate length scales with no correlations at short length scales. 

\bibliographystyle{unsrt}
\bibliography{references}

%% file: diagrams/crosslinkeranatomy.tex
\begin{tikzpicture}[scale=0.6]
\draw[<->] (xyz cs:x=-0.5) -- (xyz cs:x=5.5) node[above] {$x$};
\draw[<->] (xyz cs:y=-0.5) -- (xyz cs:y=4) node[right] {$y$};
\draw[<->] (xyz cs:z=-2.5) -- (xyz cs:z=2.5) node[above] {$z$};
\draw[->, thick] (0,0) --(2.4,2){};
\draw[->, thick] (0,0) --(4.2,2){};
\crosslinkerH{2.5}{2}{4}{2}{8pt};
\node at (1.75, 2) {$\mathbf{y}_2(t)$};
\node at (4, 1.25) {$\mathbf{y}_1(t)$};
\node at (3.25, 2.5) {$\kappa$};
\end{tikzpicture}

%% file: diagrams/crosslinkerCOM.tex
\begin{tikzpicture}[scale=0.6]
\draw[<->] (xyz cs:x=-0.5) -- (xyz cs:x=5.5) node[above] {$x$};
\draw[<->] (xyz cs:y=-0.5) -- (xyz cs:y=4) node[right] {$y$};
\draw[<->] (xyz cs:z=-2.5) -- (xyz cs:z=2.5) node[above] {$z$};
\crosslinkerH{2}{2}{3.5}{2}{8pt};
\node at (2.75, 2.75) {$\mathbf{y}(t)$};
\node at (2.75,1.25) {$\mathbf{Y}(t)$};
\draw[thick, ->] (2,2.4) --(3.5,2.4);
\draw[thick, ->] (0,0) --(2.75,2);
\end{tikzpicture}

%% file: diagrams/crosslinker_tracksDiagram.tex
\begin{tikzpicture}[scale=0.9]
\draw[thick,red] (-6,-1) --(10,-1){};

\draw  node[fill= red,circle,scale=2] at (-6,-1) {};
\draw  node[fill= red,circle,scale=2] at (-4,-1) {};
\draw  node[fill= red,circle,scale=2] at (-2,-1) {};
\draw  node[fill= red,circle,scale=2] at (0,-1) {};
\draw  node[fill= red,circle,scale=2] at (2,-1) {};
\draw  node[fill= red,circle,scale=2] at (4,-1) {};
\node[above,red] at (6, -1) {\Large $...$};
\draw  node[fill= red,circle,scale=2] at (8,-1) {};
\draw  node[fill= red,circle,scale=2] at (10,-1) {};

\node[red] at (-7,-1) {$\mathrm{A}:$};
\node[red] at (-6,-1.7) {$\phi^*(\mathbf{r}_1,t)$};
\node[red] at (-4,-1.7) {$\phi^*(\mathbf{r}_2,t)$};
\node[red] at (-2,-1.7) {$\phi^*(\mathbf{r}_3,t)$};
\node[red] at (0,-1.7) {$\phi^*(\mathbf{r}_4,t)$};
\node[red] at (2,-1.7) {$\phi^*(\mathbf{r}_5,t)$};
\node[red] at (4,-1.7) {$\phi^*(\mathbf{r}_6,t)$};
\node[above,red] at (5.5, -1.9) {\Large $...$};
\node[red] at (7.5,-1.7) {$\phi^*(\mathbf{r}_{N-1},t)$};
\node[red] at (10,-1.7) {$\phi^*(\mathbf{r}_N,t)$};

\draw[thick,blue] (-6,2) --(10,2){};
\draw  node[fill= blue,circle,scale=2] at (-6,2) {};
\draw  node[fill= blue,circle,scale=2] at (-4,2) {};
\draw  node[fill= blue,circle,scale=2] at (-2,2) {};
\draw  node[fill= blue,circle,scale=2] at (0,2) {};
\draw  node[fill= blue,circle,scale=2] at (2,2) {};
\draw  node[fill= blue,circle,scale=2] at (4,2) {};
\node[blue] at (6, 1.8) {\Large $...$};
\draw  node[fill= blue,circle,scale=2] at (8,2) {};
\draw  node[fill= blue,circle,scale=2] at (10,2) {};

\node[blue] at (-7,2) {$\mathrm{B}:$};
\node[blue] at (-6,2.7) {$\psi^*(\mathbf{R}_1,t)$};
\node[blue] at (-4,2.7) {$\psi^*(\mathbf{R}_2,t)$};
\node[blue] at (-2,2.7) {$\psi^*(\mathbf{R}_3,t)$};
\node[blue] at (0,2.7) {$\psi^*(\mathbf{R}_4,t)$};
\node[blue] at (2,2.7) {$\psi^*(\mathbf{R}_5,t)$};
\node[blue] at (4,2.7) {$\psi^*(\mathbf{R}_6,t)$};
\node[above, blue] at (5.5, 2.5) {\Large $...$};
\node[blue] at (7.5,2.7) {$\psi^*(\mathbf{R}_{N'-1},t)$};
\node[blue] at (10,2.7) {$\psi^*(\mathbf{R}_{N'},t)$};

{\color{green}\crosslinkerV{-6}{-0.5}{-6}{1.5}{18pt};}
\node[green, thick] at (-6.5,0.5) {\Large $\mathrm{e}^\epsilon$};
\crosslinkerV{-1}{0}{-1}{1}{8pt};
{\color{green}\crosslinkerV{10}{-0.5}{8}{1.5}{25pt};}
\node[green, thick] at (9.5,0.7) {\Large $\mathrm{e}^\epsilon$};

{\color{cyan}\crosslinkerH{-3.5}{2}{-2.5}{2}{8pt};}
\node[cyan, thick] at (-3,1.5) {\Large $\mathrm{e}^{\mu_\mathrm{B}}$};
{\color{magenta}\crosslinkerH{0.5}{-1}{1.5}{-1}{8pt};}
\node[magenta, thick] at (1,-0.5) {\Large $\mathrm{e}^{\mu_\mathrm{A}}$};
\crosslinkerH{3.8}{0.6}{5}{1}{8pt};

\node at (6, 1) {\Large $...$};

\end{tikzpicture}

%% file: diagrams/crosslinker_tracksDiagramIntra.tex
\begin{tikzpicture}[scale=0.9]
\draw[thick,red] (-6,-1) --(10,-1){};

\draw  node[fill= red,circle,scale=2] at (-6,-1) {};
\draw  node[fill= red,circle,scale=2] at (-4,-1) {};
\draw  node[fill= red,circle,scale=2] at (-2,-1) {};
\draw  node[fill= red,circle,scale=2] at (0,-1) {};
\draw  node[fill= red,circle,scale=2] at (2,-1) {};
\draw  node[fill= red,circle,scale=2] at (4,-1) {};
\node[above,red] at (6, -1) {\Large $...$};
\draw  node[fill= red,circle,scale=2] at (8,-1) {};
\draw  node[fill= red,circle,scale=2] at (10,-1) {};

\node[red] at (-6,-1.7) {$\mathbf{r}_1$};
\node[red] at (-4,-1.7) {$\mathbf{r}_2$};
\node[red] at (-2,-1.7) {$\mathbf{r}_3$};
\node[red] at (0,-1.7) {$\mathbf{r}_4$};
\node[red] at (2,-1.7) {$\mathbf{r}_5$};
\node[red] at (4,-1.7) {$\mathbf{r}_6$};
\node[above,red] at (6, -1.9) {\Large $...$};
\node[red] at (8,-1.7) {$\mathbf{r}_{N-1}$};
\node[red] at (10,-1.7) {$\mathbf{r}_N$};



 \crosslinkerV{-1}{0}{-1}{1}{8pt};

{\color{magenta}\crosslinkerH{-3.5}{-1}{-2.5}{-1}{8pt};}
\node[magenta, thick] at (-3,-0.5) {\Large $\mathrm{e}^{\mu}$};
{\color{magenta}\crosslinkerH{0.5}{-1}{1.5}{-1}{8pt};}
\node[magenta, thick] at (1,-0.5) {\Large $\mathrm{e}^{\mu}$};
\crosslinkerH{3.8}{0.6}{5}{1}{8pt};

\node at (6, 1) {\Large $...$};

\end{tikzpicture}

%% file: diagrams/crosslinker_tracksDiagramInterOnly.tex
\begin{tikzpicture}[scale=0.9]
\draw[thick,red] (-6,-1) --(10,-1){};

\draw  node[fill= red,circle,scale=2] at (-6,-1) {};
\draw  node[fill= red,circle,scale=2] at (-4,-1) {};
\draw  node[fill= red,circle,scale=2] at (-2,-1) {};
\draw  node[fill= red,circle,scale=2] at (0,-1) {};
\draw  node[fill= red,circle,scale=2] at (2,-1) {};
\draw  node[fill= red,circle,scale=2] at (4,-1) {};
\node[above,red] at (6, -1) {\Large $...$};
\draw  node[fill= red,circle,scale=2] at (8,-1) {};
\draw  node[fill= red,circle,scale=2] at (10,-1) {};

\node[red] at (-6,-1.7) {$\mathbf{r}_1$};
\node[red] at (-4,-1.7) {$\mathbf{r}_2$};
\node[red] at (-2,-1.7) {$\mathbf{r}_3$};
\node[red] at (0,-1.7) {$\mathbf{r}_4$};
\node[red] at (2,-1.7) {$\mathbf{r}_5$};
\node[red] at (4,-1.7) {$\mathbf{r}_6$};
\node[above,red] at (6, -1.9) {\Large $...$};
\node[red] at (8,-1.7) {$\mathbf{r}_{N-1}$};
\node[red] at (10,-1.7) {$\mathbf{r}_N$};

\draw[thick,blue] (-6,2) --(10,2){};
\draw  node[fill= blue,circle,scale=2] at (-6,2) {};
\draw  node[fill= blue,circle,scale=2] at (-4,2) {};
\draw  node[fill= blue,circle,scale=2] at (-2,2) {};
\draw  node[fill= blue,circle,scale=2] at (0,2) {};
\draw  node[fill= blue,circle,scale=2] at (2,2) {};
\draw  node[fill= blue,circle,scale=2] at (4,2) {};
\node[blue] at (6, 1.8) {\Large $...$};
\draw  node[fill= blue,circle,scale=2] at (8,2) {};
\draw  node[fill= blue,circle,scale=2] at (10,2) {};

\node[blue] at (-6,2.7) {$\mathbf{R}_1$};
\node[blue] at (-4,2.7) {$\mathbf{R}_2$};
\node[blue] at (-2,2.7) {$\mathbf{R}_3$};
\node[blue] at (0,2.7) {$\mathbf{R}_4$};
\node[blue] at (2,2.7) {$\mathbf{R}_5$};
\node[blue] at (4,2.7) {$\mathbf{R}_6$};
\node[above, blue] at (6, 2.5) {\Large $...$};
\node[blue] at (8,2.7) {$\mathbf{R}_{N'-1}$};
\node[blue] at (10,2.7) {$\mathbf{R}_{N'}$};

{\color{green}\crosslinkerV{-6}{-0.5}{-6}{1.5}{18pt};}
\node[green, thick] at (-6.5,0.5) {\Large $\mathrm{e}^\epsilon$};
\crosslinkerV{-1}{0}{-1}{1}{8pt};
{\color{green}\crosslinkerV{10}{-0.5}{8}{1.5}{25pt};}
\node[green, thick] at (9.5,0.7) {\Large $\mathrm{e}^\epsilon$};

\crosslinkerH{3.8}{0.6}{5}{1}{8pt};

\node at (6, 1) {\Large $...$};

\end{tikzpicture}

%% file: diagrams/BondLegend.tex
\begin{tikzpicture}[scale=0.5]

\node at (2.5, 4) {\textbf{Bond type Labels:}};
\node at (18,4) {\textbf{Atom type Labels:}};

\matrix [anchor=west, column sep=10pt, row sep=4pt] at (0, 0)
{
  \draw[ultra thick, violet, dashed] (0,0) -- (0.8,0); & \node {bondPolyACL}; & & \\

  \draw[ultra thick, red] (0,0) -- (0.8,0); & \node {bondPolyA}; &
  \node[fill=red, circle, scale=1.5] at (2,0) {}; & \node {beadPolyA}; \\

  \draw[ultra thick, orange] (0,0) -- (0.8,0); & \node {bondCL}; &
  \node[fill=orange, circle, scale=1.5] at (2,0) {}; & \node {beadCL}; \\

  \draw[ultra thick, blue] (0,0) -- (0.8,0); & \node {bondPolyB}; &
  \node[fill=blue, circle, scale=1.5] at (2,0) {}; & \node {beadPolyB}; \\

  \draw[ultra thick, cyan, dashed] (0,0) -- (0.8,0); & \node {bondPolyBCL}; & & \\
};

\end{tikzpicture}

%% file: diagrams/step1A.tex
\begin{tikzpicture}[scale=0.7]

\draw[ultra thick,red] (0,5) --(0,-5){};
\draw  node[fill= red,circle,scale=2.5] at (0, 4) {};
\draw  node[] at (-1,4) {1};
\draw  node[fill= red,circle,scale=2.5] at (0, 2) {};
\draw  node[] at (-1,2) {2};
\draw  node[fill= red,circle,scale=2.5] at (0, 0) {};
\draw  node[] at (-1,0) {3};
\draw  node[fill= red,circle,scale=2.5] at (0, -2) {};
\draw  node[] at (-1,-2) {4};
\draw  node[fill= red,circle,scale=2.5] at (0, -4) {};
\draw  node[] at (-1,-4) {5};

\draw[ultra thick,orange] (2,0) --(4,0){};
\draw  node[fill =orange,circle,scale=2.5] at (2, 0) {};
\draw  node[] at (2,1) {6};
\draw  node[fill =orange,circle,scale=2.5] at (4, 0) {};
\draw  node[] at (4,1) {7};

\draw node[] at (9,2.5){step 1};
\draw[thick,->] (7,2) --(11,2){};

\draw node[] at (9,-2.5){step 1 reverse};
\draw[thick,<-] (7,-2) --(11,-2){};


\draw[ultra thick, violet, dashed] (14,0)--(16,0){};
\draw[ultra thick,red] (14,5) --(14,-5){};
\draw  node[fill= red,circle,scale=2.5] at (14, 4) {};
\draw  node[] at (13,4) {1};
\draw  node[fill= red,circle,scale=2.5] at (14, 2) {};
\draw  node[] at (13,2) {2};
\draw  node[fill= red,circle,scale=2.5] at (14, 0) {};
\draw  node[] at (13,0) {3};
\draw  node[fill= red,circle,scale=2.5] at (14, -2) {};
\draw  node[] at (13,-2) {4};
\draw  node[fill= red,circle,scale=2.5] at (14, -4) {};
\draw  node[] at (13,-4) {5};

\draw[ultra thick,orange] (16,0) --(18,0){};
\draw  node[fill =orange,circle,scale=2.5] at (16, 0) {};
\draw  node[] at (16,1) {6};
\draw  node[fill =orange,circle,scale=2.5] at (18, 0) {};
\draw  node[] at (18,1) {7};

\end{tikzpicture}

%% file: diagrams/step1B.tex
\begin{tikzpicture}[scale=0.7]

\draw[ultra thick,blue] (0,5) --(0,-5){};
\draw  node[fill= blue,circle,scale=2.5] at (0, 4) {};
\draw  node[] at (-1,4) {1};
\draw  node[fill= blue,circle,scale=2.5] at (0, 2) {};
\draw  node[] at (-1,2) {2};
\draw  node[fill= blue,circle,scale=2.5] at (0, 0) {};
\draw  node[] at (-1,0) {3};
\draw  node[fill= blue,circle,scale=2.5] at (0, -2) {};
\draw  node[] at (-1,-2) {4};
\draw  node[fill= blue,circle,scale=2.5] at (0, -4) {};
\draw  node[] at (-1,-4) {5};

\draw[ultra thick,orange] (2,0) --(4,0){};
\draw  node[fill =orange,circle,scale=2.5] at (2, 0) {};
\draw  node[] at (2,1) {6};
\draw  node[fill =orange,circle,scale=2.5] at (4, 0) {};
\draw  node[] at (4,1) {7};

\draw node[] at (9,2.5){step 1};
\draw[thick,->] (7,2) --(11,2){};

\draw node[] at (9,-2.5){step 1 reverse};
\draw[thick,<-] (7,-2) --(11,-2){};


\draw[ultra thick, cyan, dashed] (14,0)--(16,0){};
\draw[ultra thick,blue] (14,5) --(14,-5){};
\draw  node[fill= blue,circle,scale=2.5] at (14, 4) {};
\draw  node[] at (13,4) {1};
\draw  node[fill= blue,circle,scale=2.5] at (14, 2) {};
\draw  node[] at (13,2) {2};
\draw  node[fill= blue,circle,scale=2.5] at (14, 0) {};
\draw  node[] at (13,0) {3};
\draw  node[fill= blue,circle,scale=2.5] at (14, -2) {};
\draw  node[] at (13,-2) {4};
\draw  node[fill= blue,circle,scale=2.5] at (14, -4) {};
\draw  node[] at (13,-4) {5};

\draw[ultra thick,orange] (16,0) --(18,0){};
\draw  node[fill =orange,circle,scale=2.5] at (16, 0) {};
\draw  node[] at (16,1) {6};
\draw  node[fill =orange,circle,scale=2.5] at (18, 0) {};
\draw  node[] at (18,1) {7};

\end{tikzpicture}

%% file: diagrams/step2interAB.tex
\begin{tikzpicture}[scale=0.7]

\draw[ultra thick, violet, dashed] (0,0)--(2,0){};
\draw[ultra thick,red] (0,5) --(0,-5){};
\draw  node[fill= red,circle,scale=2.5] at (0, 4) {};
\draw  node[] at (-1,4) {1};
\draw  node[fill= red,circle,scale=2.5] at (0, 2) {};
\draw  node[] at (-1,2) {2};
\draw  node[fill= red,circle,scale=2.5] at (0, 0) {};
\draw  node[] at (-1,0) {3};
\draw  node[fill= red,circle,scale=2.5] at (0, -2) {};
\draw  node[] at (-1,-2) {4};
\draw  node[fill= red,circle,scale=2.5] at (0, -4) {};
\draw  node[] at (-1,-4) {5};

\draw[ultra thick,orange] (2,0) --(4,0){};
\draw  node[fill =orange,circle,scale=2.5] at (2, 0) {};
\draw  node[] at (2,1) {6};
\draw  node[fill =orange,circle,scale=2.5] at (4, 0) {};
\draw  node[] at (4,1) {7};

\draw[ultra thick,blue] (6,5) --(6,-5){};
\draw  node[fill= blue,circle,scale=2.5] at (6, 4) {};
\draw  node[] at (7,4) {8};
\draw  node[fill= blue,circle,scale=2.5] at (6, 2) {};
\draw  node[] at (7,2) {9};
\draw  node[fill= blue,circle,scale=2.5] at (6, 0) {};
\draw  node[] at (7,0) {10};
\draw  node[fill= blue,circle,scale=2.5] at (6, -2) {};
\draw  node[] at (7,-2) {11};
\draw  node[fill= blue,circle,scale=2.5] at (6, -4) {};
\draw  node[] at (7,-4) {12};

\draw node[] at (10,2.5){step 2};
\draw[thick,->] (8,2) --(12,2){};

\draw node[] at (10,-2.5){step 2 reverse};
\draw[thick,<-] (8,-2) --(12,-2){};


\draw[ultra thick, violet, dashed] (14,0)--(16,0){};
\draw[ultra thick,cyan, dashed] (18,0)--(20,0){};
\draw[ultra thick,red] (14,5) --(14,-5){};
\draw  node[fill= red,circle,scale=2.5] at (14, 4) {};
\draw  node[] at (13,4) {1};
\draw  node[fill= red,circle,scale=2.5] at (14, 2) {};
\draw  node[] at (13,2) {2};
\draw  node[fill= red,circle,scale=2.5] at (14, 0) {};
\draw  node[] at (13,0) {3};
\draw  node[fill= red,circle,scale=2.5] at (14, -2) {};
\draw  node[] at (13,-2) {4};
\draw  node[fill= red,circle,scale=2.5] at (14, -4) {};
\draw  node[] at (13,-4) {5};

\draw[ultra thick,orange] (16,0) --(18,0){};
\draw  node[fill =orange,circle,scale=2.5] at (16, 0) {};
\draw  node[] at (16,1) {6};
\draw  node[fill =orange,circle,scale=2.5] at (18, 0) {};
\draw  node[] at (18,1) {7};

\draw[ultra thick,blue] (20,5) --(20,-5){};
\draw  node[fill= blue,circle,scale=2.5] at (20, 4) {};
\draw  node[] at (21,4) {8};
\draw  node[fill= blue,circle,scale=2.5] at (20, 2) {};
\draw  node[] at (21,2) {9};
\draw  node[fill= blue,circle,scale=2.5] at (20, 0) {};
\draw  node[] at (21,0) {10};
\draw  node[fill= blue,circle,scale=2.5] at (20, -2) {};
\draw  node[] at (21,-2) {11};
\draw  node[fill= blue,circle,scale=2.5] at (20, -4) {};
\draw  node[] at (21,-4) {12};

\end{tikzpicture}

%% file: diagrams/step2interBA.tex
\begin{tikzpicture}[scale=0.7]

\draw[ultra thick, cyan, dashed] (0,0)--(2,0){};
\draw[ultra thick,blue] (0,5) --(0,-5){};
\draw  node[fill= blue,circle,scale=2.5] at (0, 4) {};
\draw  node[] at (-1,4) {1};
\draw  node[fill= blue,circle,scale=2.5] at (0, 2) {};
\draw  node[] at (-1,2) {2};
\draw  node[fill= blue,circle,scale=2.5] at (0, 0) {};
\draw  node[] at (-1,0) {3};
\draw  node[fill= blue,circle,scale=2.5] at (0, -2) {};
\draw  node[] at (-1,-2) {4};
\draw  node[fill= blue,circle,scale=2.5] at (0, -4) {};
\draw  node[] at (-1,-4) {5};

\draw[ultra thick,orange] (2,0) --(4,0){};
\draw  node[fill =orange,circle,scale=2.5] at (2, 0) {};
\draw  node[] at (2,1) {6};
\draw  node[fill =orange,circle,scale=2.5] at (4, 0) {};
\draw  node[] at (4,1) {7};

\draw[ultra thick,red] (6,5) --(6,-5){};
\draw  node[fill= red,circle,scale=2.5] at (6, 4) {};
\draw  node[] at (7,4) {8};
\draw  node[fill= red,circle,scale=2.5] at (6, 2) {};
\draw  node[] at (7,2) {9};
\draw  node[fill= red,circle,scale=2.5] at (6, 0) {};
\draw  node[] at (7,0) {10};
\draw  node[fill= red,circle,scale=2.5] at (6, -2) {};
\draw  node[] at (7,-2) {11};
\draw  node[fill= red,circle,scale=2.5] at (6, -4) {};
\draw  node[] at (7,-4) {12};

\draw node[] at (10,2.5){step 2};
\draw[thick,->] (8,2) --(12,2){};

\draw node[] at (10,-2.5){step 2 reverse};
\draw[thick,<-] (8,-2) --(12,-2){};


\draw[ultra thick, cyan, dashed] (14,0)--(16,0){};
\draw[ultra thick,violet, dashed] (18,0)--(20,0){};
\draw[ultra thick,blue] (14,5) --(14,-5){};
\draw  node[fill= blue,circle,scale=2.5] at (14, 4) {};
\draw  node[] at (13,4) {1};
\draw  node[fill= blue,circle,scale=2.5] at (14, 2) {};
\draw  node[] at (13,2) {2};
\draw  node[fill= blue,circle,scale=2.5] at (14, 0) {};
\draw  node[] at (13,0) {3};
\draw  node[fill= blue,circle,scale=2.5] at (14, -2) {};
\draw  node[] at (13,-2) {4};
\draw  node[fill= blue,circle,scale=2.5] at (14, -4) {};
\draw  node[] at (13,-4) {5};

\draw[ultra thick,orange] (16,0) --(18,0){};
\draw  node[fill =orange,circle,scale=2.5] at (16, 0) {};
\draw  node[] at (16,1) {6};
\draw  node[fill =orange,circle,scale=2.5] at (18, 0) {};
\draw  node[] at (18,1) {7};

\draw[ultra thick,red] (20,5) --(20,-5){};
\draw  node[fill= red,circle,scale=2.5] at (20, 4) {};
\draw  node[] at (21,4) {8};
\draw  node[fill= red,circle,scale=2.5] at (20, 2) {};
\draw  node[] at (21,2) {9};
\draw  node[fill= red,circle,scale=2.5] at (20, 0) {};
\draw  node[] at (21,0) {10};
\draw  node[fill= red,circle,scale=2.5] at (20, -2) {};
\draw  node[] at (21,-2) {11};
\draw  node[fill= red,circle,scale=2.5] at (20, -4) {};
\draw  node[] at (21,-4) {12};

\end{tikzpicture}

%% file: diagrams/step2intraA.tex
\begin{tikzpicture}[scale=0.7]
\draw[ultra thick, violet, dashed] (0,0)--(2,0){};
\draw[ultra thick,red] (0,5) --(0,-5){};
\draw  node[fill= red,circle,scale=2.5] at (0, 4) {};
\draw  node[] at (-1,4) {1};
\draw  node[fill= red,circle,scale=2.5] at (0, 2) {};
\draw  node[] at (-1,2) {2};
\draw  node[fill= red,circle,scale=2.5] at (0, 0) {};
\draw  node[] at (-1,0) {3};
\draw  node[fill= red,circle,scale=2.5] at (0, -2) {};
\draw  node[] at (-1,-2) {4};
\draw  node[fill= red,circle,scale=2.5] at (0, -4) {};
\draw  node[] at (-1,-4) {5};

\draw[ultra thick,orange] (2,0) --(4,0){};
\draw  node[fill =orange,circle,scale=2.5] at (2, 0) {};
\draw  node[] at (2,1) {6};
\draw  node[fill =orange,circle,scale=2.5] at (4, 0) {};
\draw  node[] at (4,1) {7};

\draw[ultra thick,red] (6,5) --(6,-5){};
\draw  node[fill= red,circle,scale=2.5] at (6, 4) {};
\draw  node[] at (7,4) {8};
\draw  node[fill= red,circle,scale=2.5] at (6, 2) {};
\draw  node[] at (7,2) {9};
\draw  node[fill= red,circle,scale=2.5] at (6, 0) {};
\draw  node[] at (7,0) {10};
\draw  node[fill= red,circle,scale=2.5] at (6, -2) {};
\draw  node[] at (7,-2) {11};
\draw  node[fill= red,circle,scale=2.5] at (6, -4) {};
\draw  node[] at (7,-4) {12};

\draw node[] at (10,2.5){step 2};
\draw[thick,->] (8,2) --(12,2){};

\draw node[] at (10,-2.5){step 2 reverse};
\draw[thick,<-] (8,-2) --(12,-2){};

\draw[ultra thick, violet, dashed] (14,0)--(16,0){};
\draw[ultra thick,violet, dashed] (18,0)--(20,0){};
\draw[ultra thick,red] (14,5) --(14,-5){};
\draw  node[fill= red,circle,scale=2.5] at (14, 4) {};
\draw  node[] at (13,4) {1};
\draw  node[fill= red,circle,scale=2.5] at (14, 2) {};
\draw  node[] at (13,2) {2};
\draw  node[fill= red,circle,scale=2.5] at (14, 0) {};
\draw  node[] at (13,0) {3};
\draw  node[fill= red,circle,scale=2.5] at (14, -2) {};
\draw  node[] at (13,-2) {4};
\draw  node[fill= red,circle,scale=2.5] at (14, -4) {};
\draw  node[] at (13,-4) {5};

\draw[ultra thick,orange] (16,0) --(18,0){};
\draw  node[fill =orange,circle,scale=2.5] at (16, 0) {};
\draw  node[] at (16,1) {6};
\draw  node[fill =orange,circle,scale=2.5] at (18, 0) {};
\draw  node[] at (18,1) {7};

\draw[ultra thick,red] (20,5) --(20,-5){};
\draw  node[fill= red,circle,scale=2.5] at (20, 4) {};
\draw  node[] at (21,4) {8};
\draw  node[fill= red,circle,scale=2.5] at (20, 2) {};
\draw  node[] at (21,2) {9};
\draw  node[fill= red,circle,scale=2.5] at (20, 0) {};
\draw  node[] at (21,0) {10};
\draw  node[fill= red,circle,scale=2.5] at (20, -2) {};
\draw  node[] at (21,-2) {11};
\draw  node[fill= red,circle,scale=2.5] at (20, -4) {};
\draw  node[] at (21,-4) {12};

\end{tikzpicture}

%% file: diagrams/step2intraB.tex
\begin{tikzpicture}[scale=0.7]
\draw[ultra thick, cyan, dashed] (0,0)--(2,0){};
\draw[ultra thick,blue] (0,5) --(0,-5){};
\draw  node[fill= blue,circle,scale=2.5] at (0, 4) {};
\draw  node[] at (-1,4) {1};
\draw  node[fill= blue,circle,scale=2.5] at (0, 2) {};
\draw  node[] at (-1,2) {2};
\draw  node[fill= blue,circle,scale=2.5] at (0, 0) {};
\draw  node[] at (-1,0) {3};
\draw  node[fill= blue,circle,scale=2.5] at (0, -2) {};
\draw  node[] at (-1,-2) {4};
\draw  node[fill= blue,circle,scale=2.5] at (0, -4) {};
\draw  node[] at (-1,-4) {5};

\draw[ultra thick,orange] (2,0) --(4,0){};
\draw  node[fill =orange,circle,scale=2.5] at (2, 0) {};
\draw  node[] at (2,1) {6};
\draw  node[fill =orange,circle,scale=2.5] at (4, 0) {};
\draw  node[] at (4,1) {7};

\draw[ultra thick,blue] (6,5) --(6,-5){};
\draw  node[fill= blue,circle,scale=2.5] at (6, 4) {};
\draw  node[] at (7,4) {8};
\draw  node[fill= blue,circle,scale=2.5] at (6, 2) {};
\draw  node[] at (7,2) {9};
\draw  node[fill= blue,circle,scale=2.5] at (6, 0) {};
\draw  node[] at (7,0) {10};
\draw  node[fill= blue,circle,scale=2.5] at (6, -2) {};
\draw  node[] at (7,-2) {11};
\draw  node[fill= blue,circle,scale=2.5] at (6, -4) {};
\draw  node[] at (7,-4) {12};

\draw node[] at (10,2.5){step 2};
\draw[thick,->] (8,2) --(12,2){};

\draw node[] at (10,-2.5){step 2 reverse};
\draw[thick,<-] (8,-2) --(12,-2){};;

\draw[ultra thick, cyan, dashed] (14,0)--(16,0){};
\draw[ultra thick,cyan, dashed] (18,0)--(20,0){};
\draw[ultra thick,blue] (14,5) --(14,-5){};
\draw  node[fill= blue,circle,scale=2.5] at (14, 4) {};
\draw  node[] at (13,4) {1};
\draw  node[fill= blue,circle,scale=2.5] at (14, 2) {};
\draw  node[] at (13,2) {2};
\draw  node[fill= blue,circle,scale=2.5] at (14, 0) {};
\draw  node[] at (13,0) {3};
\draw  node[fill= blue,circle,scale=2.5] at (14, -2) {};
\draw  node[] at (13,-2) {4};
\draw  node[fill= blue,circle,scale=2.5] at (14, -4) {};
\draw  node[] at (13,-4) {5};

\draw[ultra thick,orange] (16,0) --(18,0){};
\draw  node[fill =orange,circle,scale=2.5] at (16, 0) {};
\draw  node[] at (16,1) {6};
\draw  node[fill =orange,circle,scale=2.5] at (18, 0) {};
\draw  node[] at (18,1) {7};

\draw[ultra thick,blue] (20,5) --(20,-5){};
\draw  node[fill= blue,circle,scale=2.5] at (20, 4) {};
\draw  node[] at (21,4) {8};
\draw  node[fill= blue,circle,scale=2.5] at (20, 2) {};
\draw  node[] at (21,2) {9};
\draw  node[fill= blue,circle,scale=2.5] at (20, 0) {};
\draw  node[] at (21,0) {10};
\draw  node[fill= blue,circle,scale=2.5] at (20, -2) {};
\draw  node[] at (21,-2) {11};
\draw  node[fill= blue,circle,scale=2.5] at (20, -4) {};
\draw  node[] at (21,-4) {12};

\end{tikzpicture}